\documentclass[twocolumn,aps,amssymb,noshowpacs,superscriptaddress,nofootinbib]{revtex4}

\usepackage{amsmath,amssymb}
\usepackage{graphicx}
\usepackage{verbatim}
\usepackage{amsfonts}
\usepackage{dcolumn}
\usepackage{bm}
\usepackage{color}
\usepackage[colorlinks,citecolor=blue]{hyperref}

\usepackage{mathrsfs}

\newcommand{\beq}{\begin{eqnarray} }
\newcommand{\eeq}{\end{eqnarray} }
\newcommand{\Beq}{\begin{eqnarray*} }
\newcommand{\Eeq}{\end{eqnarray*} }
\newcommand{\Bmat}{\left(\begin{matrix}}
\newcommand{\Emat}{\end{matrix}\right)}

\begin{document}

\title{Irreducible Projective Representations and Their Physical Applications}

\author{Jian Yang} 
\affiliation{International Center for Quantum Materials, School of Physics, Peking
University, Beijing, 100871, China}
\affiliation{Department of physics, Tsinghua University, Beijing 100084, China.}
\author{Zheng-Xin Liu} \email{liuzxphys@ruc.edu.cn}
\affiliation{Department of physics, Renmin University, Beijing 100876, China.}

\date{\today}
\begin{abstract}

An eigenfunction method is applied to reduce the regular projective representations (Reps) of finite groups to obtain their irreducible projective Reps. 
Anti-unitary groups are treated specially, where the decoupled factor systems and modified Schur's lemma are introduced. 
We discuss the applications of irreducible Reps in many-body physics. It is shown that in symmetry protected topological phases, geometric defects or symmetry defects may carry projective Rep of the symmetry group; while in symmetry enriched topological phases, intrinsic excitations (such as spinons or visons) may carry projective Rep of the symmetry group. We also discuss the applications of projective Reps in problems related to spectrum degeneracy, such as in search of models without sign problem in quantum Monte Carlo Simulations.


\end{abstract}

\maketitle

\tableofcontents

\section{Introduction}\label{sec1}

Group theory has wide applications in contemporary physics, including particle physics, quantum field theory, gravitational theory and condensed matter physics. There are mainly two types of groups in physics, the symmetry groups and the gauge groups, where the group elements correspond to global or local operations respectively. The linear Rep theory of groups is one of the fundamental mathematical tools in quantum physics. For example, the Hilbert space of orbital (or integer spin) angular momentum forms linear Rep space of $SO(3)$ rotational symmetry group, while charged particles carry Reps of $U(1)$ gauge group. On the other hand, projective Reps of groups were less known to physicists \cite{Schur11,MorHam,Rudra65,BradleyCrack,Janssen72,Birman,M.Saeed85,Aguilar99,Stephen04,BasJanssens15,Dangniam15,TMats16}.
In projective Reps, the representing matrices of group elements obey the group multiplication rule up to $U(1)$ phase factors. These $U(1)$ phase factors are called factor systems of the corresponding projective Rep (see \ref{Factorsys}). In some sense, the theory of projective Reps has closer relationship to quantum theory since quantum states are defined up to global $U(1)$ phase factors. The well known Kramers degeneracy owning to time reversal symmetry is actually a typical example of projective Rep\cite{Kramers30, Winger32}. When acting on operators, the square of time-reversal operator is equal to identity $T^2=E$ and thus defines a two-element group $Z_2^T=\{E,T\}$; however, when acting on a Hilbert space of odd number of electrons, the  square of (the Rep of) the time reversal operator is no longer identity: $\widehat {T}^2=-1$. This nontrivial phase factor $-1$ stands for a nontrivial projective Rep of $Z_2^T$, which guarantees that each energy level is (at least) doubly degenerate. Another typical example is that half-integer spins carry projective representations of $SO(3)$ group.

Projective Rep is a natural tool to describe symmetry fractionalization and is widely used in recently devoleped theories such as  
 Symmetry Protected Topological (SPT) phases\cite{GuWen09,CGLW1,CGLW2} and Symmetry Enriched Topological (SET) phases \cite{HungWen,ChengWang}. These exotic quantum phases are beyond the Laudau symmetry breaking paradigm because they exhibit no long-range correlations of local order parameters. The SPT states are short range entangled and thus carry trivial topological order, while the SET states are long-range entangled\cite{CGW10LU} and thus carry nontrivial topological orders. In these novel quantum states, certain defects (such as boundaries or symmetry fluxes) or elementary excitations (called anyons) carry projective Reps of the symmetry group.
Especially, one-dimensional SPT phases with an on-site symmetry group $G$ are characterized by projective Reps \cite{Pollmann10} of $G$ and therefore classified by the second group cohomology $\mathcal H^2(G, U(1))$ \cite{CGW1D1,CGW1D2}.

Most of the groups in quantum physics are unitary. A group is unitary if every group element stands for a unitary operator which keep the inner product of any two states $\langle\psi_1|\psi_2\rangle$ invariant, namely,
\[
\langle g\psi_1|g\psi_2\rangle=\langle\psi_1|\psi_2\rangle.
\]
where $|g\psi_{1,2}\rangle =\hat g|\psi_{1,2}\rangle$ for any group element $g$. 
On the other hand, a group is called anti-unitary (also called `antiunitary'\cite{Isao1974,Kim1984} or `nonunitary'\cite{JohnODim63}) if it contains at least one anti-unitary element $g$ (such as the time reversal operator $T$) which transforms the inner product of two states into its complex conjugate,
\[
\langle g\psi_1|g \psi_2\rangle=\langle\psi_1|\psi_2\rangle^*=\langle\psi_2|\psi_1\rangle.
\]
Generally, the anti-unitary operator $\hat g$ corresponding to the anti-unitary element $g$ can be written as $\hat{g}=\widehat U K$, where $\widehat U$ is a unitary operator and $K$ is the complex-conjugate operator satisfying
\beq\label{K}
Ka=a^{\ast}K
\eeq
for any complex number $a$. Anti-unitary elements also differ from unitary ones by their nontrivial module as defined later in Eq.(\ref{module}). In the case that an anti-unitary group has the same group table with that of a unitary group, one can distinguish them by checking the existence of anti-unitary elements (such as the anti-unitary time-reversal group $Z_2^T=\{E,T\}$ and unitary spatial-inversion group $Z_2^P=\{E, P\}$). The simplest anti-unitary group is the two-element group $Z_2^T=\{E,T\}$, where the time reversal operator $T$ is anti-unitary.

Anti-unitary groups play important roles in quantum theory. For example, the Schr\"{o}dinger equations and the Dirac equations for free particles are  invariant under the time-reversal symmetry group $Z_2^T$ if there are no magnetic fields. Under time-reversal transformation $T$, in addition to $t\rightarrow -t$, the equations should be transformed into their complex conjugation.
The importance of anti-unitary groups can also be seen from the well known CPT theorem\cite{CPT}. In condensed matter physics,  anti-unitary symmetry groups are also important. For instance, the magnetic point groups and magnetic space groups are anti-unitary groups\cite{magnetic}; the well studied topological insulators\cite{TI1,TI2,TI3,TI4,TI5, TI6} are SPT phases protected by $U(1)$ charge conservation symmetry and time reversal symmetry. Many properties of unitary groups cannot be straightforwardly generalized to anti-unitary groups. For example, the classification of SPT phases with anti-unitary groups in three dimensions is different from that with unitary groups\cite{senthilT, Kapustin}. Comparing to unitary groups \cite{LevinGu}, it is also difficult to gauge a global anti-unitary symmetry into a local symmetry\cite{ChenXieT}. Since anti-unitary groups are special, a systematic method of obtaining their linear Reps (also called co-representations \cite{JohnODim63}) and projective Reps (also called co-ray representations\cite{MVMurthy66}) is urgent.

Similar to the linear Rep theory, only irreducible projective Reps are of physical interest. So it is an important issue to find a systematic way to obtain all the irreducible projective Reps. 
The irreducible projective Reps of a group $G$ can be obtained from the linear Reps of its covering groups (also called representation groups\cite{BYLGreen}). The covering groups are central extensions of $G$ and are also classified by the second group cohomology. For example, the double point group in spin-orbit coupled systems is a covering group of the corresponding point group. For a given factor system, the corresponding covering group can be constructed in the following steps: (1) transform the factor system into its standard form where all the $U(1)$ phase factors are the $N$th roots of 1, here $N$ is the period of the factor system (see Sec.\ref{app:Dim_Period}); (2) split each group element $g$ in $G$ into $N$ elements, $\eta g, \eta^2 g, ... , \eta^N g$, where $\eta^N=1$; (3) taking into account the factor system, the new elements obey a new multiplication rule under which the $G\times N$ elements (here $G$ is the order of group $G$) form a new group $G'$ satisfying
$G'/Z_N=G$. This extended group $G'$ is the covering group corresponding to the given factor system. Then from the irreducible linear Rep of $G'$ and the many-to-one mapping from $G'$ to $G$, one can obtain irreducible projective Rep of the quotient group $G$ with the given factor system. Since the covering group $G'$ is larger than $G$, this method is very indirect and will not be discussed in detail.

Alternatively, we can obtain the irreducible projective Reps from the regular projective Reps without referring to the covering group. Useful knowledge can be learned from linear Reps of groups, where all irreducible Reps can be obtained by reducing the regular Reps. A remarkable eigenfunction method was introduced by J.-Q. Chen\cite{ChenJQRMP85,ChenJQ02} to reduce the regular Reps. In this approach the Rep theory is handled in a physical way. The main idea is to label the irreducible bases by non-degenerate quantum numbers, namely, the eigenvalues of a set of commuting operators. So the main task is to find the Complete Set of Commuting Operators (CSCO) of the Rep space.  By making use of the class operators of the group $G$ and those of the canonical subgroup chain of $G$, together with the class operators of the canonical subgroup chain of the intrinsic group $\bar G$, the regular Reps of finite groups and the tensor Reps of compact Lie groups (such as $U(n)$) are successfully reduced. Chen \cite{ChenJQ02} applied this method to obtain irreducible Reps of space groups, and as a byproduct, part of the projective Reps of point groups were obtained. In this paper, we will generalize the eigenfunction method to reduce the regular projective Reps of finite groups, especially {\it anti-unitary groups}. The factor systems of projective Reps are classified by group cohomology and can be obtained by solving the 2-cocycle equations. For anti-unitary groups we show that the factor system can be decoupled into two parts, one contains the information of the quotient group $Z_2^T$ while other part contains the information of the unitary normal subgroup.  The introduction of decoupled factor systems is an important step which greatly simplifies the calculations. The regular projective Reps are then obtained by acting the group elements on the group space itself, and the matrix elements are the 2-cocycles.
After finding out the CSCO and their eigenfunctions of the regular projective Reps, all the irreducible projective Reps are obtained. In this approach, we only need to treat matrices with maximum dimension $G$, so it is much simpler than the covering group method.

The physical applications of projective Reps are then discussed. Some new viewpoints are presented, for instance, the Majorana zero modes in topological superconductors are explained as projective Reps of some symmetry groups; some models which are free of sign problem under quantum Monte Carlo simulations are interpreted as the properties of projective Reps of anti-unitary groups, and generalizations (i.e. possible new classes of sign-free models) are proposed.




The paper is organized as follows. In section \ref{regularProj}, we introduce the projective Reps and the factor systems, and the regular projective Reps of finite groups. In section \ref{sec4}, we apply the eigenfunction method to reduce the regular projective Reps into irreducible ones. Unitary groups and anti-unitary groups are discussed separately and the results of some finite groups are listed in Table \ref{tb1}. Readers who are more interested in the applications of projective Reps in concrete physical problems can skip this part and go to section \ref{sec:apply} directly, where  
symmetry fractionalizations in topological phases of matter, sign problems in quantum Monte Carlo simulations, and other topics related to spectrum degeneracy are discussed.
Section \ref{sec:sum} is devoted to the conclusion and discussion.


\section{Regular Projective Reps}\label{regularProj}

\subsection{Factor systems and 2-cocycles}\label{Factorsys}

{\bf Projective Reps and the factor systems}. We first consider $U(1)$-coefficient projective Reps of a finite unitary group $G$. Later we will also discuss $\mathcal A$-coefficient projective Reps, where $\mathcal A$ is a finite subgroup of $U(1)$.

A projective Rep of $G$ is a map from the  element $g\in G$ to a matrix $M(g)$ such that for any pair of elements $g_1, g_2\in G$,
 \begin{eqnarray}\label{Def_Proj}
 M(g_1)M(g_2)=M(g_1g_2)e^{i\theta_2(g_1,g_2)},
 \end{eqnarray}
where the $U(1)$ phase factor $\omega_2(g_1,g_2) = e^{i\theta_2(g_1,g_2)}$ is a function of two group variables and is called the factor system. If $\omega_2(g_1,g_2) = 1$ for any $g_1,g_2\in G$, then above projective Rep is trivial, namely, it is a linear Rep.

The associativity relation of matrix multiplication yields constraints on the factor system. For any three elements $g_1,g_2,g_3\in G$,
\Beq
M(g_1)M(g_2)M(g_3)&=&[M(g_1)M(g_2)]M(g_3)\\
&=&M(g_1g_2g_3)e^{i\theta_2(g_1,g_2)}e^{i\theta_2(g_1g_2,g_3)}\\
&=&M(g_1)[M(g_2)M(g_3)]\\
&=&M(g_1g_2g_3)e^{i\theta_2(g_2,g_3)}e^{i\theta_2(g_1,g_2g_3)},
\Eeq
which requires that
\beq\label{factor}
\omega_2(g_1,g_2)\omega_2(g_1g_2,g_3) = \omega_2(g_2,g_3)\omega_2(g_1,g_2g_3),
\eeq
or equivalently
\Beq
\theta_2(g_1,g_2) + \theta_2(g_1g_2,g_3) = \theta_2(g_2,g_3) + \theta_2(g_1,g_2g_3),
\Eeq
where the equal sign means equal mod $2\pi$. The factor system of a projective Rep must satisfy equation (\ref{factor}). Conversely, all the solutions of above equation correspond to the factor system of a projective Rep.

If we introduce a `gauge transformation' to the projective Rep,
\beq\label{gauge}
M'(g) = M(g)e^{i\theta_1(g)},
\eeq
where the phase factor $\Omega_1(g)=e^{i\theta_1(g)}$ depends on a single group variable, then
\Beq
M'(g_1)M'(g_2) &=& M(g_1g_2)e^{i\theta_2(g_1,g_2)}e^{i\theta_1(g_1)+i\theta_1(g_2)}\\
&=& M'(g_1g_2)e^{i\theta'_2(g_1,g_2)},
\Eeq
where the factor system transforms into
\Beq\label{gaugetheta}
e^{i\theta'_2(g_1,g_2)}=e^{i\theta_2(g_1,g_2)}{e^{i\theta_1(g_1)+i\theta_1(g_2)}\over e^{i\theta_1(g_1g_2)}},
\Eeq
namely,
\beq\label{gaugeomega}
\omega'_2(g_1,g_2)=\omega_2(g_1,g_2)\Omega_2(g_1,g_2),
\eeq
with
\beq\label{2cob}
\Omega_2(g_1,g_2) = {\Omega_1(g_1)\Omega_1(g_2)\over \Omega_1(g_1g_2)}
\eeq
where $\Omega_1(g)=e^{i\theta_1(g)}$. Since $M'(g)$ can be adiabatically transformed into $M(g)$ by continuously adjusting the phase $\Omega_1(g)$, the two projective Reps are considered to be equivalent. Generally, if the factor systems of any two projective Reps $M(g)$ and $M'(g)$ are related by the relation (\ref{gaugeomega}), then $M(g)$ and $M'(g)$ are considered to belong to the same class of projective Reps (even if their dimensions are different).

Now we consider {\it anti-unitary} groups. If a group $G$ is anti-unitary, then half of its group elements are anti-unitary, and the remaining unitary elements form a normal subgroup $H$,
\[
G/H\simeq Z_2^T
\]
with $Z_2^T=\{E,\mathbb T\}$ and $\mathbb T^2=E$. There are two types of anti-unitary groups. In Type-I anti-unitary groups, $\mathbb T$ can be chosen as an element of $G$, in this case the group $G$ is either a product group $G=H\times Z_2^T$ or a semi-product group $G=H\rtimes Z_2^T$. In type-II anti-unitary groups, $\mathbb T\notin G$ and the period of any anti-unitary elements in $G$ is at least 4. More rigorous definitions of type-I and type-II anti-unitary groups are given in appendix \ref{standardcyc}. In this work, we will mainly focus on type-I anti-unitary groups. Only some simple fermionic groups of type-II will be discussed.

Supposing $g$ is a group element in $G$, then it is represented by $M(g)$ if $g$ is a unitary element and represented by $M(g)K$ if $g$  is anti-unitary, where $K$ is the complex-conjugate operator shown in Eq.\eqref{K}.
The multiplication of projective Reps of $g_1, g_2$ depends on if they are unitary or anti-unitary. There are four cases:

(A) both $g_1,g_2$ are unitary, then we obtain
\[
M(g_1)M(g_2)=M(g_1g_2)e^{i\theta_2(g_1,g_2)};
\]
which is the same as Eq.\eqref{Def_Proj};

(B) both $g_1,g_2$ are anti-unitary, then the result is
\[
 M(g_1)KM(g_2)K=M(g_1)M^*(g_2)=M(g_1g_2)e^{i\theta_2(g_1,g_2)};
\]

(C) if $g_1$ is unitary while $g_2$ is anti-unitary, then the result is
\[
 M(g_1)M(g_2)K=M(g_1g_2)e^{i\theta_2(g_1,g_2)}K;
\]

(D) if $g_1$ is anti-unitary while $g_2$ is unitary, then the result is
\[
 M(g_1)KM(g_2) = M(g_1)M^*(g_2)K= M(g_1g_2)e^{i\theta_2(g_1,g_2)}K.
\]

If we define
\[
s(g)=\left\{
\begin{aligned}
&1,& &{\ \rm if} \ g \ {\rm is\ unitary,\ \ \ }   \\
&-1,& &{\ \rm if}\ g \ {\rm is\ antiunitary,\ \ \ }
\end{aligned}
\right.
\]
and define the corresponding operator $K_{s(g)}$ as
\[
K_{s(g)}=\left\{
\begin{aligned}
&I,& &{\ \rm if\ } s(g)=1,\  \  \\
&K,& &{\ \rm if\ }s(g)=-1,
\end{aligned}
\right.
\]
then the above four cases (A)$\sim$(D) can be unified as a single equation
\[
M(g_1)K_{s(g_1)}M(g_2)K_{s(g_2)} = M(g_1g_2)e^{i\theta_2(g_1,g_2)}K_{s(g_1g_2)}.
\]

Substituting above results into the associativity relation of the sequence of operations $g_1\times g_2\times g_3$, similar to (\ref{factor}), we can obtain
\Beq
&&M(g_1)K_{s(g_1)}M(g_2)K_{s(g_2)}M(g_3)K_{s(g_3)} \ \\
&=& M(g_1g_2g_3)\omega_2(g_1,g_2)\omega_2(g_1g_2,g_3) K_{s(g_1g_2g_3)}\\
&=& M(g_1g_2g_3)\omega_2(g_1,g_2g_3)\omega_2^{s(g_1)}(g_2,g_3)K_{s(g_1g_2g_3)},
\Eeq
namely,
\beq\label{2cocyl}
\omega_2(g_1,g_2)\omega_2(g_1g_2,g_3) = \omega_2^{s(g_1)}(g_2,g_3)\omega_2(g_1,g_2g_3).
\eeq
Eq.(\ref{2cocyl}) is the general relation that the factor systems of any finite group (no matter unitary or anti-unitary) should satisfy.

Similar to Eq.(\ref{gauge}), if we introduce a gauge transformation $M'(g)K_{s(g)}=M(g)\Omega_1(g)K_{s(g)}$, then the factor system changes into
\beq\label{gaugeomega2}
\omega'_2(g_1,g_2)=\omega_2(g_1,g_2)\Omega_2(g_1,g_2),
\eeq
with
\beq\label{2cob2}
\Omega_2(g_1,g_2) = {\Omega_1(g_1)\Omega_1^{s(g_1)}(g_2)\over \Omega_1(g_1g_2)}.
\eeq
The equivalent relations (\ref{gaugeomega2}) and (\ref{2cob2}) define the equivalent classes of the solutions of (\ref{2cocyl}). The number of equivalent classes for a finite group is usually finite.

{\bf The 2$^{\rm nd}$ group cohomology and the 2-cocycles.}  Actually, the equivalent classes of the factor systems associated with the projective Reps of group $G$ form a group, called the second group cohomology. The group cohomology $\{{\rm Kernel}\ d/{\rm Image}\ d\}$ is defined by the coboundary operator $d$ (for details see appendix \ref{app1}),
\begin{eqnarray}
& & (d\omega_{n})(g_{1},\ldots,g_{n+1}) \nonumber\\
&& =[g_{1}\cdot\omega_{n}(g_{2},\ldots,g_{n+1})]\omega^{(-1)^{n+1}}_{n}(g_{1},\ldots,g_{n})\times  \nonumber\\
&&
\prod^{n}_{i=1}\omega^{(-1)^{i}}_{n}(g_{1},\ldots,g_{i-1},g_{i}g_{i+1},g_{i+2},\ldots,g_{n+1}).
\end{eqnarray}
where $g_1,...,g_{n+1}\in G$ and the variables $\omega_{n}(g_{1},\ldots,g_{n})$ take value in an Abelian group $\mathcal A$ [usually $\mathcal A$ is a subgroup of $U(1)$]. 
The set of variables $\omega_{n}(g_{1},\ldots,g_{n})$ is called a $\mathcal A$-cochain. The module $g\cdot$ is defined by
\begin{eqnarray}\label{module}
g\cdot\omega_{n}(g_{1},\ldots,g_{n}) =\omega_{n}^{s(g)}(g_{1},\ldots,g_{n}).
\end{eqnarray}

With this notation, Eq. (\ref{2cocyl}) can be rewritten as
\Beq
(d\omega_2)(g_1,g_2,g_3)=1,
\Eeq
the solutions of above equations are called 2-cocycles with $\mathcal A$-coefficient. Similarly, Eq. (\ref{2cob2}) can be rewritten as
\Beq
\Omega_2(g_1,g_2)=(d\Omega_1)(g_1,g_2),
\Eeq
where $\Omega_1(g_1),\Omega_1(g_2)\in \mathcal A$ and thus defined $\Omega_2(g_1,g_2)$ are called 2-coboundaries. Two 2-cocycles $\omega_2'(g_1,g_2)$ and $\omega_2(g_1,g_2)$ are equivalent if they differ by a 2-coboundary, see Eq. (\ref{gaugeomega2}). The equivalent classes of the 2-cocycles $\omega_2(g_1,g_2)$ form the second group cohomology $\mathcal H^2(G,\mathcal A)$.

 A projective Rep whose factor system is a $\mathcal A$-coefficient 2-cocycle is called $\mathcal A$-coefficient projective Rep. {\it By default, projective Reps are defined with $U(1)$ coefficient}, namely, $\mathcal A=U(1)$. In this case, $\omega_{2}(g_{1},g_{2})\in U(1)$ so we can write  $\omega_{2}(g_{1},g_{2})=e^{i\theta_{2}(g_{1},g_{2})}$, where $\theta_{2}(g_{1},g_{2})\in[0,2\pi)$. The cocycle equations $(d\omega_{2})(g_{1},g_2,g_{3})=1$ can be written in terms of linear equations,
\begin{eqnarray}
&&s(g_{1})\theta_{2}(g_{2},g_{3})-\theta_{2}(g_{1}g_{2},g_{3})+\theta_{2}(g_{1},g_{2}g_{3})\nonumber\\
&&  -\theta_{2}(g_{1},g_{2})=0.
\label{2cocycle}
\end{eqnarray}
Similarly, if we write $\Omega_1(g_1)=e^{i\theta_1(g_1)}$ and $\Omega_2(g_1,g_2)=e^{i\Theta_2(g_1,g_2)}$, then the 2-coboundary (\ref{2cob2})
can be written as
\beq
\Theta_{2}(g_{1},g_{2})= s(g_{1})\theta_{1}(g_{2})-\theta_{1}(g_{1}g_{2})+\theta_{1}(g_{1}) .
  \label{2coboundary}
\end{eqnarray}
The equal sign in (\ref{2cocycle}) and (\ref{2coboundary}) means equal mod $2\pi$. From these linear equations, we can obtain the classification and the cocycles solutions of each class (for details see appendix \ref{app2}). The classification of the factor system also gives a constraint on the dimensions of corresponding projective Reps, as discussed in appendix \ref{app:Dim_Period}.

{\bf Gauge fixing.} For each class of 2-cocyles satisfying (\ref{2cocycle}), there are infinite number of solutions which differ by 2-coboundaries.
We will adopt the canonical gauge by fixing \[\omega_2(E,g)=\omega_2(g,E)=1.\] However, above canonical gauge condition doesn't completely fix the coboundaries and there are still infinite number of solutions. In later discussion, we will select one of the 2-cocycle solutions in each class as the factor system of the corresponding projective Rep. In that case, the gauge degrees of freedom (i.e. the 2-coboundaries) are completely fixed.

{\bf The case where $\mathcal A$ is a finite abelian group}. Mathematically, the second group cohomology $\mathcal H^2(G,\mathcal A)$ with $\mathcal A$-coefficient classifies the central extensions of $G$ by $\mathcal A$. 
In the default case, $\mathcal A=U(1)$ and the projective Reps of $G$ are classified by $\mathcal H^2(G,U(1))$ and correspond to linear Reps of $U(1)$ extensions of $G$.

Physically, we are also interested in the cases where $\mathcal A$ is a finite abelian group, such as $Z_N$. For example, the topological phases with $Z_2$ topological order and symmetry group $G$ are (partially) classified by $\mathcal H^2(G,Z_2)$ (see section \ref{sec:SET}). The classification of $Z_N$ coefficient projective Reps is usually different from the $U(1)$ coefficient projective Reps (see appendix \ref{app2}), for instance, certain $U(1)$ coboundaries may be nontrivial $Z_N$ cocycles.

Once the factor systems are obtained by solving the cocycle equaitons and coboundary equations (see appendix \ref{app2}), the method of obtaining $\mathcal A$-coefficient irreducible projective Reps is the same with the cases of $U(1)$-coefficient. In the following we will discuss the method of obtaining $U(1)$ coefficient irreducible projective Reps by reducing the regular projective Reps.

\subsection{Regular projective Reps}\label{Regular}

{\bf Regular projective Reps.} For a given factor system, we can easily construct the corresponding regular projective Rep (the regular Rep twisted by the 2-cocycles) using the group space as the Rep space. The group element $g$ is not only an operator $\hat{g}$, but also a basis $|g\rangle$.
The operator $\hat{g}_{1}$ acts on the basis $|g_{2}\rangle$ as the following,
\begin{equation}
\hat{g}_{1}|g_{2}\rangle=e^{i\theta_2(g_{1},g_{2})}K_{s(g_1)}|g_{1}g_{2}\rangle,
\end{equation}
or in matrix form
\beq\label{gaction1}
\hat g_1=M(g_1)K_{s(g_1)},
\eeq
with matrix element
\begin{equation}\label{gmatrix}
M(g_{1})_{g,g_{2}}=\langle g|\hat g_1|g_2\rangle=e^{i\theta_2(g_{1},g_{2})}\delta_{g,g_{1}g_{2}}.
\end{equation}

For any group element $g_{3}$, we have
\begin{eqnarray*}
\hat{g}_{1}\hat{g}_{2}|g_{3}\rangle&=& \hat{g}_{1}\left[e^{i\theta_2(g_{2},g_{3})}K_{s(g_2)}|g_{2}g_{3}\rangle\right] \\
&=&e^{i\theta_2(g_{1},g_{2}g_{3})}K_{s(g_1)}\left[e^{i\theta_2(g_{2},g_{3})}K_{s(g_2)}|g_{1}g_{2}g_{3}\rangle\right] \\
&=&e^{i\theta_2(g_{1},g_{2}g_{3})}e^{is(g_1)\theta_2(g_{2},g_{3})}K_{s(g_1g_2)}|g_{1}g_{2}g_{3}\rangle,
\Eeq
and
\Beq
\widehat{g_{1}g_{2}}|g_{3}\rangle&=&e^{i\theta_2(g_{1}g_{2},g_{3})}K_{s(g_1g_2)}|g_{1}g_{2}g_{3}\rangle.
\end{eqnarray*}
Comparing with the $2$-cocycle equation (\ref{2cocycle}), it is easily obtained that $\hat g_1\hat g_2=e^{i\theta_2(g_1,g_2)} \widehat{g_1g_2}$.  In matrix form, this relation reads
\begin{eqnarray}\label{MMProj}
M(g_{1})K_{s(g_1)}M(g_{2})K_{s(g_2)}= e^{i\theta_2(g_{1},g_{2})}M(g_{1}g_{2})K_{s(g_1g_2)}.\nonumber\\
\end{eqnarray}
Eq.(\ref{MMProj}) indicates that $M(g)K_{s(g)}$ is indeed a projective Rep of the group $G$. For the trivial 2-cocycle where $e^{i\theta_2(g_{1},g_{2})}=1$ for all $g_1, g_2\in G$, $M(g)K_{s(g)}$ reduces to the regular linear Rep of $G$.

{\bf Intrinsic Regular projective Reps.} In order to reduce the regular projective Rep of $G$, we will make use of the intrinsic group $\bar G$. Each element $g$ in $G$ corresponds to an intrinsic group element $\bar g$ in $\bar G$, and the intrinsic elements obey right-multiplication rule with $\bar g_1 g_2=g_2g_1$ and $\bar g_1\bar g_2=\overline{g_2g_1}$.
Obviously, the intrinsic group $\bar G$ commutes with $G$ since $(\bar g_1g_2)g_3 = (g_2\bar g_1) g_3 =  g_2g_3g_1$.
Corresponding to $\bar{G}$, we can define the intrinsic regular projective Rep,
\begin{eqnarray}\label{intrgaction}
\hat{\bar{g}}_{1} |g_{2}\rangle =  e^{i\theta_2(g_{2},g_{1})}|g_{2}g_{1}\rangle,
\end{eqnarray}
or in matrix form $\hat {\bar g}_1=M(\bar g_1)$ with
\begin{equation}\label{intrgmatrix}
M(\bar{g}_{1})_{g,g_{2}}=e^{i\theta_2(g_{2},g_{1})}\delta_{g,g_{2}g_{1}}.
\end{equation}
The complex-conjugate operator $K$ is absent in (\ref{intrgaction}), indicating that {\it the intrinsic projective Reps for anti-unitary groups are essentially unitary.}

For any group element $g_{3}$, we have
\begin{eqnarray*}
\hat{\bar{g}}_{1}\hat{\bar{g}}_{2} |g_{3}\rangle&=& \hat{\bar{g}}_{1}e^{i\theta_2(g_{3},g_{2})}|g_{3}g_{2}\rangle\\
&=& e^{i\theta_2(g_{3},g_{2})}e^{i\theta_2(g_{3}g_{2},g_{1})}|g_{3}g_{2}g_{1}\rangle
\Eeq
and
\Beq
\widehat{\overline{g_{2}g_{1}}} |g_{3}\rangle&=& e^{i\theta_2(g_{3},g_{2}g_{1})}|g_{3}g_{2}g_{1}\rangle
\end{eqnarray*}
Comparing with the $2$-cocycle equation (\ref{2cocycle}), we find that
\beq\label{intr_proj}
\hat{\bar{g}}_1\hat{\bar{g}}_2|g_3\rangle =\widehat{\overline{g_2g_1}}e^{is(g_{3})\theta_2(g_{2},g_{1})}|g_3\rangle,
\eeq
which means that if the state $|g_3\rangle$ is anti-unitary (unitary), then the coefficients of the operators on its left will take a complex conjugate (remain unchanged).
In matrix form, above relation can be written as
\begin{eqnarray*}
 M(\bar{g}_{1})M(\bar{g}_{2})  =  M(\overline{g_{2}g_{1}})\left[e^{i\theta_2(g_{2},g_{1})}P_u + e^{-i\theta_2(g_{2},g_{1})}P_a\right],
\end{eqnarray*}
where $P_u (P_a)$ is the projection operator projecting onto the Hilbert space formed by unitary (anti-unitary) bases.

Now we show that {\it the regular projective Rep of $G$ commutes with its intrinsic regular projective Rep}. On the one hand,
\Beq
\hat{\bar g}_1\hat g_2 |g_3\rangle &=& \hat{\bar g}_1 \omega_2(g_2,g_3)K_{s(g_2)}|g_2g_3\rangle\\
&=&\omega_2(g_2g_3,g_1) \omega_2(g_2,g_3)K_{s(g_2)}|g_2g_3g_1\rangle.
\Eeq
On the other hand,
\Beq
\hat g_2\hat{\bar g}_1|g_3\rangle &=& \hat g_2\omega_2(g_3,g_1)|g_3g_1\rangle\\
&=&\omega_2^{s(g_2)}(g_3,g_1)\omega_2(g_2,g_3g_1)K_{s(g_2)}|g_2g_3g_1\rangle.
\Eeq
Comparing with the cocycle equation $(d\omega_2)(g_2,g_3,g_1)=1$, we have
\[
\hat{\bar g}_1\hat g_2 =\hat g_2\hat{\bar g}_1.
\]
In matrix form, above equation reads
\[
M(\bar g_1)M(g_2)K_{s(g_2)}=M(g_2)K_{s(g_2)}M(\bar g_1).
\]

\section{CSCO and irreducible projective Reps }\label{sec4}

Supposing a group element $g\in G$ is represented by $M(g)K_{s(g)}$, we assume that all the Rep matrices $M(g)$ are unitary (for finite groups all the Reps can be transformed into this form). If a projective Rep can not be reduced by unitary transformations into direct sum of lower dimensional Reps, then it is called irreducible projective Rep. If two irreducible projective Reps $M(g)K_{s(g)}$ and $M'(g)K_{s(g)}$ can be transformed into each other by a unitary matrix $U$,
\beq\label{UnitTrsf}
M'(g)K_{s(g)}=U^\dag M(g)K_{s(g)} U,
\eeq
then $M'(g)K_{s(g)}$ and $M(g)K_{s(g)}$ are said to be the same projective Rep; otherwise, they are two different Reps.

On the other hand, if two different irreducible projective Reps have the same (class of) factor systems, they belong to the same class. Obviously, any one-dimensional Rep of a group must be a trivial projective Rep since it is gauge equivalent to the identity Rep (except for the cases where the coefficient group $\mathcal A$ is a finite group).

If $M(g)K_{s(g)}$ is a projective Rep of group $G$ with factor system $\omega_2(g_1,g_2)$, then its complex conjugate $M^*(g)K_{s(g)}$ is also a projective Rep whose factor system is $\omega_2^*(g_1,g_2)=\omega_2^{-1}(g_1,g_2)$.

It is known that all the irreducible linear Reps of a finite group $G$ can be obtained by reducing its regular Rep. Similar idea can be applied for projective Reps.  In section \ref{regularProj}, we have constructed the regular projective Rep with a given factor system(or 2-cocycle). In this section we will generalize the eigenfunction method to reduce the regular projective Reps into irreducible ones.

As mentioned, we will completely fix the 2-coboundary by selecting one solution in each class of 2-cocyles (for anti-unitary groups we
will transform the 2-cocyles into the decoupled factor system as defined in Appendix \ref{standardcyc}).

Suppose that we choose a different 2-coboundary [see(\ref{gaugeomega2}) and (\ref{2cob2})],
then the corresponding regular projective Rep is given by $\hat g_1=W(g_1)K_{s(g_1)}$ with
\beq\label{W_regular}
W(g_1)_{g,g_2}=\delta_{g,g_1g_2}\omega_2(g_1,g_2){\Omega_1(g_1)\Omega_1^{s(g_1)}(g_2)\over \Omega_1(g_1g_2)}.
\eeq
Above $W(g_1)K_{s(g_1)}$ is equivalent to $M(g_1)K_{s(g_1)}$ defined in (\ref{gaction1}) since they are related by a unitary transformation $U$ followed by a gauge transformation $\Omega_1(g_1)$,
\beq\label{U_transf}
W(g_1)K_{s(g_1)}=\Omega_1(g_1)[U^\dag M(g_1)K_{s(g_1)}U],
\eeq
where $U$ is a diagonal matrix with entries $U_{g,g'}=\delta_{g,g'}\Omega_1(g)$.  Similarly, the intrinsic regular projective Rep (\ref{intrgmatrix}) becomes $\hat{\bar g}_1=W(\bar g_1)$ with
\beq \label{intr U trans}
W(\bar g_1) = U^\dag M(\bar g_1) U\left[\Omega_1(g_1) P_u +\Omega_1^*(g_1) P_a\right] .
\eeq
Therefore, we can safely select one 2-cocyle in each class to construct the regular projective Rep.

Some tools used in linear Reps, such as class operators and characters, can be introduced for the projective Reps. For unitary groups, the character $\chi_i^{(\nu)}$ of an irreducible (projective) Rep is a function of class operators (each conjugate class gives rise to a class operator)
\beq\label{ClassO}
C_i = \sum_{g\in G}g^{-1}g_i g,
\eeq
here we have ignored a normalization constant. On the other hand, for a given class operator $C_i$ the character $\chi_i^{(\nu)}$ is a function of irreducible (projective) Reps $(\nu)$ since the unitary transformation (\ref{UnitTrsf}) will not change the character. The characters $\chi_i^{(\nu)}$ form complete bases for both the class space $\{C_i\}$ and the irreducible (projective) Rep space $\{(\nu)\}$. As a result, the number of different irreducible (projective) Reps is equal to  $N_c$, the number of independent class operators.  Group theory also tells us that for a finite group $G$, a $n_\nu$ dimensional irreducible Rep appears $n_\nu$ times in the reduced regular Rep. Consequently, \[\sum_{\nu=1}^{N_c} n_\nu^2=G,\] where $G$ is the order of the group $G$. This result also holds for the projective Reps of unitary groups.

However, we should carefully use these tools for anti-unitary groups. For example, the `character' of the anti-unitary element $g$ may be changed under unitary transformation
\Beq
{\rm Tr}\ [M(g)K]\neq {\rm Tr}\ [U^\dag M(g)KU]={\rm Tr} [U^\dag M(g)U^*]
\Eeq
for arbitrary unitary matrix $U$. Similarly, the two unitary elements $g$ and $\tilde g=T^{-1}gT$, which belong to the same class, may have different characters. 
Therefore we need to redefine the conjugate classes and the class operators such that the number of different irreducible Reps (of the same class) is still equal to the number of independent class operators.
Similarly, some other conclusions of unitary groups should be modified for anti-unitary groups (see Sec.\ref{sec:antiU}).

\subsection{Brief review of the reduction of regular linear Reps for unitary groups}

First we introduce Schur's lemma and its corollary for  unitary groups without proving them. They are also valid for irreducible projective Reps of unitary groups.

{\it\bf Schur's lemma:} If a nonzero matrix $C$ commutes with all the irreducible Rep matrices $M^{(\nu)}(g)$ of a unitary group $G$, namely,
\[
CM^{(\nu)}(g)=M^{(\nu)}(g)C,
\]
for $g\in G$, then $C$ must be a constant matrix $C=\lambda I$.

{\it\bf Corollary:} If a unitary operator $\widehat C$ commutes with a unitary group $G$ with $[\hat g,\widehat C]=0$ for all $g\in G$,
then an eigenspace of $C$ (or the surporting space of Jordan blocks with the same eigenvalue) is a Rep space of $G$.

In most cases, the operator $C$ (such as the class operators discussed below) is diagonalizable and the Hilbert space can be reduced as direct sum of its eigenspaces. However, it is possible that $C$ cannot be completely diagonalized, in this case $C$ can be transformed into its Jordan normal form\cite{CJWu05} and we can just replace each `eigenspace' by a `supporting space of Jordan blocks with the same eigenvalue'.

Reducing the regular Rep is equivalent to identifying all the bases of the irreducible Reps. In Ref.~\onlinecite{ChenJQ02}, the authors use a series of quantum numbers, i.e. the eigenvalues of the complete set of commuting operators (CSCO), to distinguish the irreducible bases.  This idea comes from quantum mechanics. For example, for spin systems we use two quantum numbers $|S,m\rangle$ to label a state, where $S$ is the main quantum number labeling the irreducible Rep and $m$ is the magnetic quantum number labeling different bases. Noticing that $S(S+1)$ and $m$ are eigenvalues of the operators $(\hat S^2, \hat S_z)$ respectively, where  $\hat S^2$ is the invariant quantity (Casimir operator) of $SO(3)$ and $\hat S_z$ is that of its subgroup $SO(2)$. The two commuting operators $(\hat S^2, \hat S_z)$ form the CSCO of the Hilbert space of a spin, so we can use their eigenvalues to label different bases.

Similarly, for any unitary group $G$, we can use invariant quantities (that commute with the group $G$) to provide the main quantum numbers to label different irreducible Reps. The class operators defined in (\ref{ClassO})
are ideal candidates since they commute with each other and commute with all the group elements in $G$. According to the corollary of Schur's lemma, every irreducible Rep space $(\nu)$ is an eigenspace of all the class operators $C_i$ and can be labeled by the eigenvalues (it can be shown that the eigenvalues are proportional to the corresponding characters $\chi_i^{(\nu)}$). It turns out that the $N_c$ independent class operators can provide exactly $N_c$ sets of different quantum numbers and can completely identify the $N_c$ different irreducible Reps. The reason is that the $N_c$ linearly independent class operators form an algebra since they are closed under multiplication. The natural Rep of the class algebra provides $N_c$ different sets of eigenvalues (corresponding to the eigenstates in the class space). These $N_c$ different eigenvalues are nothing but the quantum numbers labeling the irreducible Reps. Generally, these quantum numbers are degenerate and can not distinguish each of the irreducible bases. Therefore, the class operators of $G$ form a subset of the commuting operators and are called the CSCO-I.

Actually, we can linearly combine the $N_c$ independent class operators to form a single operator $C=\sum_ik_iC_i$ (where $k_i$ are some real constants) as the CSCO-I, as long as the operator $C$ has $N_c$ different eigenvalues.

Secondly, we need some `magnetic' quantum numbers to distinguish different bases in one copy of irreducible Rep. We will make use of the chain of subgroups $G(s)=G_1\supset G_2\supset ...$, the set of their CSCO-I $C(s)=(C(G_1),C(G_2),...)$ commute with the CSCO-I of $G$ and provide the required quantum numbers.  The set $\left(C,C(s)\right)$ formed by $C$ (the CSCO-I of $G$) and  $C(s)$ is called the CSCO-II \cite{ChenJQ02}.

Finally, since the $n_\nu$ dimensional irreducible Rep $(\nu)$ appears $n_\nu$ times in the regular Rep, the quantum numbers of CSCO-II are still degenerate. To lift this degeneracy, we need more quantum numbers (i.e. the multiplicity quantum numbers). Noticing that the intrinsic group $\bar G$ commutes with $G$, we can use the CSCO-Is of the subgroup chain $\bar G(s)=\bar G_1\supset \bar G_2\supset ...$, namely, $\bar C(s)=(\bar C(G_1), \bar C(G_2), ...)$. The operators $\bar C(s)$ commute with the CSCO-II of $G$ and can completely lift the degeneracy. So we obtain the complete set $(C, C(s), \bar C(s))$, called the CSCO-III. The CSCO-III provides $G$ sets of non-degenerate quantum numbers which completely label all the irreducible bases in the regular Rep space.

As an example, 
we apply this method to reduce the regular Rep of the permutation group $S_3$.

For 
 $S_3$, there are $6$ group elements and the canonical subgroup chain is $S_3\supset S_2$. The group elements $P=(12),Q=(23),R=(13)$ belong to the same class. So we may choose class operator $C=(12)+(23)+(13)$, the class operator of subgroup $S_2$ $C(s)=(12)$ and the class operator of intrinsic subgroup $\overline{S_2}$ $\bar{C}(s)=(\overline{12})$ to construct CSCO. In the regular Rep space (group space), the Rep matrices of these class operators can be written as,respectively
\begin{eqnarray}
C&=&\left(\begin{array}{cccccc}
0&1&1&1&0&0\\1&0&0&0&1&1\\
1&0&0&0&1&1\\1&0&0&0&1&1\\0&1&1&1&0&0\\0&1&1&1&0&0\end{array}\right),
C(s) =\left(\begin{array}{cccccc}
0&1&0&0&0&0\\1&0&0&0&0&0\\
0&0&0&0&0&1\\0&0&0&0&1&0\\0&0&0&1&0&0\\0&0&1&0&0&0\end{array}\right),\nonumber\\
\bar{C}(s)&=&\left(\begin{array}{cccccc}
0&1&0&0&0&0\\1&0&0&0&0&0\\
0&0&0&0&1&0\\0&0&0&0&0&1\\0&0&1&0&0&0\\0&0&0&1&0&0\end{array}\right).
\end{eqnarray}

The eigenvalues of $C$ are $3,0,-3$, in which $3$ and $-3$ are non-degenerate but $0$ is four-fold degenerate. We may use the complete set $(C,C(s),\bar{C}(s))$ to lift the degeneracy. Then a unitary transformation matrix $U$ can be formed via the common eigenvectors of $(C,C(s),\bar{C}(s))$ and the regular linear Rep matrices of all group elements can be block diagonalized simultaneously. We only give the results of two generators:
\beq
U^{\dag}M(P)U&=&\left(\begin{array}{cccccc}
-1&0&0&0&0&0\\0&-1&0&0&0&0\\
0&0&1&0&0&0\\0&0&0&-1&0&0\\0&0&0&0&1&0\\0&0&0&0&0&1\end{array}\right),\\
U^{\dag}M(Q)U&=&\left(\begin{array}{cccccc}
-1&0&0&0&0&0\\0&\frac{1}{2}&\frac{\sqrt{3}}{2}&0&0&0\\
0&\frac{\sqrt{3}}{2}&-\frac{1}{2}&0&0&0\\0&0&0&\frac{1}{2}&\frac{\sqrt{3}}{2}&0\\0&0&0&\frac{\sqrt{3}}{2}&-\frac{1}{2}&0
\\0&0&0&0&0&1\end{array}\right).
\eeq

From above results, we find two inequivalent one-dimensional Reps occur only once, but a two-dimensional Rep occurs twice. The quantum numbers of $\bar{C}(s)$ are used to distinguish the two equivalent irreducible Reps.


\subsection{Reduction of regular projective Reps}\label{Reduce}

Now we generalize the eigenfunction method to reduce the regular projective Reps. Unitary groups and anti-unitary groups will be discussed separately.

\subsubsection{  Unitary groups}
First of all, we need to define the class operators. For a regular projective Rep of any unitary group $G$, the class operator $C_1$ corresponding to $g_1\in G$ is defined as \cite{ChenJQRMP85}
\begin{equation}\label{CSCO1}
C_1=\sum_{g_{2}\in G}\hat{  g}^{-1}_{2}\hat  {g}_{1}\hat{ g}_{2}=\sum_{g_{2}\in G}M^{-1}(g_{2})M(g_{1})M(g_{2}).
\end{equation}

The class operator $C_1$ commutes with all the regular projective Rep matrices since
\Beq
C_1\hat { g}_3&=&\sum_{g_{2}\in G}\hat { g}_3[\hat { g}_2\hat { g}_3]^{-1}\hat { g}_1\hat { g}_2\hat { g}_3\\
&=&\hat { g}_3\sum_{g_{2}\in G}\widehat { {g_2g_3}}^{-1}\hat { g}_1\widehat { {g_2g_3}}\\
&=&\hat { g}_3C_1
\Eeq
for all $g_3\in G$. It can be shown that if $g_1'=g^{-1}g_1g$ then $C_{g_1'}\propto C_{g_1}$, namely, each conjugate class gives at most one linearly independent class operator. Since a nontrivial projective Rep $M(g)$ is not a faithful Rep of $G$, the class operators for some conjugate classes may be zero. Consequently, the number of linearly independent class operators is generally less than the number of classes of the group $G$. Since character is still a good quantity, the number of different irreducible projective Reps (of the same class) is still equal to the number of independent class operators. Due to the corollary of Schur's lemma, we can use the eigenvalues of the class operators to label different Rep spaces. Therefore, the class operators (or their linear combination $C=\sum_i k_i C_i$) form CSCO-I of the regular projective Rep.

The eigenvalues of CSCO-I are degenerate. To distinguish the bases of the same irreducible Rep, we can make use of the class operators of the subgroup chain $G(s)=G_1\supset G_2\supset ...$. For instance, the class operators of the subgroup $G_1$ are defined as
\begin{equation}\label{CSCO2}
C_1(G_1)=\sum_{g_{2}\in G_1}M^{-1}(g_{2})M(g_{1})M(g_{2}) ,\ \ \  g_{1}\in G_1,
\end{equation}
and $C(G_1)=\sum_i k_i C_i(G_1)$ is the CSCO-I of $G_1$. Repeating the procedure we obtain the set of CSCO-Is for the subgroup chain $C(s)=(C(G_1),C(G_2),...)$. The operator set $(C,C(s))$ is called CSCO-\textrm{II}, which can be used to distinguish all the bases if every irreducible Rep occurs only once in the reduced projective Rep.

However, a $n_\nu$-dimensional irreducible projective Rep $(\nu)$ may occur more than once and this will cause degeneracy in the eigenvalues of CSCO-\textrm{II}. To lift this degeneracy we need more quantum numbers to label the mutiplicity. Similar to the linear Reps, we can use the class operators of the intrinsic group $\bar{G}$. It can be shown that the class operators of $\bar{G}$ is identical to those of $G$: for any $g_3\in G$, we have
\beq\label{barCg3}
\bar C_1|g_3\rangle &=& \left[ \sum_{g_2}\hat {\bar g}_2\hat {\bar g}_1\hat {\bar g}_2^{-1}\right] \hat{\bar g}_3|E\rangle\nonumber\\
			       &=&\sum_{g_2}e^{-i\theta_2(g_2,g_2^{-1})}\left\{\left[(\hat {\bar g}_2\hat {\bar g}_1)\widehat {\overline {g_2^{-1}}}\right] \hat{\bar g}_3\right\}|E\rangle\nonumber\\
			       &=&\sum_{g_2}e^{-i\theta_2(g_2,g_2^{-1})}e^{i\theta_2(g_1,g_2)}e^{i\theta_2(g_2^{-1},g_1g_2)}\nonumber\\
&&\times
e^{i\theta_2(g_3,g_2^{-1}g_1g_2)}|g_3g_2^{-1}g_1g_2\rangle,
\eeq
where we have used $\widehat {\overline {g_2^{-1}}}=e^{i\theta_2(g_2,g_2^{-1})}\hat {\bar g}_2^{-1}$.
On the other hand,
\beq\label{Cg3}
C_1|g_3\rangle &=& \left[ \sum_{g_2}\hat {g}^{-1}_2\hat { g}_1\hat { g}_2\right] \hat{ g}_3|E\rangle\nonumber\\
			&=& \hat{ g}_3\left[ \sum_{g_2} e^{-i\theta_2(g_2,g_2^{-1})}\widehat {g^{-1}_2}\hat { g}_1\hat { g}_2\right]|E\rangle\nonumber\\
			       &=&\sum_{g_2}e^{-i\theta_2(g_2,g_2^{-1})}e^{i\theta_2(g_1,g_2)}e^{i\theta_2(g_2^{-1},g_1g_2)}\nonumber\\
&&\times
e^{i\theta_2(g_3,g_2^{-1}g_1g_2)}|g_3g_2^{-1}g_1g_2\rangle,
\eeq
where $\widehat {g^{-1}_2}=e^{i\theta_2(g_2,g_2^{-1})}\hat {g}^{-1}_2$. Therefore, $\bar C_1=C_1$, namely, the class operators of $\bar G$ do not provide any new invariant quantities.

However, the class operators $\bar C(s)$ [$\bar C(s)$ can be obtained from (\ref{CSCO2})  by replacing $g$ with $\bar g$] of the chain of subgroups $\bar G(s)=\bar G_1\supset \bar G_2\supset ...$ are different from $C(s)$ and can lift the degeneracy of the eigenvalues of CSCO-II. Now we obtain the complete set of class operators $\left(C,C(s),\bar{C}(s)\right)$, called CSCO-\textrm{III}. {The eigenvalues of $\bar C(s)$ are the same as those of $C(s)$, consequently the number of times an irreducible projective Rep $(\nu)$ occurs is equal to its dimension $n_\nu$, and $\sum_\nu n_\nu^2=G$ is still valid for a given class of irreducible projective Reps.}

For unitary groups, the common eigenvectors of the operators in CSCO-\textrm{III} are the orthonormal bases of the irreducible projective Reps, and each eigenvector has unique `quantum numbers'. These eigenvectors form a unitary matrix $U$ which block diagonalizes all regular projective Rep matrices $M(g)$ simultaneously.

Now we summarize the steps to obtain all the irreducible projective Reps with the eigenfunction method:

1, Solve the 2-cocycle equations, obtain the solutions and their classification (see appendix \ref{app2});

2, Select one solution of a given class as the factor system and obtain the corresponding regular projective Rep (see section \ref{Regular});

3, Construct the CSCO-III with $\left(C,C(s),\bar{C}(s)\right)$, where $C, C(s), \bar C(s)$ are class operators of $G$, of the subgroups $G(s)\subset G$, and of the subgroups $\bar G(s)\subset \bar G$ respectively. Then reduce the regular projective Reps into irreducible ones;

4, Change the class and repeat above procedure, until the irreducible projective Reps of all classes are obtained.

\subsubsection{ Anti-unitary groups}\label{sec:antiU}

Now denote the time reversal conjugate of $g$ as $\tilde g$, namely, $T^{-1}gT=\tilde g$ for $g\in G$. Since $G/H=Z_2^T$, any group element in $G$ either belongs to the unitary normal subgroup $H$ or belongs to its coset $TH=HT$. This means that any anti-unitary group element must be written in forms of $hT$ or $T\tilde h$ with $h,\tilde h\in H$.

Some conclusions of unitary groups should be modified for anti-unitary groups. We first generalize Schur's lemma to anti-unitary groups.

{\it\bf Generalized Schur's lemma:} If a nonzero matrix $C$ commutes with the irreducible (projective or linear) Rep $(\nu)$ of an anti-unitary group $G$ with $C\hat g=\hat gC$, namely,
\[
CM^{(\nu)}(g)K_{s(g)}=M^{(\nu)}(g)K_{s(g)}C,
\]
for all $g\in G$, then $C$ has at most two eigenvalues which are complex conjugate to each other, including the special case that $C$ is a real constant matrix $C=\lambda I$ with a single real eigenvalue $\lambda$.


The proof is simple. No matter $C$ is completely diagonalized or not, $C$ has at least one eigenstate $|\lambda\rangle$ with
\Beq
C|\lambda\rangle &=& \lambda|\lambda\rangle.
\Eeq
For any unitary element $h\in H$ and anti-unitary element $Th$, we have
\Beq
&&C\hat h|\lambda\rangle = \hat hC|\lambda\rangle =\lambda\left(\hat h|\lambda\rangle\right),\\
&&C\widehat{Th}|\lambda\rangle=\widehat{Th}C|\lambda\rangle=\widehat{Th}(\lambda|\lambda\rangle)= \lambda^* \left(\widehat{Th}|\lambda\rangle\right).
\Eeq
Above relations indicate that unitary operators reserve the eigenvalue of $C$, while anti-unitary operators switch the eigenspace of $\lambda$ to the eigenspace of its complex conjugate $\lambda^*$. If $(\nu)$ is irreducible, then $C$ is completely diagonalizable and has at most two eigenvalues $\lambda$ and $\lambda^*$. Specially, if $C$ is Hermitian, then it must be a real constant matrix $C=\lambda I$ with a single real eigenvalue $\lambda$ \cite{JohnODim63}. The generalized Schur's lemma yields the following corollary:

{\it\bf Corollary:} If a linear operator $\widehat C$ commutes with an anti-unitary group $G$ with $[\hat g,\widehat C]=0$ for all $g\in G$, namely,
\[
M(C)M(g)K_{s(g)}=M(g)K_{s(g)}M(C),
\]
then a direct sum of two eigenspaces whose eigenvalues are mutually complex conjugate is a Rep space of $G$ (If $C$ is not completely diagonalizable, we can transform it into Jordan normal form and replace each `eigenspace' by a `supporting space of Jordan blocks with the same eigenvalue').

Our aim is to use the eigenvalues of a set of {\it linear operators} to label the irreducible bases of anti-unitary groups.

{\bf Decoupled factor systems.} For type-I anti-unitary groups, we can adopt the following decoupled factor system (see Appendix \ref{standardcyc} for details of tuning the coboundary), which satisfy
\beq\label{fixgauge}
&&\omega_2(E,g_1)=\omega_2(g_1,E)= 1,\nonumber \\
&&\omega_2(T,T)=\pm1,\nonumber\\  &&\omega_2(T,g_1)=\omega_2(g_1,T)=1,\nonumber\\
&&\omega_2(Tg_1,g_2)=\omega_2^{-1}(g_1,g_2),\nonumber\\
&&\omega_2(g_1,Tg_2)=\omega_2(g_1,\tilde g_2), \nonumber\\
&&\omega_2(Tg_1,Tg_2)=\omega_2(\tilde g_1,g_2)\omega_2(T,T),
\eeq
for any $g_1,g_2\in H$. The first equation is the canonical gauge condition. When above relations are satisfied, it can be shown that \[\omega_2(g_1, g_2)\omega_2(\tilde g_1,\tilde g_2)=1\] is automatically satisfied. From (\ref{fixgauge}), only the cocycles $\omega_2(T,T)$ and $\omega_2(g_1,g_2)$ with $g_1,g_2\in H$ are important, therefore we only need to focus on $\widehat{T}$ and the normal subgroup $H$. Once the representation matrices of $T$ and the unitary elements in $H$ are reduced, we then obtain the irreducible Rep for the full group $G$.

For type-II anti-unitary groups, above discussion of decoupled factor system should be slightly modified. In Appendix \ref{app:type-II}, we provide the procedure to obtain the decoupled factor system in the fermionic anti-unitary groups. Since the discussions for the two types of anti-unitary groups are similar, we will mainly focus on type-I anti-unitary groups in the following.

{\bf Class operators and CSCO-I.} In order to reserve the relation between irreducible representations and classes, we should redefine the classes and class operators for an anti-unitary group $G$. The class corresponds to unitary element $g_i$ is defined as
\[
\mathrm{Class}(g_i)=\{ h^{-1}g_ih, (Th)^{-1} g_i^{-1} (Th); h\in H\},
\]
while for anti-unitary element $g_iT$, the conjugate class is defined as
\[
\mathrm{Class}(g_iT)=\{ h^{-1}(g_iT)\tilde h, (Th)^{-1} ( \tilde{g}_iT)^{-1}T^2(T\tilde h); h\in H\}.
\]
Obviously, $\{T\}$ forms a class itself.

Before defining class operators, we define the following operators,
\beq
&&D_{g_i}=\sum_{h\in H}\left( \hat h^{-1}\hat g_i\hat h + \widehat{Th}^{-1} \hat g_i^{-1} \widehat{Th}\right)
,\nonumber\\
&&D_{g_i^{-1}}=\sum_{h\in H}\left( \hat h^{-1}\hat g_i^{-1}\hat h + \widehat{Th}^{-1} \hat g_i \widehat{Th}\right)
,\nonumber\\
&&D_{g_iT}=\sum_{h\in H}\left( \hat {h}^{-1}\widehat {g_iT}\hat {\tilde h} + \widehat{Th}^{-1} \widehat{\tilde{g}_i T}^{-1}\widehat T^2 \widehat {T\tilde h} \right),\nonumber\\
&&D_{(g_iT)^{-1}}=\sum_{h\in H}\left( \hat {h}^{-1}\widehat{\tilde{g}_i T}^{-1}\widehat T^2\hat {\tilde h} + \widehat{Th}^{-1}  \widehat {g_iT}\widehat {T\tilde h} \right).\label{DT}\nonumber\\
\eeq
For type-I anti-unitary groups $T^2=E$, and $\omega_2(g_i,T)=\omega_2(T,\tilde g_i)=1$, we have $\hat g_i\widehat T = \widehat T\hat{\tilde g}_i$. Therefore,
\beq
&&D_{g_i}=\sum_{h\in H}\left( \hat h^{-1}\hat g_i\hat h + \hat h^{-1}\hat{\tilde{g}}_i^{-1}\hat h \right)
= C_{g_i}^H + C_{\tilde g_i^{-1}}^H,\nonumber\\
&&D_{g_i^{-1}}=\sum_{h\in H}\left( \hat h^{-1}\hat g_i^{-1}\hat h + \hat h^{-1}\hat{\tilde{g}}_i\hat h \right)
= C_{g_i^{-1}}^H + C_{\tilde g_i}^H,\nonumber\\
&&D_{g_iT}
=\sum_{h\in H}\left( \hat h^{-1}\hat g_i\hat h + \hat h^{-1}\hat{\tilde{g}}_i^{-1}\hat h \right)\widehat{T}=D_{g_i}\widehat{T},\nonumber\\
&&D_{(g_iT)^{-1}}=\sum_{h\in H}\left( \hat h^{-1}\hat g_i^{-1}\hat h + \hat h^{-1}\hat{\tilde{g}}_i\hat h \right)\widehat{T}=D_{g_i^{-1}}\widehat{T},\nonumber\\\label{DT1}
\eeq
where $C_{g_i}^{H}$, $C_{\tilde g_i^{-1}}^{H}$, $C_{g_i^{-1}}^{H}$ and $C_{\tilde g_i}^{H}$  denote the class operators of the normal subgroup $H$ corresponding to unitary elements $g_i$, $\tilde{g}_{i}^{-1}$, $g_i^{-1}$ and $\tilde{g}_{i}$  respectively.


It is easy to prove that $$\widehat T^{-1}D_{g_i}\widehat T=D_{g_i^{-1}},$$ or $D_{g_i}\widehat T=\widehat T D_{g_i^{-1}}$. Namely, the operators $D_{g_i}$ do not commute with all the group elements. We can define the following class operators to solve this problem:

1) if $g_i$ and $g_i^{-1}$ belong to different classes, namely, if $g_i\neq g_i^{-1}$ and $g_i\neq \tilde g_i$, then we define two class operators corresponding to the classes of $g_i$ and $g_i^{-1}$,
\beq
&& C_{i^+}=D_{g_i}+D_{g_i^{-1}}=C_{g_i}^H +C_{\tilde g_i^{-1}}^H+C_{g_i^{-1}}^H +C_{\tilde g_i}^H ,   \label{classG1}\\
&&C_{i^-}=i(D_{g_i}-D_{g_i^{-1}})=i(C_{g_i}^H +C_{\tilde g_i^{-1}}^H-C_{g_i^{-1}}^H -C_{\tilde g_i}^H ).   \label{classG2}\nonumber\\
 \eeq
Obviously $C_{i^{\pm}}$ commute with $\hat g_j$ for any $g_j\in H$, $C_{i^{\pm}}\hat g_j=\hat g_j C_{i^{\pm}}$.
Since
\Beq
&&\widehat TC_{g_i}^H=C_{\tilde{g} _i}^H\widehat T ~,~
 \widehat{T}C_{\tilde{g} _i}^H=C_{g_i}^H\widehat{T},\\
&&\widehat TC_{g_i^{-1}}^H=C_{\tilde{g} _i^{-1}}^H\widehat T ~,~
\widehat{T}C_{\tilde{g} _i^{-1}}^H=C_{g_i^{-1}}^H\widehat{T},
\Eeq
$C_{i^{\pm}}$ commute with the time reversal operator,
\Beq
&&\widehat{T}C_{i^+}=(C_{\tilde{g}_i}^H +C_{g_i^{-1}}^H+C_{\tilde{g}_i^{-1}}^H +C_{ g_i}^H)\widehat{T}=C_{i^+}\widehat{T},\\
&&\widehat{T}C_{i^-}=-i(C_{\tilde{g}_i}^H +C_{ g_i^{-1}}^H-C_{\tilde{g}_i^{-1}}^H -C_{ g_i}^H )\widehat{T}=C_{i^-}\widehat{T}.
\Eeq
Therefore, $C_{i^{\pm}}$ commute with all the operators in $G$.

2) if $g_i$ and $g_i^{-1}$ belong to the same class, namely, if $g_i=g_i^{-1}$ or $g_i=\tilde g_i$, then the class operator corresponding to $g_i$ is $C_i=C_{i^+}=D_{g_i}$ (obviously $C_{i^-}=0$ in this case). 

It can be easily checked that above class operators $C_{i^\pm}$ are Hermitian if the Rep is unitary.

Owing to relation (\ref{DT1}), the class operators for anti-unitary elements $g_iT$ and $(g_iT)^{-1}$ are
\begin{eqnarray}
C_{i^{+}T}&=&D_{g_i T}+D_{(g_i T)^{-1}}\nonumber\\
&=&(C_{g_i}^H +C_{\tilde g_i}^H+C_{g_i^{-1}}^H+C_{\tilde g_i^{-1}}^H)\widehat{T} =C_{i^{+}}\widehat{T},  \label{Ci+T}\\
C_{i^{-}T}&=&i(D_{g_i T}-D_{(g_i T)^{-1}})\nonumber\\
&=&i[(C_{g_i}^H -C_{\tilde g_i}^H)-(C_{g_i^{-1}}^H-C_{\tilde g_i^{-1}}^H  )]\widehat{T}=C_{i^{-}}\widehat{T}.  \label{Ci-T}\nonumber\\
\end{eqnarray}
This gives a one-to-one correspondence between the anti-unitary class operators $C_{i^{\pm}T}$ and the unitary class operators $C_{i^{\pm}}$,  where the unitary class operators can be obtained solely from the subgroup $H$ by the equations \eqref{classG1} and \eqref{classG2}. However, it can be shown that {\it the anti-unitary class operators do not provide any meaningful quantum numbers for $U(1)$ coefficient projective Reps. }

To see why the anti-unitary class operators do not correspond to more irreducible Reps, we focus on the class operator $\widehat T$ first. Noticing that $M(T)$ is a real matrix, so $\widehat T^2=[M(T)]^2=\omega_2(T,T)=\pm1$. If $\widehat T^2=1$, then $M(T)$ has eigenvalues $\pm1$ with eigenstates $|\phi^+\rangle$ and $|\phi^-\rangle$ satisfying $\widehat T|\phi^\pm\rangle=\pm|\phi^\pm\rangle$ respectively. Under the unitary transformation $|\phi^+\rangle'=i|\phi^+\rangle$,
\[\widehat T|\phi^+\rangle'=-i|\phi^+\rangle=-|\phi^+\rangle',\]
which gives $M^{+'}(T)=-1=M^-(T)$. So the two one-dimensional Reps $(+)$ and $(-)$ are equivalent, and the quantum numbers corresponding to the class operator $\widehat T$ are redundant (The two Reps are non-equivalent   when one considers $Z_2$ coefficient projective Reps). On the other hand, if $\widehat T^2=-1$, the bases form Kramers doublets and $\widehat T$ cannot reduce to one dimensional Reps. In other words, the operator $\widehat T$ has no `eigenvalues' at all. In this case, the class operator $\widehat T$ doesn't contribute any quantum numbers to label different Reps.

Since $C_{i^\pm T}=C_{i^\pm}\widehat T=\widehat T C_{i^\pm}$, the eigenvalues of $C_{i^\pm T}$, if exist, are equal to the product of eigenvalues of $C_{i^\pm}$ and $\widehat T$. As a result, all the anti-unitary class operators do not contribute useful quantum numbers. Therefore, the number of different irreducible projective Reps is determined by the unitary class operators $C=\{C_{i^\pm}\}$.  Namely, $C=\{C_{i^\pm}\}$ are the CSCO-I of the anti-unitary group $G$.


It can be shown that unitary elements belonging to the same class have the same character in an irreducible projective Rep. Therefore, the characters of the unitary group elements as functions of the irreducible Reps, are also functions of the unitary class operators. As a consequence, the number of different irreducible projective Reps is equal to the number of linearly independent unitary class operators in $\{C_{i^\pm}\}$. This result is similar to the unitary groups. Since all the class operators corresponding to unitary classes are Hermitian, their eigenvalues are all real numbers, each set of eigenvalues corresponds to an irreducible Rep space.


{\bf CSCO-II and CSCO-III.} The class operators $C(s)$ of the subgroups $G(s)\subset H$ are defined as usual unitary groups. Then we obtain the CSCO-II $(C, C(s))$, where the quantum numbers of $C(s)$ can be used to distinguish different bases of one copy of irreducible Rep.

Since an irreducible Rep may occur more than once, we need to make use of the class operators of the intrinsic group to label the multiplicity.

For the unitary elements $\bar{g}_i$ and $\bar{g}^{-1}_{i}$, we define the following operators
\beq
&&\bar{D}_{g_i}=\sum_{\bar{h}\in \overline{H}}\left( \hat {\bar{h}}^{-1}\hat {\bar{g}}_i\hat {\bar{h}} + \hat {\bar{h}}^{-1}\hat{\bar{\tilde{g}}}_i^{-1}\hat {\bar{h}} \right)
= \bar{C}_{g_i}^H + \bar{C}_{\tilde g_i^{-1}}^H,\nonumber\\
&&\bar{D}_{g_i^{-1}}=\sum_{\bar{h}\in \overline{H}}\left( \hat {\bar{h}}^{-1}\hat {\bar{g}}_i^{-1}\hat{ \bar{h}} + \hat {\bar{h}}^{-1}\hat {\bar{\tilde{g}}}_i\hat {\bar{h}}\right)=\bar{C}_{g_i^{-1}}^H + \bar{C}_{\tilde g_i}^H\label{Dgbar}\nonumber\\
\eeq

For the anti-unitary elements $\overline{Tg_i }$ 
the corresponding conjugate class 
is defined as
\[
\mathrm{Class}(\overline{Tg_i})=\{ \bar{h}^{-1}\overline{Tg_i}\bar{\tilde h}, \overline{hT}^{-1}  \overline{T\tilde{g}_i}^{-1}\overline{T}^2\overline{\tilde h T}; \overline{h}\in \overline{H}\},
\]
we have
\beq
\bar{D}_{Tg_i}&=&\sum_{\bar{h}\in \overline{H}}\left( \hat {\bar{h}}^{-1}\widehat {\overline{Tg_i}}\hat {\bar{\tilde h}} + \widehat{\overline{hT}}^{-1} \widehat{ \overline{T\tilde{g}_i}}^{-1}\widehat{ \overline{T}}^2 \widehat {\overline{\tilde h T}} \right)\nonumber\\
&=&\sum_{\bar{h}\in \overline{H}}\left( \hat {\bar{h}}^{-1}\hat {\bar{g}}_i\hat {\bar{h}} + \hat {\bar{h}}^{-1}\hat{\bar{\tilde{g}}}_i^{-1}\hat {\bar{h}} \right)\widehat{\overline{T}}=\bar{D}_{g_i}\widehat{\overline{T}},\nonumber
\eeq
\beq
\bar{D}_{(Tg_i)^{-1}}&=&\sum_{\bar{h}\in \overline{H}}\left( \hat {\bar{h}}^{-1}\widehat{ \overline{T\tilde{g}_i}}^{-1}\widehat{ \overline{T}}^2\hat {\bar{\tilde h}} + \widehat{\overline{hT}}^{-1}  \widehat {\overline{Tg_i}} \widehat {\overline{\tilde h T}} \right)\nonumber\\
&=&\sum_{\bar{h}\in \overline{H}}\left( \hat {\bar{h}}^{-1}\hat {\bar{g}}_i^{-1}\hat{ \bar{h}} + \hat {\bar{h}}^{-1}\hat {\bar{\tilde{g}}}_i\hat {\bar{h}}\right)\widehat{\overline{T}}=\bar{D}_{g_i^{-1}}\widehat{\overline{T}}.\label{DTbar}\nonumber\\
\eeq

The class operators corresponding to $\bar g_i, \bar{g}^{-1}_{i}\in \overline H$ , $\overline{Tg_i }$ and $\overline{Tg_{i}}^{-1}$ are defined following \eqref{classG1} and \eqref{classG2} as
\beq
\bar C_{i^+} &=&\bar{D}_{g_i}+\bar{D}_{g_i^{-1}}\nonumber\\
&=&\bar C_{g_i}^{H} + \bar C_{\tilde g_i^{-1}}^{H} + \bar C_{g_i^{-1}}^{H}  + \bar C_{\tilde g_i}^{H},
\label{bargi}\\
\bar C_{i^-} &=&\left(\bar{D}_{g_i}-\bar{D}_{g_i^{-1}}\right)S\nonumber\\
&=&(\bar C_{g_i}^{H} + \bar C_{\tilde g_i^{-1}}^{H} - \bar C_{g_i^{-1}}^{H}  - \bar C_{\tilde g_i}^{H})S,
\eeq
where Eq.~(\ref{intr_proj}) has been used, and $S=i(P_u-P_a)=\Bmat iI&0\\0&-iI \Emat$ (here $I$ stands for a $H$-dimensional identity matrix, $H$ is the order of normal subgroup $H$) 
is a diagonal matrix with entries $i$ for unitary bases or $-i$ for anti-unitary bases, which satisfies $S^*=-S, S^2=-1$ and
\Beq
&&M(g_i) S=SM(g_i),\ \ M(g_iT)S^*=SM(g_iT),\\
&&M(\bar g_i)S=SM(\bar g_i),\ \ M(\overline{g_i T})S^\ast=SM(\overline{g_i T}).
\Eeq
Above relations can be easily verified since $M(g_i)$ and $M(\bar g_i)$ are block diagonalized with the form $\Bmat A&0\\0&B \Emat $, where the blocks $A$  and $B$ are $H\times H$ 
nonzero matrices, while  $M(g_iT)$ and $M(\overline{g_iT})$ are  block-off-diagonalized with the form $\Bmat 0&A\\B&0 \Emat $.

Furthermore, it can be checked (see Appendix \ref{app: prove}) that
\beq
&&\bar C_{i^+}=C_{i^+}, \ \ \ \ \ \ \ \  \bar C_{i^-}=C_{i^-},\nonumber
\\
&&\bar C_{i^+ T}=\bar {C}_{i^+}\widehat {\overline {T}}, \ \ \ \ \bar C_{i^- T}=\bar {C}_{i^-}\widehat {\overline {T}},\label{bargiT}
\eeq
where $\bar C_{i^+ T}=\bar{D}_{Tg_i}+\bar{D}_{(Tg_i)^{-1}}$ and $\bar C_{i^- T}=S(\bar{D}_{Tg_i}-\bar{D}_{(Tg_i)^{-1}})=(\bar{D}_{Tg_i}-\bar{D}_{(Tg_i)^{-1}})S^{\ast}$.

Again, $\widehat {\overline T}$ forms a class operator itself. And for any $g\in G$ we have
\Beq
\widehat{\overline T}\hat{\bar g}=\hat{\bar{\tilde g}}\widehat{\overline T},\ \ \ \widehat{\overline T}\hat{g}=\hat{ g}\widehat{\overline T},
\Eeq
or equivalently,
\beq \label{commut2}
&&M(\overline T)M(\bar g)=M(\bar {\tilde g})M(\overline T),\nonumber\\
&&M(\overline T)M(g)K_{s(g)}=M(g)K_{s(g)}M(\overline T)\nonumber\\
&&=M(g)M(\overline T)K_{s(g)},
\eeq
where $M(\overline T)$ is a real matrix. $\widehat {\overline T}$ doesn't contain complex-conjugate operator $K$ and commutes with all the Rep matrices $M(g)$ of $G$.

Noticing that $\widehat T|g\rangle =|Tg\rangle$,  $\widehat{\overline {T}}|g\rangle =|gT\rangle =|T\tilde g\rangle$, and $\tilde g$ is generally different from $g$ (except that $G=H\times Z_2^T$), so generally $M(T)\neq M({\overline T})$. This implies $\bar C_{i^{+}T}\neq C_{i^{+}T}$, namely, the anti-unitary class operator is generally different from its intrinsic partner. This is a difference between unitary and anti-unitary groups.

Since $G$ is anti-unitary while all the operators in CSCO-II are obtained from the unitary normal subgroup $H$ (and its subgroups),  we need to use at least one anti-unitary class operator of $\bar G$ (or its subgroup chain $\bar G(s')$) to construct CSCO-III. In most cases, we can adopt the class operator $\widehat{\overline T}$ as a member of $\bar C(s')$ in CSCO-III (an exception is given in Appendix \ref{Z2Z2Z2T}).
If an intrinsic subgroup $\bar G_1\in \bar G(s)$ is anti-unitary, and $\bar G_1/\bar H_1=\overline{Z_2^T}$ where $\bar H_1$ is the unitary normal subgroup with $\bar H_1\subset \bar H$, then the corresponding class operators are given by
\beq\label{BarCs}
&&\bar C_{g_i}^{H_1} + \bar C_{\tilde g_i^{-1}}^{H_1}+ \bar C_{ g_i^{-1}}^{H_1}+ \bar C_{\tilde g_i}^{H_1}, \nonumber\\
&& (\bar C_{g_i}^{H_1} + \bar C_{\tilde g_i^{-1}}^{H_1}- \bar C_{ g_i^{-1}}^{H_1}- \bar C_{\tilde g_i}^{H_1}) S,
\eeq
where $\bar{C}_{g_i}^{H_1}$ (or $\bar{C}_{\tilde g_i}^{H_1}$) is the class operator of $\bar H_1$ corresponding to the element $\bar{g}_i$ (or $\bar{\tilde g}_i$).
{\textit  On the other hand, if an intrinsic subgroup $\bar G_2\in \bar G(s)$ is unitary, then its (intrinsic) class operators are defined in the same way as usual unitary groups, the only constraint is that all these class operators should commute with all the class operators in CSCO-II and should be mutually commuting.}  After carefully choosing the operators of $\bar C(s')$ 
such that  the degeneracy of the quantum numbers are completely lifted, we obtain the CSCO-III $\left(C,C(s), \bar C(s') \right)$, where $C=\{C_{i^\pm}\}$.

Before going to examples, we summarize some special properties of anti-unitary groups which are different from unitary groups:

1, The anti-unitary class operators do not contribute any meaningful quantum numbers to CSCO-I. The simplest example is $Z_2^T$, which has only one 1-dimensional linear Rep \footnote{When acting on a Hilbert space, the 1-dimensional Reps of
  $Z_2^T$ is classified by $\protect \mathcal H^1(Z_2^T, U(1))=\protect \mathbb
  Z_1$, so there is only one 1-dimensional Rep. However, when acting on
  Hermitian operators, then the 1-dimensional Reps is classified by $\protect
  \mathcal H^1(Z_2^T, Z_2)=\protect \mathbb Z_2$, namely, there are TWO
  different Reps characterized by $T\protect \mathaccentV {widehat}05EOT^{-1} =\pm
  \protect \mathaccentV {widehat}05EO$, where $\protect \mathaccentV {widehat}05EO$ is
  an Hermitian operator. This result can be generalized to any anti-unitary
  groups.}  and one 2-dimensional irreducible projective Rep.

2, An irreducible Rep (after lifting the multiplicity) may be either labeled by a real quantum number, or labeled by a pair of complex conjugating quantum numbers. Since we redefined the conjugate classes for anti-unitary groups and all the class operators in CSCO-I are Hermitian, the main quantum numbers are real. But the multiplicity quantum numbers are generally complex numbers, so an irreducible Rep is generally labeled by a pair of complex conjugating quantum numbers.

\subsection {Some examples}

In this section, we list the nontrivial irreducible projective Reps of a few finite groups. The group elements, the generators, the classification labels and the coboundary variables are given below. The results of the projective Reps are shown in Table \ref{tb1}. For simplicity,  we only list the irreducible projective Rep with the lowest dimension in each nontrivial class, and only the Rep matrices of the generators are given.

{\bf Unitary groups:}

$Z_2\times Z_2=\{E, P\}\times \{E, Q\}$ with $P^2=Q^2=E, QP=PQ$. There are two generators $P, Q$. The classification is labeled by $\omega_{2}(Q,PQ)$. The coboundary variables ($11,15,16$) are set to be $1$.

$Z_2\times Z_2\times Z_2 = \{E, P\}\times \{E, Q\}\times \{E, R\}$ with three generators $P, Q,R$.  The classification is labeled by $\omega_{2}(PR,QR)$, $\omega_{2}(PR,PQR)$, $\omega_{2}(QR,PQR)$.The coboundary variables ($46,54,55,61\sim 64$) are set to be $1$.


$Z_3\times Z_3=\{E, P,P^2\}\times \{E, Q,Q^2\}$ with two generators $P, Q$. The classification is labeled by $\omega_{2}(PQ^{2},P^{2}Q^{2})$. The coboundary variables ($70,71,76\sim 81$) are set to be $1$.

$Z_3\times Z_3\times Z_3=\{E, P,P^2\}\times \{E, Q,Q^2\}\times \{E, R,R^2\}$ with three generators $P, Q, R$. The classification is labeled by
$\omega_{2}(P^{2}QR^{2},PQ^{2}R^{2})$, $\omega_{2}(P^{2}QR^{2},P^{2}Q^{2}R^{2})$, $\omega_{2}(PQ^{2}R^{2},P^{2}Q^{2}R^{2})$. The coboundary variables ($642,645,695,696,698,699\sim701,
712 \sim 729$) are set to be $1$.

$Z_4\times Z_8 = \{E, P,P^2,P^3\}\times\{E,Q,Q^2, Q^3,
 Q^4, Q^5,
 Q^6, Q^7\}$, with two generators $P, Q$. The classification is labeled by $\omega_{2}(P^{2}Q^{7},P^{3}Q^{7})$. The coboundary variables
($989,990,991,997\sim1024$) are set to be $1$.

$A_4=\{E,(123),(132),(124),(142),(134), (143), (234),\\ (243), (12)(34),(13)(24),(14)(23)\}$, the normal subgroup of
$S_4=\{E,\ (123),\ (132),\ (124),\ (142), ...,\ (12),\ (13),\\ (23), (14), (24), ... \}$ formed by even-parity permutations with $S_4/A_4\simeq Z_2=\{E,(12)\}$. The group $A_4$ has generators $P=(123), Q=(124)$ with $P^3=Q^3=E,\ PQ=Q^{2}P^{2}$.
The classification is labeled by $\omega_{2}(P^2Q^2,P^2Q^2)$. The coboundary variables ($104,105,107,108,128,129,136,138,140,141,143$) are set to be $1$.

$Z_4\rtimes Z_2=\{E, P, P^2, P^3\}\rtimes\{E, Q\}$ with $P^mQ=QP^{4-m}$, there are two generators $P$ and $Q$. The classification is labeled by $\omega_{2}(P^{2}Q,Q)$. The coboundary variables ($54\sim56, 61\sim64$) are set to be $1$.


$Z_2\times Z_2\times Z_2\times Z_2=\{E, P\}\times \{E, Q\}\times \{E, R\}\times \{E, S\}$ with four generators $P, Q, R, S$. The classification is labeled by $\omega_{2}(PQS,PRS),\ \omega_{2}(PQS,QRS),\ \omega_{2}(PQS,PQRS)$, $\omega_{2}(PRS,QRS),\ \omega_{2}(PRS,PQRS),\ \omega_{2}(QRS,PQRS)$. The coboundary variables ($205, 221,222,234, 237\sim239, 249\sim256$) are set to be $1$.

{\bf Anti-unitary groups:}

$Z_2\times Z_2^T=\{E, P\}\times\{E, T\}$ with two generators $P$ and $T$. The classification is labeled by ($\omega_{2}(T,T)$, $\omega_{2}(PT,PT)$). The coboundary variables ($12,15$) are set to be $1$.

$Z_2\times Z_2\times Z_2^T = \{E, P\}\times \{E, Q\}\times \{E, T\}$ with three generators $P, Q,T$. The classification is labeled by $\omega_{2}(PT,PT)$, $\omega_{2}(PT,QT)$, $\omega_{2}(QT,QT)$, $\omega_{2}(PQT,PQT)$. The coboundary variables ($48,54,56,61\sim63$) are set to be $1$.

$(Z_2\times Z_2)\rtimes Z_2^T = (\{E, P\}\times \{E, Q\})\rtimes \{E, T\}$ with $PQ=QP,\ TP=PT$ and $TQ=PQT$. This group can be generated by two generators $Q,T$ since $(TQ)^2=(QT)^2=P$. The classification is labeled by $\omega_{2}(PT,PT)$, $\omega_{2}(QT,Q)$. The coboundary variables ($48,60\sim64$) are set to be $1$.

$Z_4\times Z_2^T=\{E, P, P^2, P^3\}\times\{E, T\}$, there are two generators $P$ and $T$. The classification is labeled by $\omega_{2}(P^{2}T,P^{2}T)$, $\omega_{2}(P^{3}T,P^{2})$. The coboundary variables ($56,60\sim64$) are set to be $1$.

$Z_4\rtimes Z_2^T=\{E, P, P^2, P^3\}\rtimes\{E, T\}$ with $P^mT=TP^{4-m}$, there are two generators $P$ and $T$. The classification is labeled by $\omega_{2}(P^{2}T,P^{2}T)$ , $\omega_{2}(P^{3}T,P^{3}T)$. The coboundary variables ($53,54,56,61\sim63$) are set to be $1$.

$Z_{3}\times(Z_{3}\rtimes Z_{2}^{T}) \simeq(Z_3\times Z_3)\rtimes Z_2^T=\{E,P,P^2\}\times (\{E,Q,Q^2\}\rtimes \{E,T\})$ with $TP=PT, TQ=Q^2T,PQ=QP$, there are three generators $P,Q,T$. The classification is labeled by $\omega_{2}(P^{2}QT,PQ^{2}T)$. The coboundary variables ($264,270,309,311,312,314\sim324$) are set to be $1$.

$S_4^T=A_4\rtimes Z_2^T$, there are three generators $P=(123), Q=(124)$ and $T=(12)K$ with $P^3=Q^3=T^2=E$ and $TP=P^{2}T, TQ=Q^2T,PQ=Q^{2}P^{2}$. The classification is labeled by
$\omega_{2}(PQT,QP^{2}T)$, $\omega_{2}(P^{2}QT,PQ)$. The coboundary variables ($548\sim552, 557, 559\sim561, 564\sim576$) are set to be $1$.

$A_4\times Z_2^T$, there are three generators $P=(123), Q=(124)$ and $T$ with $P^3=Q^3=T^2=E$ and $TP=PT, TQ=QT,PQ=Q^{2}P^{2}$. The classification is labeled by $\omega_{2}(P^2,P)$, $\omega_{2}(P^{2},Q^{2}T)$. The coboundary variables ($34,35,37,45,84,93,148,   152,260,286,296,334,351,373$,
$390,399,408,427,459,477,537,549$) are set to be $1$.

{\bf Fermionic Anti-unitary groups:}

$Z_4^T=\{E,P_f,T,P_f T\}$ with $T^4=E$ and $T^2=P_f$. There is only one generator $T$. The classification is labeled by $\omega_{2}(P_f T,P_f)$. The coboundary variables ($15,16$) are set to be $1$. The projective Reps can also be characterized by the invariant $[M(T)K]^4=\pm1$ [see eqs. (\ref{Z4T}) and (\ref{Z4TInv})].

$Z_2\times Z_4^T=\{E,P\}\times\{E,P_f,T,P_f T\}$ with two generators $P$ and $T$. The classification is labeled by $\omega_{2}(P_f T,P_f T)$ , $\omega_{2}(PP_f T,P_f)$. The coboundary variables ($56,60\sim64$) are set to be $1$.

$Z_2\ltimes Z_4^T=\{E,P\}\ltimes\{E,T,T^2,T^3\}$ with $TP=PT^3$ is a type-I anti-unitary group since $(TP)^2=E$. It is easy to verify that $Z_2\ltimes Z_4^T$ is isomorphic to $D_{2d}^T \simeq (Z_2\times Z_2)\rtimes Z_2^T$.

$G_-^+(Z_4,T)$ with two generators $P$ and $T$ where $Z_4=\{E,P,P^2,P^3\}$ and $P^2=T^2=P_f$, $PT=TP$. 
Since $(TP)^2=E$ and $P^4=E$, $G_-^+(Z_4,T)\simeq Z_4\times Z_2^T$, they have the same representations.

$G_-^-(Z_4,T)$ with two generators $P$ and $T$ where $Z_4=\{E,P,P^2,P^3\}$ and $P^2=T^2=P_f$, $P^mT=TP^{4-m}$. 
The classification is labeled by $\omega_{2}(P_f T,P_f T)$. The coboundary variables ($56,60\sim64$) are set to be $1$.

{\bf Remarks}:

1) Some of the above groups are isomorphic to point groups. For example,$Z_2\times Z_2 \simeq \mathscr D_2$; $Z_2\times Z_2\times Z_2\simeq \mathscr D_{2h}$; $Z_4\rtimes Z_2 \simeq \mathscr C_{4v}$ (or $\mathscr D_{2d}$); $A_4\simeq \mathscr T$ (the symmetry group of tetrahedron);  $S_4\simeq \mathscr T_d$ (or $\mathscr O$). For anti-unitary groups, we interpret the operations containing mirror reflection as anti-unitary elements, for example, we regard the horizontal mirror reflection in the group $Z_2\times Z_2^T\simeq \mathscr C_{2h}^T$ or the vertical mirror reflection in the group $(Z_2\times Z_2)\rtimes Z_2^T\simeq \mathscr D_{2d}^T$, as the generator $T$ of $Z_2^T$.  It should be noted that although $\mathscr C_{4v}\simeq \mathscr D_{2d}$, the anti-unitary groups $\mathscr C_{4v}^T$ and $\mathscr D_{2d}^T$ are NOT isomorphic since their unitary normal subgroups are different. 

2) When solving the cocycle equations (\ref{2cocycle}) to obtain the factor systems, we have set some coboundary variables to be 1. In order to label these variables, we first label the $G$ group elements as $1, 2, ..., G$. For direct product (or semi-direct product) groups $G_1\times G_2$, the group elements are sorted by the coset of the first group $G_1$, for example, for the $Z_2\times Z_2$ group the elements are sorted by $\{E, P\}\times \{E, Q\}= \{ \{E, P\}, \{E, P\}Q\}=\{E,P, Q,PQ\}$.  Then we sort the $G\times G$ variables of the 2-cocycle with the order $\omega_2(1,1), \omega_2(1,2),..., \omega_2(1,G),\omega_2(2,1),...,\omega_2(G,G)$ and further label them by numbers $1,2, ..., G^2$. The values of the classification labels and the coboundary variables completely fix the factor system (see appendix \ref{app2}).

3) For anti-unitary groups, we adopt the decoupled factor system (\ref{fixgauge}) by multiplying a coboundary $\Omega_{1}(g)$ (see Appendix \ref{standardcyc}). After the reduction, we divide the irreducible Rep matrices $M(g)$ by the coboundary $\Omega_{1}(g)$ to go back to the original factor system.

4) In Table \ref{tb2} in the appendix, we list all the irreducible linear Reps of several anti-unitary groups. We also give the number of independent unitary class operators and the multiplicity of each irreducible Rep. From the table we can see that different from unitary groups, the number of times an irreducible Rep occurs in the regular Rep is not always equal to its dimension.

\begin{table*}[htbp]
\caption{Irreducible projective Reps of some simple finite groups, we only give the lowest dimensional Rep, and only list the representation matrices of the generators.  The symbols $\sigma_{x,y,z}$ denote the Pauli matrices, and $\omega=e^{i\frac{2\pi}{3}}$,\ $\Omega=e^{i\frac{2\pi}{9}}$,\ $\omega^{\frac{1}{2}}=e^{i\frac{\pi}{3}}$. } \label{tb1}
\centering
 \begin{tabular}{ |c||ccc|c| }
 \hline
 Group & & Rep of generators & & label of classification\\
 \hline
 $ Z_{2}\times Z_{2}\simeq \mathscr D_2$ &$P$&$Q$&& \\
 \hline
 & $-i\sigma_{y}$ & $\sigma_{z}$& &$-1$\\
 \hline
$ Z_{2}\times Z_{2}\times Z_{2}\simeq\mathscr D_{2h}$ &$P$&$Q$&$R$ &\\
 \hline
 & $-i\sigma_{z}$&$\sigma_{x}$ & $i\sigma_{z}$&($+1,+1,-1$)\\
 &$-\sigma_{z}$ &$i\sigma_{y}$ &$-i\sigma_{y}$ &($+1,-1,+1$)\\ %
 & $-i\sigma_{z}$&$-i\sigma_{z}$ & $\sigma_{x}$&($+1,-1,-1$)\\
  &$-\sigma_{z}$ &$\sigma_{x}$ &$-i\sigma_{y}$ &($-1,+1,+1$)\\ %
 &$i\sigma_{z}$ &$I$ &$-\sigma_{x}$ &($-1,+1,-1$)\\
  &$I$ &$i\sigma_{z}$ &$\sigma_{x}$ &($-1,-1,+1$)\\ %
 &$-i\sigma_{z}$ &$-i\sigma_{x}$ &$iI$ &($-1,-1,-1$)\\
 \hline
  $Z_{3}\times Z_{3}$ &$P$&$Q$& &\\ 
 \hline
 &$\Omega^2\left(\begin{array}{ccc}
0&0&1\\ 1&0&0\\0&1&0 \end{array}\right) $
 &$\left( \begin{array}{ccc}\Omega&0&0 \\ 0 &\Omega^{7}& 0\\0 & 0 &\Omega^4\end{array} \right)$& &$\omega$\\
 &$\Omega^7\left(\begin{array}{ccc}
0&0&1\\ 1&0&0\\0&1&0 \end{array}\right) $
 &$\left( \begin{array}{ccc}\Omega^8&0&0 \\ 0 &\Omega^{2}& 0\\0 & 0 &\Omega^5\end{array} \right)$& &$\omega^{2}$\\
 \hline
 $Z_{3}\times Z_{3}\times Z_{3}$ &$P$&$Q$&$R$& \\
 \hline
 &$\left( \begin{array}{ccc}
\Omega^5&0&0 \\ 0 &\Omega^{2}& 0\\
0 & 0 &\Omega^8\end{array} \right)$
 &$\left( \begin{array}{ccc}
0&0&\omega^2\\ \omega^2&0&0\\
0&\omega^2&0\end{array} \right)$
 &$\left( \begin{array}{ccc}
0&0&\Omega\\ \Omega^4&0&0\\
0&\Omega^7&0\end{array} \right)$&($1,1,\omega$)\\
&$\left( \begin{array}{ccc}
\Omega^4&0&0 \\ 0 &\Omega^{7}& 0\\
0 & 0 &\Omega\end{array} \right)$
 &$\left( \begin{array}{ccc}
0&0&\omega\\ \omega&0&0\\
0&\omega&0\end{array} \right)$
 &$\left( \begin{array}{ccc}
0&0&\Omega^{8}\\ \Omega^5&0&0\\
0&\Omega^2&0\end{array} \right)$&($1,1 ,\omega^{2} $)\\

&$\left( \begin{array}{ccc}
1&0&0 \\ 0 &\omega& 0\\
0 & 0 &\omega^{2}\end{array} \right)$
 &$\left( \begin{array}{ccc}
0&0&\omega^2\\ \omega&0&0\\
0&\omega^2&0\end{array} \right)$
 &$\left( \begin{array}{ccc}
0&\omega&0\\0&0&\omega\\
\omega^2&0&0\end{array} \right)$&($1,\omega ,1 $)\\%
&$\left( \begin{array}{ccc}
\Omega^{8}&0&0 \\ 0 &\Omega^{5}& 0\\
0& 0 &\Omega^{2}\end{array} \right)$
 &$\left( \begin{array}{ccc}
\Omega^{2}&0&0\\ 0&\Omega^{8}&0\\
0&0&\Omega^{5}\end{array} \right)$
 &$\left( \begin{array}{ccc}
0&\Omega^{2}&0\\ 0&0&\Omega^{8}\\
\Omega^{5}&0&0\end{array} \right)$&($1,\omega ,\omega $)\\
&$\left( \begin{array}{ccc}
\Omega^{4}&0&0 \\ 0 &\Omega^{7}& 0\\
0 & 0 &\Omega\end{array} \right)$
 &$\left( \begin{array}{ccc}
0&\omega&0\\ 0&0&\omega^2\\
\omega^2&0&0\end{array} \right)$
 &$\left( \begin{array}{ccc}
\omega&0&0\\0&\omega&0\\
0&0&\omega\end{array} \right)$ &($1,\omega , \omega^2$)\\

&$\left( \begin{array}{ccc}
1&0&0 \\ 0 &\omega^{2}& 0\\
0 & 0 &\omega\end{array} \right)$
 &$\left( \begin{array}{ccc}
0&0&\omega\\ \omega^{2}&0&0\\
0&\omega&0\end{array} \right)$
 &$\left( \begin{array}{ccc}
0&\omega^{2}&0\\0&0&\omega^{2}\\
\omega&0&0\end{array} \right)$&($1,\omega^2 ,1 $)\\ %
&$\left( \begin{array}{ccc}
\Omega^{5}&0&0 \\ 0 &\Omega^{2}& 0\\
0 & 0 &\Omega^{8}\end{array} \right)$
 &$\left( \begin{array}{ccc}
0&\omega^{2}&0\\ 0&0&\omega\\
\omega&0&0\end{array} \right)$
 &$\left( \begin{array}{ccc}
\omega^{2}&0&0\\0&\omega^{2}&0\\
0&0&\omega^{2}\end{array} \right)$&($1,\omega^2 ,\omega $)\\
&$\left( \begin{array}{ccc}
\Omega&0&0 \\ 0 &\Omega^{4}& 0\\
0& 0 &\Omega^{7}\end{array} \right)$
 &$\left( \begin{array}{ccc}
\Omega^{7}&0&0\\ 0&\Omega&0\\
0&0&\Omega^{4}\end{array} \right)$
 &$\left( \begin{array}{ccc}
0&\Omega^{7}&0\\ 0&0&\Omega\\
\Omega^{4}&0&0\end{array} \right)$&($1,\omega^2 ,\omega^2 $)\\

&$\left( \begin{array}{ccc}
\omega&0&0 \\ 0 &1& 0\\
0 & 0 &\omega^{2}\end{array} \right)$
 &$\left( \begin{array}{ccc}
0&0&\omega\\ \omega&0&0\\
0&\omega&0\end{array} \right)$
 &$\left( \begin{array}{ccc}
0&\omega^{2}&0\\0&0&1\\
\omega&0&0\end{array} \right)$&($\omega,1 ,1 $)\\ %
&$\left( \begin{array}{ccc}
\Omega^{2}&0&0 \\ 0 &\Omega^{8}& 0\\
0 & 0 &\Omega^{5}\end{array} \right)$
 &$\left( \begin{array}{ccc}
0&\omega&0\\ 0&0&\omega\\
\omega&0&0\end{array} \right)$
 &$\left( \begin{array}{ccc}
\Omega^{7}&0&0\\0&\Omega&0\\
0&0&\Omega^{4}\end{array} \right)$&($\omega, 1, \omega$)\\
 & $\left( \begin{array}{ccc}
\Omega^{4}&0&0 \\ 0 &\Omega& 0\\
0 & 0 &\Omega^{7}\end{array} \right)$
 &$\left( \begin{array}{ccc}
\omega&0&0\\ 0&\omega&0\\
0&0&\omega\end{array} \right)$
 &$\left( \begin{array}{ccc}
0&0&\omega^{2}\\1&0&0\\
0&1&0\end{array} \right)$&($\omega, 1, \omega^2$)\\

&$\left( \begin{array}{ccc}
1&0&0 \\ 0 &1&0\\
0 & 0 &1\end{array} \right)$
&$\left( \begin{array}{ccc}
\Omega^{5}&0&0\\ 0&\Omega^{2}&0\\
0&0&\Omega^{8}\end{array} \right)$
& $\left( \begin{array}{ccc}
0&\Omega^{5}&0\\0&0&\Omega^{2}\\
\Omega^{5}&0&0\end{array} \right)$&($\omega, \omega,1 $)\\ %
&$\left( \begin{array}{ccc}
\Omega^5&0&0 \\ 0 &\Omega^{2}& 0\\
0 & 0 &\Omega^{8}\end{array} \right)$
 &$\left( \begin{array}{ccc}
0&0&\Omega^{4}\\ \Omega&0&0\\
0&\Omega&0\end{array} \right)$
 &$\left( \begin{array}{ccc}
0&0&\Omega^{4}\\ \Omega&0&0\\
0&\Omega&0\end{array} \right)$ &($\omega, \omega, \omega$)\\
 \hline
\end{tabular}
\end{table*}

\begin{table*}[htbp]
 \centering
 \begin{tabular}{ |c||ccc|c| }
 \hline
 $Z_{3}\times Z_{3}\times Z_{3}$ &$P$&$Q$&$R$&label of classification
 \\{\rm (continue)}& & & & \\
 \hline
 &$\left( \begin{array}{ccc}
\Omega^{7}&0&0 \\ 0 &\Omega& 0\\
0 & 0 &\Omega^{4}\end{array} \right)$
 &$\left( \begin{array}{ccc}
0&0&1\\ \omega&0&0\\
0&\omega&0\end{array} \right)$
 &$\left( \begin{array}{ccc}
0&0&\Omega^{4}\\ \Omega&0&0\\
0&\Omega^{4}&0\end{array} \right)$ &($\omega,\omega,\omega^2$)\\

&$\left( \begin{array}{ccc}
\omega&0&0 \\ 0 &\omega^{2}& 0\\
0 & 0 &1\end{array} \right)$
 &$\left( \begin{array}{ccc}
0&0&\omega\\ \omega&0&0\\
0&\omega^{2}&0\end{array} \right)$
 &$\left( \begin{array}{ccc}
0&1&0\\0&0&\omega^{2}\\
1&0&0\end{array} \right)$&($\omega, \omega^2,1 $)\\ %
&$\left( \begin{array}{ccc}
\Omega^{8}&0&0 \\ 0 &\Omega^{5}& 0\\
0 & 0 &\Omega^{2}\end{array} \right)$
 &$\left( \begin{array}{ccc}
\Omega^{7}&0&0\\ 0&\Omega&0\\
0&0&\Omega^{4}\end{array} \right)$
 &$\left( \begin{array}{ccc}
0&\omega^{2}&0\\0&0&\omega^{2}\\
\omega^{2}&0&0\end{array} \right)$&($\omega, \omega^2,\omega $)\\
&$\left( \begin{array}{ccc}
\Omega&0&0 \\ 0 &\Omega^{4}& 0\\
0 & 0 &\Omega^{7}\end{array} \right)$
 &$\left( \begin{array}{ccc}
0&\Omega&0\\ 0&0&\Omega^{4}\\
\Omega^{7}&0&0\end{array} \right)$
 &$\left( \begin{array}{ccc}
\Omega^{4}&0&0\\ 0&\Omega^{7}&0\\
0&0&\Omega\end{array} \right)$&($\omega, \omega^2,\omega^2 $)\\

&$\left( \begin{array}{ccc}
\omega^{2}&0&0 \\ 0 &1& 0\\
0 & 0 &\omega\end{array} \right)$
 &$\left( \begin{array}{ccc}
0&0&\omega^{2}\\ \omega^{2}&0&0\\
0&\omega^{2}&0\end{array} \right)$
 &$\left( \begin{array}{ccc}
0&\omega&0\\0&0&1\\
\omega^{2}&0&0\end{array} \right)$&($\omega^2,1 ,1 $)\\ %

&$\left( \begin{array}{ccc}
\Omega^{5}&0&0 \\ 0 &\Omega^{8}& 0\\
0 & 0 &\Omega^{2}\end{array} \right)$
 &$\left( \begin{array}{ccc}
\omega^{2}&0&0\\ 0&\omega^{2}&0\\
0&0&\omega^{2}\end{array} \right)$
 &$\left( \begin{array}{ccc}
0&0&\omega\\1&0&0\\
0&1&0\end{array} \right)$&($\omega^2,1,\omega $)\\
&$\left( \begin{array}{ccc}
\Omega^{7}&0&0 \\ 0 &\Omega& 0\\
0 & 0 &\Omega^{4}\end{array} \right)$
 &$\left( \begin{array}{ccc}
0&\omega^{2}&0\\ 0&0&\omega^{2}\\
\omega^{2}&0&0\end{array} \right)$
 &$\left( \begin{array}{ccc}
\Omega^{2}&0&0\\0&\Omega^{8}&0\\
0&0&\Omega^{5}\end{array} \right)$&($\omega^2,1,\omega^2 $)\\

&$\left( \begin{array}{ccc}
\omega^{2}&0&0 \\ 0 &\omega& 0\\
0 & 0 &1\end{array} \right)$
 &$\left( \begin{array}{ccc}
0&0&\omega^{2}\\ \omega^{2}&0&0\\
0&\omega&0\end{array} \right)$
 &$\left( \begin{array}{ccc}
0&1&0\\0&0&\omega\\
1&0&0\end{array} \right)$&($\omega^2,\omega ,1 $)\\ %
&$\left( \begin{array}{ccc}
\Omega^{8}&0&0 \\ 0 &\Omega^{5}& 0\\
0 & 0 &\Omega^{2}\end{array} \right)$
 &$\left( \begin{array}{ccc}
0&\Omega^{8}&0\\ 0&0&\Omega^{5}\\
\Omega^{2}&0&0\end{array} \right)$
 &$\left( \begin{array}{ccc}
\Omega^{5}&0&0\\ 0&\Omega^{2}&0\\
0&0&\Omega^{8}\end{array} \right)$&($\omega^2,\omega ,\omega$)\\
&$\left( \begin{array}{ccc}
\Omega&0&0 \\ 0 &\Omega^{4}& 0\\
0 & 0 &\Omega^{7}\end{array} \right)$
 &$\left( \begin{array}{ccc}
\Omega^{2}&0&0\\ 0&\Omega^{8}&0\\
0&0&\Omega^{5}\end{array} \right)$
 &$\left( \begin{array}{ccc}
0&\omega&0\\0&0&\omega\\
\omega&0&0\end{array} \right)$&($\omega^2,\omega,\omega^2 $)\\

&$\left( \begin{array}{ccc}
1&0&0 \\ 0 &1&0\\
0 & 0 &1\end{array} \right)$
&$\left( \begin{array}{ccc}
\Omega^{4}&0&0\\ 0&\Omega^{7}&0\\
0&0&\Omega\end{array} \right)$
& $\left( \begin{array}{ccc}
0&\Omega^{4}&0\\0&0&\Omega^{7}\\
\Omega^{4}&0&0\end{array} \right)$&($\omega^2,\omega^2,1$)\\ %
&$\left( \begin{array}{ccc}
\Omega^{2}&0&0 \\ 0 &\Omega^{8}& 0\\
0 & 0 &\Omega^{5}\end{array} \right)$
 &$\left( \begin{array}{ccc}
0&0&1\\ \omega^{2}&0&0\\
0&\omega^{2}&0\end{array} \right)$
 &$\left( \begin{array}{ccc}
0&0&\Omega^{5}\\ \Omega^{8}&0&0\\
0&\Omega^{5}&0\end{array} \right)$&($\omega^2,\omega^2,\omega $)\\
&$\left( \begin{array}{ccc}
\Omega^4&0&0 \\ 0 &\Omega^{7}& 0\\
0 & 0 &\Omega\end{array} \right)$
 &$\left( \begin{array}{ccc}
0&0&\Omega^{5}\\ \Omega^8&0&0\\
0&\Omega^{8}&0\end{array} \right)$
 &$\left( \begin{array}{ccc}
0&0&\Omega^{5}\\ \Omega^8&0&0\\
0&\Omega^{8}&0\end{array} \right)$&($\omega^2,\omega^2,\omega^{2} $)\\
\hline
$ Z_{4}\rtimes Z_{2}\simeq \mathscr D_{2d}\simeq \mathscr C_{4v}$ &$P$&$Q$& & \\
\hline
&$\left(\begin{array}{cc}
 e^{i\frac{3\pi}{4}}&0\\0&e^{i\frac{5\pi}{4}}\end{array}\right)$
&$\sigma_{x}$ & &$-1$\\
\hline
 $Z_{4}\times Z_{8}$ &$P$&$Q$ &&\\ 
 \hline
&$\sigma_{z}\otimes\left( \begin{array}{cc}
e^{-i\frac{5\pi}{8}}&0\\ 0 &e^{i\frac{7\pi}{8}}\end{array} \right)$
&$\left( \begin{array}{cccc}
0&e^{i\frac{\pi}{4}}&0&0 \\ 0&0& e^{-i\frac{7\pi}{8}}&0\\
0&0&0&e^{i\frac{3\pi}{8}}\\e^{-i\frac{\pi}{4}}&0&0&0\end{array} \right)$ & & $+i$\\
& $e^{i\frac{5\pi}{4}}\sigma_{z}$& $e^{i\frac{3\pi}{4}}\sigma_{x}$& &$-1$\\
&$\sigma_{z}\otimes\left( \begin{array}{cc}
e^{i\frac{5\pi}{8}}&0\\ 0 &e^{-i\frac{7\pi}{8}}\end{array} \right)$
&$\left( \begin{array}{cccc}
0&e^{-i\frac{\pi}{4}}&0&0 \\ 0&0& e^{i\frac{7\pi}{8}}&0\\
0&0&0&e^{-i\frac{3\pi}{8}}\\e^{i\frac{\pi}{4}}&0&0&0\end{array} \right)$ & &$-i$\\
\hline
$A_{4}\simeq \mathscr T$ &$P=(123)$ &$Q=(124)$ & & \\
\hline
&$-\left(\begin{array}{cc}
\omega&0\\0& \omega^2\end{array}\right)$
& $\frac{\sqrt{3}}{3}\left(\begin{array}{cc}
e^{-i\frac{\pi}{6}} &\sqrt{2}i\\ \sqrt{2}i & e^{i\frac{\pi}{6}} \end{array}\right)$ & &$-1$\\
\hline
\end{tabular}
\end{table*}

\begin{table*}[htbp]
 \centering
 \begin{tabular}{ |c||cccc|c| }
 \hline
$Z_{2}\times Z_{2}\times Z_{2}\times Z_{2}$ &$P$&$Q$&$R$&$S$&label of classification \\
\hline
&$i\sigma_{z}$ & $\sigma_{y}$&$\sigma_{y}$ &$\sigma_{x}$ &($+1,+1,+1,+1,+1,-1$)\\
&$-\sigma_{z}$ & $i\sigma_{y}$&$\sigma_{z}$ &$\sigma_{x}$& ($+1,+1,+1,+1,-1,+1$)\\%
&$-i\sigma_{z}$ & $i\sigma_{z}$&$-\sigma_{x}$ &$I$& ($+1,+1,+1,+1,-1,-1$)\\
&$-\sigma_{z}$ & $\sigma_{x}$&$-I$ &$i\sigma_{y}$ &($+1,+1,+1,-1,+1, +1$)\\%
&$i\sigma_{z}$ & $I$&$\sigma_{x}$ &$-i\sigma_{z}$ &($+1,+1,+1,-1,+1,-1$)\\
&$I$ & $-i\sigma_{z}$&$\sigma_{y}$ &$i\sigma_{z}$ &($+1,+1,+1,-1,-1,+1$)\\%
&$-i\sigma_{z}$ & $i\sigma_{y}$&$\sigma_{x}$ &$i\sigma_{x}$ &($+1,+1,+1,-1,-1,-1$)\\
&$-\sigma_{z}$ & $-\sigma_{z}$&$i\sigma_{y}$ &$\sigma_{x}$ &($+1,+1,-1,+1,+1,+1$)\\%
&$i\sigma_{z}$ & $-\sigma_{x}$&$-i\sigma_{z}$ &$I$ &($+1,+1,-1,+1,+1,-1$)\\
&$-\sigma_{z}$ & $-i\sigma_{y}$&$i\sigma_{y}$ &$-I$ &($+1,+1,-1,+1,-1,+1$)\\%
&$i\sigma_{z}$ & $i\sigma_{z}$&$i\sigma_{z}$ &$\sigma_{x}$ &($+1,+1,-1,+1,-1,-1$)\\
&$-\sigma_{z}\otimes I$ & $\sigma_{x}\otimes\sigma_{z}$&$-i\sigma_{y}\otimes I$ &$\sigma_{z}\otimes i\sigma_{y}$ &($+1,+1,-1,-1,+ 1, +1$)\\%
&$-i\sigma_{z}\otimes I$ & $I\otimes\sigma_{z}$&$-i\sigma_{z}\otimes\sigma_{x}$ &$\sigma_{y}\otimes i\sigma_{y}$ &($+1,+1,-1,-1,+1,-1$)\\
&$\sigma_{z}\otimes I$ & $\sigma_{z}\otimes i\sigma_{z}$&$-i\sigma_{y}\otimes\sigma_{x}$ &$-i\sigma_{x}\otimes \sigma_{z}$ &($+1,+1,-1,-1,-1,+1$)\\%
&$-i\sigma_{z}\otimes I$ & $-i\sigma_{y}\otimes I$&$I\otimes i\sigma_{z}$ &$I\otimes i\sigma_{y}$ &($+1,+1,-1,-1,-1,-1$)\\
&$-\sigma_{z}$ & $I$&$-\sigma_{x}$ &$i\sigma_{y}$ &($+1,-1,+1,+1, +1, +1$)\\%
&$i\sigma_{z}$ & $-\sigma_{x}$&$I$ &$i\sigma_{z}$ &($+1,-1,+1,+1, +1,-1$)\\
&$-\sigma_{z}\otimes I$ & $-i\sigma_{y}\otimes I$&$-\sigma_{x}\otimes\sigma_{z}$ &$-\sigma_{z}\otimes i\sigma_{y}$ &($+1,-1,+1,+1, -1, +1$)\\%
&$i\sigma_{z}\otimes \sigma_{z}$ & $i\sigma_{z}\otimes I$&$\sigma_{x}\otimes\sigma_{y}$ &$iI\otimes \sigma_{x}$ &($+1,-1,+1,+1, -1,-1$)\\
&$-\sigma_{z}$ & $\sigma_{x}$&$\sigma_{x}$ &$-\sigma_{z}$ &($+1,-1,+1,-1,+ 1, +1$)\\%
&$i\sigma_{z}$ & $-I$&$-I$ &$\sigma_{y}$ &($+1,-1,+1,-1, +1, -1$)\\
&$-\sigma_{z}\otimes \sigma_{z}$ & $i\sigma_{z}\otimes I$&$\sigma_{x}\otimes I$ &$I\otimes \sigma_{x}$ &($+1,-1,+1,-1, -1, +1$)\\%
&$-i\sigma_{z}\otimes \sigma_{z}$ & $-i\sigma_{x}\otimes I$&$\sigma_{z}\otimes I$ &$\sigma_{y}\otimes \sigma_{x}$ &($+1,-1,+1,-1, -1,- 1$)\\
&$I$ & $-\sigma_{z}$&$i\sigma_{y}$ &$i\sigma_{y}$ &($+1,-1,-1,+1, +1, +1$)\\%
&$-i\sigma_{z}$ & $-\sigma_{y}$&$i\sigma_{x}$ &$-i\sigma_{y}$ &($+1,-1,-1,+1, +1, -1$)\\
&$I\otimes \sigma_{z}$ & $iI\otimes \sigma_{x}$&$i\sigma_{z}\otimes I$ &$-i\sigma_{y}\otimes \sigma_{x}$ &($+1,-1,-1,+1, -1, +1$)\\%
&$-i\sigma_{z}\otimes I$ & $iI\otimes \sigma_{z}$&$-i\sigma_{x}\otimes \sigma_{x}$ &$-I\otimes i\sigma_{y}$ &($+1,-1,-1,+1, -1, -1$)\\
&$I\otimes \sigma_{z}$ & $-\sigma_{x}\otimes \sigma_{x}$&$-i\sigma_{z}\otimes I$ &$-I\otimes \sigma_{x}$ &($+1,-1,-1,-1, +1,  +1$)\\%
&$-i\sigma_{z}\otimes I$ & $\sigma_{z}\otimes \sigma_{z}$&$-i\sigma_{x}\otimes I$ &$-\sigma_{z}\otimes \sigma_{y}$ &($+1,-1,-1,-1, +1, -1$)\\
&$-I$ & $i\sigma_{z}$&$i\sigma_{z}$ &$\sigma_{x}$ &($+1,-1,-1,-1, -1,+1$)\\%
&$i\sigma_{z}$ & $i\sigma_{x}$&$i\sigma_{x}$ &$\sigma_{x}$ &($+1,-1,-1,-1, -1, -1$)\\
&$-I$ & $\sigma_{z}$&$\sigma_{x}$ &$i\sigma_{y}$ &($-1, +1,+ 1,+1, + 1, + 1$)\\%
&$iI\otimes \sigma_{z}$ & $\sigma_{y}\otimes \sigma_{y}$&$\sigma_{x}\otimes \sigma_{y}$ &$-i\sigma_{z}\otimes \sigma_{x}$&($-1, +1, +1,+1, + 1,  -1$)\\ 
&$-\sigma_{z}$ & $-i\sigma_{y}$&$I$ &$i\sigma_{y}$ &($-1, +1, +1,+1,  -1,+  1$)\\%
&$-i\sigma_{z}\otimes I$ & $i\sigma_{z}\otimes \sigma_{z}$&$\sigma_{x}\otimes \sigma_{x}$ &$I\otimes i\sigma_{y}$&($-1, +1,+ 1,+1,  -1, -1$)\\
&$-\sigma_{z}$ & $-\sigma_{x}$&$\sigma_{z}$ &$\sigma_{x}$ &($-1,+ 1, +1,-1, + 1, + 1$)\\%
&$-i\sigma_{z}\otimes \sigma_{z}$ & $-\sigma_{z}\otimes I$&$-\sigma_{y}\otimes I$ &$\sigma_{y}\otimes\sigma_{x}$ &($-1, +1, +1,-1,  +1,  -1$)\\
&$I$ & $i\sigma_{z}$&$I$ &$\sigma_{x}$ &($-1, +1, +1,-1,  -1, + 1$)\\%
&$-i\sigma_{z}\otimes I$ & $-i\sigma_{y}\otimes \sigma_{x}$&$\sigma_{x}\otimes \sigma_{y}$ &$-\sigma_{x}\otimes\sigma_{z}$ &($-1, +1, +1,-1,  -1,  -1$)\\
&$-\sigma_{z}$ & $I$&$i\sigma_{y}$ &$i\sigma_{y}$ &($-1, +1, -1,+1,  +1, + 1$)\\%
&$-i\sigma_{z}\otimes I$ & $-\sigma_{y}\otimes \sigma_{y}$&$-i\sigma_{z}\otimes \sigma_{z}$ &$iI\otimes\sigma_{x}$ &($-1,+ 1, -1,+1, + 1,  -1$)\\
&$-\sigma_{z}$ & $-i\sigma_{y}$&$i\sigma_{x}$ &$-i\sigma_{z}$ &($-1,+ 1, -1,+1,  -1,+ 1$)\\%
&$i\sigma_{z}\otimes I$ & $-i\sigma_{z}\otimes \sigma_{z}$&$-i\sigma_{z}\otimes \sigma_{x}$ &$-i\sigma_{y}\otimes\sigma_{y}$ &($-1, +1, -1,+1,  -1,  -1$)\\
&$-I\otimes\sigma_{z}$ & $-\sigma_{z}\otimes\sigma_{y}$&$-I\otimes i\sigma_{y}$ &$\sigma_{y}\otimes I$ &($-1,+ 1, -1,-1, + 1,  +1$)\\%
&$i\sigma_{z}$ & $I$&$-i\sigma_{z}$ &$\sigma_{y}$ &($-1, +1, -1,-1, + 1,  -1$)\\
&$-\sigma_{z}\otimes I$ & $iI\otimes\sigma_{z}$&$-i\sigma_{y}\otimes\sigma_{x}$ &$\sigma_{y}\otimes\sigma_{z}$ &($-1, +1, -1,-1,  -1, + 1$)\\%
&$i\sigma_{z}$ & $-i\sigma_{y}$&$i\sigma_{z}$ &$\sigma_{z}$ &($-1,+ 1, -1,-1,  -1,  -1$)\\
&$-\sigma_{z}$ & $-\sigma_{z}$&$-\sigma_{x}$ &$\sigma_{x}$ &($-1, -1,+ 1,+1, + 1, + 1$)\\%
&$i\sigma_{z}\otimes I$ & $\sigma_{x}\otimes\sigma_{x}$&$I\otimes\sigma_{z}$ &$-I\otimes\sigma_{x}$ &($-1, -1, +1,+1, + 1,  -1$)\\
&$-\sigma_{z}\otimes \sigma_{z}$ & $I\otimes i\sigma_{y}$&$\sigma_{z}\otimes\sigma_{y}$ &$\sigma_{x}\otimes\sigma_{y}$&($-1, -1, +1,+1,  -1, + 1$)\\%
&$i\sigma_{z}$ & $-i\sigma_{z}$&$I$ &$-\sigma_{y}$ &($-1, -1, +1,+1,  -1,  -1$)\\
&$\sigma_{z}$ & $-\sigma_{y}$&$-\sigma_{x}$ &$iI$ &($-1, -1,+ 1,-1,  + 1, +1$)\\%
&$i\sigma_{z}\otimes \sigma_{z}$ & $-\sigma_{z}\otimes I$&$-\sigma_{x}\otimes\sigma_{y}$ &$i\sigma_{x}\otimes\sigma_{z}$ &($-1, -1, +1,-1,  + 1, -1$)\\
&$-\sigma_{z}\otimes \sigma_{z}$ & $-iI\otimes\sigma_{z}$&$-\sigma_{x}\otimes\sigma_{z}$ &$-iI\otimes\sigma_{x}$ &($-1, -1,  +1,-1,  -1, + 1$)\\%
&$-i\sigma_{z}$ & $i\sigma_{y}$&$-I$ &$-iI$ &($-1, -1, + 1,-1,  -1,  -1$)\\
&$I$ & $I$&$i\sigma_{z}$ &$\sigma_{x}$ &($-1, -1, -1, +1, + 1,  +1$)\\%
&$-iI\otimes \sigma_{z}$ & $-\sigma_{y}\otimes\sigma_{x}$&$\sigma_{x}\otimes i\sigma_{y}$ &$-I\otimes\sigma_{y}$ &($-1, -1, -1, +1,  +1, -1$)\\
&$-\sigma_{z}\otimes \sigma_{z}$ & $i\sigma_{y}\otimes I$&$i\sigma_{z}\otimes I$ &$\sigma_{z}\otimes\sigma_{y}$&($-1, -1, -1, +1,  -1, + 1$)\\%
&$-i\sigma_{z}$ & $-i\sigma_{z}$&$i\sigma_{x}$ &$\sigma_{z}$ &($-1, -1, -1, +1,  -1,  -1$)\\
&$I\otimes \sigma_{z}$ & $-\sigma_{x}\otimes\sigma_{x}$&$-i\sigma_{z}\otimes\sigma_{z}$ &$I\otimes i\sigma_{y}$&($-1, -1, -1,-1,   +1, + 1$)\\%
&$i\sigma_{z}$ & $I$&$i\sigma_{y}$ &$-iI$ &($-1, -1, -1,-1,  + 1, -1$)\\
&$I$ & $-i\sigma_{z}$&$i\sigma_{y}$ &$iI$& ($-1, -1, -1,-1,  -1, +1$)\\%
&$-i\sigma_{z}\otimes \sigma_{z}$ & $I\otimes i\sigma_{y}$&$-iI\otimes\sigma_{x}$ &$i\sigma_{x}\otimes\sigma_{z}$&($-1,-1,-1,-1,-1,-1$)\\
\hline
\end{tabular}
\end{table*}

\begin{table*}[htbp]
\footnotesize
 \centering
 \begin{tabular}{ |c||ccc|c| }
  \hline
Anti-unitary group & & Rep of generators & & label of classification\\
\hline
$Z_{2}^{T}$ &$T$& & &  \\
\hline
&$i\sigma_{y}K$ & & &$-1$ \\
\hline
 $Z_{2}\times Z_{2}^{T}\simeq \mathscr C_{2h}^T$ &$T$&$P$& & \\
 \hline
 &$\sigma_{x}K$ & $i\sigma_{z}$& &($+1,-1$) \\
 &$\sigma_{y}K$ & $i\sigma_{z}$& &($-1,+1$) \\%
 &$\sigma_{y}K$ & $I$& & ($-1,-1$)\\
 \hline
$Z_2\times Z_2\times Z_{2}^{T}\simeq \mathscr D_{2h}^T$ &$T$&$P$&$Q$&\\ 
 \hline
 & $\sigma_{y}K$&$i\sigma_{z}$ & $-i\sigma_{z}$&($+1,+1,+1,-1$)\\
  &$\sigma_{y}K$&$i\sigma_{z}$ & $I$&($+1,+1,-1,+1$) \\%
 &$\sigma_{x}K$&$I$ & $i\sigma_{z}$&($+1,+1,-1,-1$) \\
  &$\sigma_{y}K$&$\sigma_{z}$ & $\sigma_{x}$&($+1,-1,+1,+1$) \\%
 & $IK$&$i\sigma_{z}$ & $i\sigma_{x}$&($+1,-1,+1,-1$)\\
  &$IK$&$i\sigma_{z}$ & $\sigma_{y}$&($+1,-1,-1,+1$) \\%
 & $\sigma_{y}\otimes\sigma_{x}K$&$I\otimes \sigma_{z}$ & $iI\otimes\sigma_{x}$&($+1,-1,-1,-1$)\\
  &$\sigma_{y}K$&$I$ & $i\sigma_{z}$&($-1,+1,+1,+1$)\\%
 &$\sigma_{x}K$&$i\sigma_{z}$ & $I$&($-1,+1,+1,-1$) \\
  &$\sigma_{x}K$&$i\sigma_{z}$ & $i\sigma_{z}$&($-1,+1,-1,+1$) \\%
 &$\sigma_{y}K$&$I$ & $I$&($-1,+1,-1,-1$) \\
  &$IK$&$\sigma_{y}$ & $i\sigma_{z}$&($-1,-1,+1,+1$) \\%
 &$\sigma_{y}\otimes\sigma_{x}K$ & $i\sigma_{z}\otimes\sigma_{z}$ &$\sigma_{z}\otimes\sigma_{x}$&($-1,-1,+1,-1$) \\
 &$\sigma_{y}\otimes \sigma_{x}K$& $i\sigma_{z}\otimes\sigma_{z}$&$iI\otimes\sigma_{y}$ &($-1,-1,-1,+1$) \\%
 &$\sigma_{x}\otimes \sigma_{x}K$&$I\otimes\sigma_{z}$ & $\sigma_{z}\otimes\sigma_{y}$&($-1,-1,-1,-1$) \\
\hline
$(Z_2\times Z_2)\rtimes Z_{2}^{T}\simeq \mathscr D_{2d}^T$ &$T$&$P$&$Q$&\\
\hline
&$\left(\begin{array}{cc}e^{-i\frac{\pi}{4}}&0\\0&e^{i\frac{\pi}{4}}\end{array}\right)K$ &$-i\sigma_z$&$i\sigma_y$&$(+1,-1)$\\
&$\sigma_yK$ &$-I$&$iI$&$(-1,+1)$\\
&$\sigma_y\otimes\left(\begin{array}{cc}e^{-i\frac{\pi}{4}}&0\\0&e^{i\frac{\pi}{4}}\end{array}\right)K$ &$iI\otimes \sigma_z$&$-I\otimes\sigma_y$&$(-1,-1)$\\ %
\hline
$Z_{4}\times Z_{2}^{T}\simeq \mathscr C_{4h}^T$ &$T$ &$P$& & \\
\hline
&$\sigma_{y}K$ &$\left(\begin{array}{cc}
e^{-i\frac{\pi}{4}}&0\\0&e^{i\frac{\pi}{4}}\end{array}\right)$ & &($+1,-1$)\\
&$\sigma_{y}K$& $I$ & &($-1,+1$)\\%
&$\sigma_{x}K$ &$\left(\begin{array}{cc}
e^{-i\frac{\pi}{4}}&0\\0&e^{i\frac{\pi}{4}}\end{array}\right)$ & &($-1,-1$)\\
\hline
$Z_{4}\rtimes Z_{2}^{T}\simeq \mathscr C_{4v}^T$ &$T$ &$P$ && \\ 
\hline
&$\sigma_{x}K$ &$e^{-i\frac{\pi}{4}} \sigma_z$
& &($+1,-1$) \\
&$\sigma_{y}K$ &$e^{i\frac{5\pi}{4}} \sigma_z$ 
& &($-1,+1$) \\ %
&$\sigma_{y}K$ &$iI$& &($-1,-1$)\\
 \hline

$Z_{3}\times(Z_{3}\rtimes Z_{2}^{T})$ & & & &  \\
$\simeq(Z_3\times Z_3)\rtimes Z_2^T$ &$T$ &$P$&$Q$& \\ %
\hline
&$i\sigma_{y}\otimes\left(\begin{array}{ccc}
 \Omega^8&0&0\\
 0& \Omega^5&0\\
 0&0& \Omega^2\end{array}\right)K$
&\ \ \ $-I\otimes\left(\begin{array}{ccc}
0&0&1\\ 1&0&0\\0&1&0 \end{array}\right)$
&$-I\otimes\left(\begin{array}{ccc}
\Omega&0&0\\0&\Omega^7&0\\
0&0&\Omega^4
\end{array}\right)$ &$\omega^{\frac{1}{2}}$ \\ %
&\ \ \ \ \ \ \ \ $\left(\begin{array}{ccc}
\Omega&0&0\\0&\Omega^{4}&0\\0&0&\Omega^{7} \end{array}\right)K$
&\ \ \ \ \ \ \ \ \ \ $\left(\begin{array}{ccc}
0&0&1\\ 1&0&0\\0&1&0 \end{array}\right) $
&\ \ \ \ \ \ $\left(\begin{array}{ccc}
\Omega^{8}&0&0\\0&\Omega^2&0\\0&0&\Omega^{5} \end{array}\right) $&$\omega$\\%
&$i\sigma_{y}K$ &\ \ \ \ \ \ $ -I$ &\ \ \ \ \ \ \ $-I$ &$\omega^{3\over2}$ \\%
&\ \ \ \ \ \ $\left(\begin{array}{ccc}
\Omega^{8}&0&0\\0&\Omega^{5}&0\\0&0&\Omega^{2} \end{array}\right)K$
&\ \ \ \ \ \ \ \ \ \  $\left(\begin{array}{ccc}
0&0&1\\ 1&0&0\\0&1&0 \end{array}\right) $
&\ \ \ \ \ \ \ $\left(\begin{array}{ccc}
\Omega&0&0\\0&\Omega^7&0\\0&0&\Omega^{4} \end{array}\right) $&$\omega^2$\\%
&$i\sigma_{y}\otimes\left(\begin{array}{ccc}
 \Omega &0&0\\
 0& \Omega^4&0\\
 0&0& \Omega^7\end{array}\right)K$
&\ \ \ $-I\otimes\left(\begin{array}{ccc}
0&0&1\\ 1&0&0\\0&1&0 \end{array}\right)$
&$-I\otimes\left(\begin{array}{ccc}
\Omega^8&0&0\\0&\Omega^2&0\\
0&0&\Omega^{5}
\end{array}\right)$
&$\omega^{\frac{5}{2}}$ \\%
\hline

$A_4\rtimes Z_2^T\simeq S_4^T\simeq \mathscr T_{d}^{T}$ &$T=(12)K$&$P=(123)$ &$Q=(124)$ &\\
\hline
& $IK$
&$\left(\begin{array}{cc}
\omega&0\\0& \omega^2\end{array}\right)$
& $\frac{\sqrt{3}}{3}\left(\begin{array}{cc}
 e^{-i\frac{\pi}{6}} &\sqrt{2}i\\ \sqrt{2}i & e^{i\frac{\pi}{6}} \end{array}\right)$ &($+1,-1$)\\%
&$\sigma_{y}K$ &$-I$ & $I$&($-1,+1$)\\%
& $\sigma_{y}\otimes IK$
&$-I\otimes\left(\begin{array}{cc}
\omega&0\\0&\omega^2\end{array}\right)$
& $\frac{\sqrt{3}}{3}I\otimes\left(\begin{array}{cc}
 e^{-i\frac{\pi}{6}} &  \sqrt{2}i\\ \sqrt{2}i& e^{i\frac{\pi}{6}} \end{array}\right)$ &($-1,-1$)\\
\hline

$ A_4\times Z_2^T$ &$T $&$P=(123)$ &$Q=(124)$ &\\
\hline
& $\sigma_{y}K$
&$-I$
& $I$ &($+1,-1$)\\%
&$\sigma_{y}\otimes \sigma_{y}K$
&$I\otimes\left(\begin{array}{cc}
\omega&0\\0&\omega^2\end{array}\right)$
& $\frac{\sqrt{3}}{3}I\otimes\left(\begin{array}{cc}
 e^{-i\frac{\pi}{6}} &  \sqrt{2}i\\ \sqrt{2}i& e^{i\frac{\pi}{6}} \end{array}\right)$ &($-1,+1$)\\%
& $\sigma_{y}K$
&$-\left(\begin{array}{cc}
\omega&0\\0&\omega^2\end{array}\right)$
& $\frac{\sqrt{3}}{3}\left(\begin{array}{cc}
 e^{-i\frac{\pi}{6}} &  \sqrt{2}i\\ \sqrt{2}i& e^{i\frac{\pi}{6}} \end{array}\right)$ &($-1,-1$)\\

\hline
\end{tabular}
\end{table*}

\begin{table*}[htbp]
\centering
\begin{tabular}{ |c||ccc|c| }
\hline
Fermionic anti-unitary groups && Reps  &\ \ \ \ \ \ \ \ \ \  \ \ \ \ \ \ \ \ \ \ & label of classification\\
\hline
$Z_{4}^{T}$ &$T$&&  & \\
 \hline
 &$\left(\begin{array}{cc}
0&e^{i\frac{\pi}{4}}\\e^{-i\frac{\pi}{4}}&0\end{array}\right)K$&  & &$-1$\\
\hline

$Z_2\times Z_{4}^{T}$ &$T$&&$P$  &\\
 \hline
 &$\left(\begin{array}{cc}
0&e^{i\frac{\pi}{4}}\\e^{-i\frac{\pi}{4}}&0\end{array}\right)K$&&$I$  &($+1,-1$) \\%
&$\sigma_{y}K$&&$-i\sigma_{z}$ & ($-1,+1$) \\%
&$\left(\begin{array}{cc}
0&e^{i\frac{\pi}{4}}\\e^{-i\frac{\pi}{4}}&0\end{array}\right)K$&&$i\sigma_{z}$  &($-1,-1$) \\%
\hline
$G_-^-(Z_4,T)$& $T$&& $P$&\\
\hline
&$\sigma_{y}K$&&$i\sigma_{z}$ &$-1$ \\
\hline
\end{tabular}
\end{table*}

\subsection{CG coefficients for projective Reps}

The direct product of two projective Reps $(\nu_1)$ and $(\nu_2)$ of group $G$ is usually a reducible (projective) Rep. Once the direct product Rep is reduced into irreducible (projective) Reps,
\[
\mathscr C^\dag M^{(\nu_1)}(g) \otimes M^{(\nu_2)}(g) K_{s(g) }\mathscr C =\bigoplus_{\nu_3} M^{(\nu_3)}(g)K_{s(g)}
\]
we can obtain the corresponding Clebsch-Gordan (CG) coefficients $\mathscr C$. The eigenfunction method can also be applied to calculate the CG coefficients. Here we will not give details of the calculations.


\section{Applications of projective Reps}\label{sec:apply}

In this section, we will summarize some physical applications of irreducible projective Reps.

\subsection{ 1-D SPT phases}\label{sec2}

The well studied spin-1 Haldane phase\cite{HaldaneA, HaldaneL,AKLT} was known for its disordered gapped ground state and its spin-1/2 edge states at the open boundaries\cite{White93,Ng94}. The exotic properties of the Haldane phase is protected by $Z_2\times Z_2$ spin rotation symmetry or time reversal symmetry\cite{GuWen09, Pollmann10}. The edge states vary projectively under the action of the symmetry group. The Haldane phase was later generalized to other 1D SPT phases with different symmetries classified by the second group cohomology\cite{CGW1D1,CGW1D2}. In Ref.\onlinecite{CGLW2}, topological nonlinear sigma model (NLSM) was generalized to finite symmetry group $G$ in discrete space-time to describe SPT phases in any spatial dimensions. In the following, based on irreducible projective Reps we will give a minimal field theory description of 1D SPT phases.

Traditionally, the spin-$1$ Haldane phase is described by the $O(3)$ NLSM with the following topological $\theta$-term \cite{HaldaneA, HaldaneL},
\[
\mathcal L_{\theta}[\pmb {n}(\pmb x, \tau)] =\theta{i\over 4\pi}  \pmb{n}\cdot (\partial_x\pmb {n}\times \partial_\tau\pmb{n}),
\]
where $\theta= {2\pi }$ and $|\pmb{n}\rangle$ is the spin coherent state $\langle\pmb{n}|\hat {\pmb S}|\pmb{n}\rangle= \pmb{n}$ which varies under rotation in the following way,
\[\hat R|\pmb n\rangle=e^{i\varphi}|R\pmb n\rangle,\]
the phase factor $e^{i\varphi}$ depends on the axis of the rotation $R$ and the gauge choice of the bases $|\pmb n\rangle$. The collection of the end points of the vector $\pmb n$ form a sphere, i.e. the symmetric space of the $SO(3)$ group $S^2=SO(3)/SO(2)$.  The Lagrange density ${i\theta\over 4\pi}  \pmb{n}\cdot (\partial_x\pmb {n}\times \partial_\tau\pmb{n})dxd\tau$ describes the `Berry phase' of a spin-$1/2$ particle evolving among three states at $(x,\tau)$,  $(x+dx,\tau)$ and  $(x,\tau+d\tau)$. If the space-time is closed, then the action amplitude $e^{-\int d {x} d\tau \mathcal{L}_{\theta}[\pmb{n}(\pmb{x},\tau)]}$ equals 1. If spacial boundary condition is open, then the Berry phase on the boundary explains the existence of spin-1/2 edge state\cite{Ng94}.

The spin-1/2 edge state varries as $M[R_{\pmb m}(\theta)] =e^{-i\pmb m\cdot\pmb{\sigma\over2} \theta}$ under $SO(3)$ spin rotation of angle $\theta$ along direction $\pmb m$. Since $M[R_{\pmb m}(\pi)]^2=M[R_{\pmb m}(2\pi)]=-1$ and the minus sign can not be gauged away,  the edge state carries a nontrivial projective Rep of the symmetry group $SO(3)$.

Above picture can be generalized even if the symmetry is a finite group $G$. Similar to the spin coherent state, we introduce the following group element labeled  bases
\Beq
&&|g^{r}\rangle = \hat g^r|\alpha_1^{(\nu)}\rangle,\\
%
&&|g^{l}\rangle = \hat g^l|\beta_1^{( {\nu}^*)}\rangle,
\Eeq
where $|\alpha_1^{(\nu)}\rangle$ (or $|\beta_1^{( {\nu}^*)}\rangle$) is one of the irreducible bases of projective Rep $(\nu)$ [or $( {\nu}^*)$, the complex conjugate Rep of $(\nu)$], $g^r$ and $g^l$ are different group elements with
\Beq
&&\hat g |g_1^{r}\rangle =\omega_2(g ,g_1^r)|g g_1^{r}\rangle,\\
&&\hat g |g_1^{l}\rangle =\omega^{-1}_2(g ,g_1^l)| g g_1^{l}\rangle.
\Eeq
The states $\{|g^{r}\rangle\}$ (or $\{|g^{l}\rangle\}$) are not necessarily orthogonal, but they form over complete bases for the Rep $(\nu)$ [or $( {\nu}^*)$] since (see appendix  \ref{app6})
\[
\sum_g |g^{r}\rangle\langle g^{r}|\propto I =\sum_{i=1}^{n_\nu}|\alpha_i^{(\nu)}\rangle\langle\alpha_i^{(\nu)}|.
\]
So summing over $|g^{r}\rangle$ is equivalent to summing over all the $n_\nu$ bases $\alpha_1^{(\nu)},...,\alpha_{n_\nu}^{(\nu)}$ of the irreducible Rep space $(\nu)$ (and similar result holds for $|g^{l}\rangle$) .

The physical degrees of freedom at site $i$ are combinations of the bases $|g^{l}_i\rangle$ and $|g^{r}_i\rangle$
\[
|g^l_i,g^r_i\rangle^{\rm p} = \omega_2(g^l_i, (g_i^l) ^{-1}g_i^r )|g_i^l,g_i^r\rangle,
\]
such that $|g^l_i,g^r_i\rangle^{\rm p}$ varies under group action in a way similar to $|\pmb n\rangle$ varies under rotation (the difference is that there is no phase factor $e^{i\varphi}$ here),
\[
\hat g|g^l_i,g^r_i\rangle^{\rm p} = \omega_2(g^l_i,(g_i^l)^{-1}g_i^r ){\omega_2(g,g_i^r)\over\omega_2(g,g_i^l)}|gg_i^l,gg_i^r\rangle=|gg_i^l, gg_i^r\rangle^{\rm p}.
\]
In the bond between neighboring sites $i$ and $i+1$, the degrees of freedom $g^r_i$ and $g^l_{i+1}$ are locked by the constraint $g^r_i=g^l_{i+1}$ owning to strong interactions. 

\begin{figure}[t]
\centering
\includegraphics[width=3.4in]{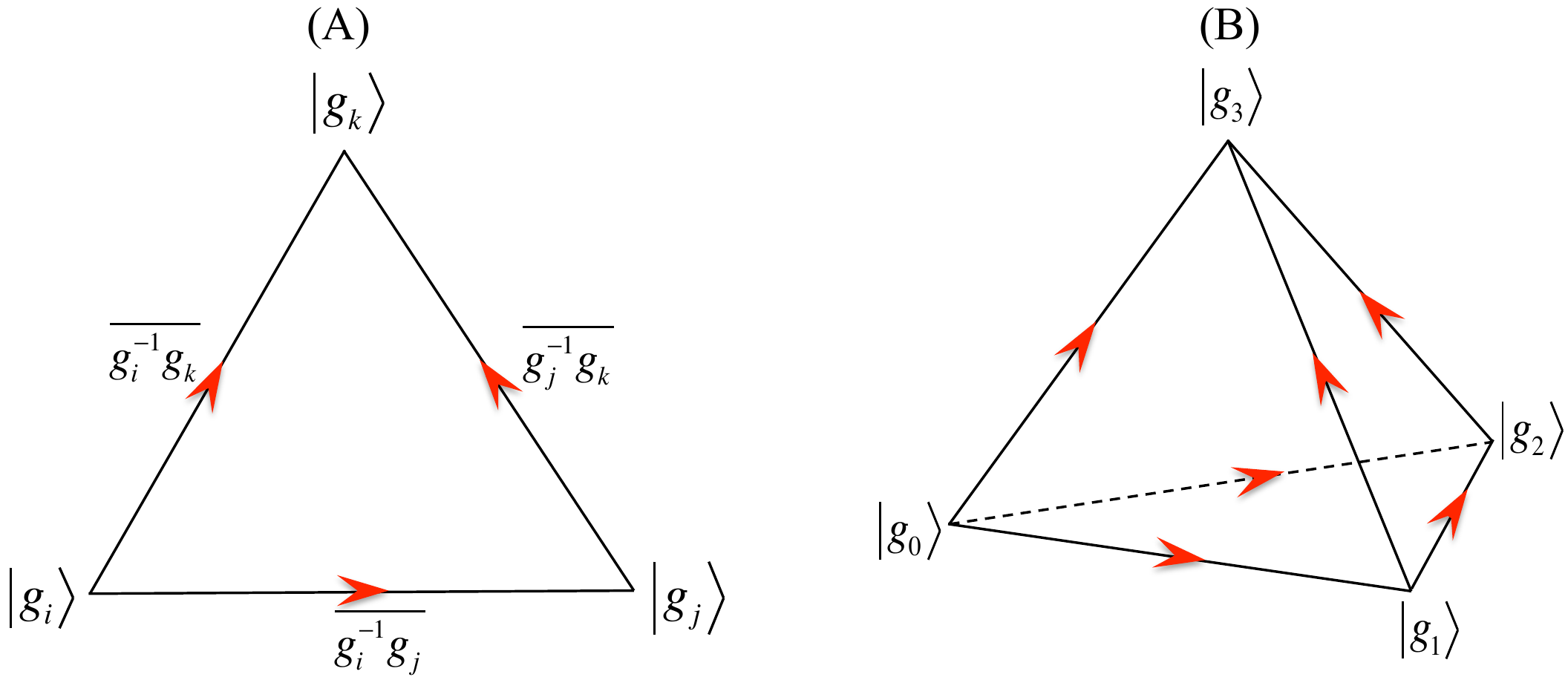}
\caption{(A)Aharonov-Bohm phase as the topological term; (B)Quantized topological action amplitude on closed space-time.} \label{fig:AB}
\end{figure}

We discretize the space-time and put a variable $g_i^r$ 
at each space-time point $i$ and omit $g_i^l$ 
since it is the same as $g_{i-1}^r$. 
(Without causing confusion, we can eliminate the superscript $r$ in the following.) The action amplitude for a space-time unit $(ijk)$ is defined as the Aharonov-Bohm phase of the basis $|g_i\rangle$ evolving between the two paths $\overline{(g_j^{-1}g_k)}\ \overline{(g_i^{-1}g_j)}|g_i\rangle$ and $\overline{(g_i^{-1}g_k)}|g_i\rangle$, see Fig.\ref{fig:AB}(A),
\Beq
\varphi(|{g}_i \rangle, |{g} _j\rangle, |{g} _k\rangle) &=& M(\overline{g_j^{-1}g_k})M(\overline{g_i^{-1}g_j})\left[M(\overline{g_i^{-1}g_k})\right]^{-1}\\
&=&\omega_2^{s(g_i)}(g_i^{-1}g_j, g_j^{-1}g_k).
\Eeq
The total action amplitude stands for the topological phase factor (the `Berry phase') of the physical degrees of freedom and can be used to describe the low energy effective field theory of the system.


Neglecting dynamic terms, we obtain the fixed point partition function of the corresponding SPT phase,
\begin{eqnarray}
Z\propto\sum_{\{g_{i}\}}e^{ - S(\{g_{i}\}) }
&=&\sum_{\{g_{i}\}}\prod_{\{ijk\}}\varphi^{s_{ijk}}(|{g} _{i}\rangle,|{g} _{j}\rangle,|{g} _{k}\rangle)\nonumber\\
\end{eqnarray}
where $s_{ijk}$ is the orientation of the triangle $(ijk)$, which is equal to 1 if it is pointing outside and $-1$ otherwise.

Similar to the $\theta$-term of the $O(3)$ NLSM, the total action amplitude is normalized if the space-time is closed (the simplest case is the surface of a tetrahedron, see Fig.\ref{fig:AB}(B)),
\Beq
&&\varphi^{s_{012}}(|{g} _0\rangle, |{g}_1 \rangle, |{g}_2 \rangle) \varphi^{s_{013}}(|{g}_0 \rangle, |{g}_1 \rangle, |{g}_3 \rangle) \\
&&\times \varphi^{s_{023}}(|{g}_0 \rangle, |{g}_2 \rangle, |{g}_3 \rangle) \varphi^{s_{123}}(|{g}_1 \rangle, |{g}_2 \rangle, |{g}_3 \rangle) \\
&&=1
\Eeq
owning to the cocycle equation
\Beq
(d\omega_2)(g_0^{-1}g_1, g_1^{-1}g_2, g_2^{-1}g_3)=1.
\Eeq

The partition function can be regarded as imaginary time evolution operator $Z=U(0,\tau)=|\psi^\tau\rangle\langle\psi^0|$, where $|\psi^\tau\rangle$ is the ground state at time $\tau$. After some calculation we can write out the ground state (under periodic boundary condition) as
\Beq
|\psi\rangle &=&\sum_{\{g_ig_jg_k\}} \prod_{\{ijk\}}B^{-1}(|E\rangle,|{g}^r _{i}\rangle,|{g}^r _{j}\rangle)|{g} _{i}^l{g} _{i}^r{g} _{j}^l{g} _j^r{g} _{k}^l{g} _k^r...\rangle^{\rm p}
\\
&=&\sum_{\{g_ig_jg_k\}} \prod_{\{ijk\}}\omega_2^{-1} ({g}_i^r,(g_i^r)^{-1}g_j^r)|{g} _{i}^l{g} _{i}^r{g} _{j}^l{g} _j^r{g} _{k}^l{g} _k^r...\rangle^{\rm p}.
\Eeq
Noticing that in 1D $j=i+1,\ k=j+1,\ ....$,  the wave function $\omega_2^{-1} ({g}_i^r,(g_i^r)^{-1}g_j^r)=\omega_2^{-1} ({g}_j^l,(g_j^l)^{-1}g_j^r)$ is the CG coefficient that fractionalize the physical degrees of freedom  into two projective Reps
\[
|g_j^lg_j^r\rangle=\omega_2^{-1}({g}_j^l,(g_j^l)^{-1}g_j^r)|g_j^lg_j^r\rangle^{\rm p},
\]
the ground state wave function can also be written in forms of product of dimers
\Beq
|\psi\rangle =\sum_{\{g_i^lg_i^r...\}} |...{g} _{i}^l)({g} _{i}^r{g} _{j}^l)({g} _j^r{g} _{k}^l)({g} _k^r...\rangle,
\Eeq
where $g_i^r=g_j^l, g_j^r=g_k^l,...$ and each bracket means a singlet (or a dimer) on a bond between neighboring sites. The dangling degrees of freedom on the ends stand for the edge states which carry projective Reps of the symmetry group.

From the above  ground  state wave function of SPT phase, it is easily seen that the fixed point parent Hamiltonian is constructed by projector onto the bond singlets\cite{CGLW2}. If we further project the physical degrees of freedom to its subspace, then we can obtain an AKLT-type state, and the parent Hamiltonian can also be constructed using projection operators\cite{LLW12}.

\subsection{Defects in 2D topological phases}

Except for 1D SPT phases, projective Reps also have applications in 2D topological phases, including SPT phases and intrinsic topological phases. In the following we will give several examples.

\subsubsection{Vortices in topological superconductors (fermionic SPT phases)}

It is known that the vortices of $p+ip$ topological superconductor\cite{ReadGreen2000} carry Majorana zero modes and the degeneracy of the wave function depends on the number of spatially separated vortices. In the following, we will show that if we interpret each Majorana zero mode as a `symmetry' operation, then the set of symmetry operations form an Abelian group, and the degeneracy of the Majorana Hilbert space can be understood as the projective Rep of this group.

Supposing the majorana zero mode $\gamma_1$ is an eigen state of a Hamiltonian $H$ with $[\gamma_1, H]=0$, then we can define an operation $\hat \Gamma_1$ corresponding to the majorana mode $\gamma_1$,
\[\hat\Gamma_1(O)=\gamma_1O\gamma_1,\]
where $O$ is an arbitrary operator.  Obviously,
\Beq
&&\hat\Gamma_1(\gamma_1)=\gamma_1 \gamma_1 \gamma_1 = \gamma_1,\\
&&\hat\Gamma_1(\gamma_i)=\gamma_1 \gamma_i \gamma_1 = -\gamma_i, \ \ {\rm for \ any\ }\langle \gamma_1|\gamma_i\rangle=0.
\Eeq
Since $\hat\Gamma_1(H)=\gamma_1 H \gamma_1 = H$, the operation $\hat\Gamma_1$ is a `symmetry operation' of the system. The eigenvalues of $\hat \Gamma_1$ are $\pm1$ and $\hat \Gamma_1^2=I$, so $\hat \Gamma_1$ generates a $Z_2$ `symmetry group'.

It should be mentioned that the form of $\hat \Gamma_i$ ( or $\gamma_1$) depends on the details of $H$, this $Z_2$ `symmetry group' is not a symmetry in the usual sense. The existence of this special $Z_2$ symmetry is a consequence of the nontrivial winding number of the $p+ip$ superconductor and the presence of topological defect (i.e. vortex) where the Majorana mode $\gamma_1$ locates\cite{ReadGreen2000}.

If $\gamma_2$ is another majorana zero mode of the Hamiltonian, then it defines another $Z_2$ symmetry group generated by \[\hat\Gamma_2(O)=\gamma_2O\gamma_2,\] and it is easily checked that $\hat\Gamma_1\hat\Gamma_2(O)=\hat\Gamma_2\hat\Gamma_1(O)$, namely,
\[\hat \Gamma_1\hat \Gamma_2=\hat \Gamma_2\hat \Gamma_1.\]
So, if $H$ contains two Majorana zero modes $\gamma_1$ and $\gamma_2$, then it has a $Z_2\times Z_2$ `symmetry group'. The two majorana zero modes result in two-fold degeneracy of the ground states. In the ground state subspace, the operators $\hat \Gamma_{1}, \hat \Gamma_{2}$ act projectively and their Rep matrices are
\[
\hat \Gamma_1\to M(\Gamma_1)= \sigma_x,\ \ \ \hat \Gamma_2\to M( \Gamma_2)=\sigma_y,
\]
with $M(\Gamma_1)^2=M(\Gamma_2)^2=1$ and the fermionic anti-commuting relation $\{M(\Gamma_1),M(\Gamma_2)\}=0$.

Now suppose the system has four Majorana zero modes $\gamma_1,\gamma_2,\gamma_3, \gamma_4$, then the corresponding symmetry operations $\hat\Gamma_1,\hat\Gamma_2,\hat\Gamma_3, \hat\Gamma_4$ generate a $Z_2\times Z_2\times Z_2\times Z_2$ symmetry group. The degenerate ground states carry an irreducible projective Rep of the class $(-1,-1,-1,-1,-1,-1)$ (up to a gauge transformation comparing with Table \ref{tb1})
\Beq
M(\Gamma_{1})&=&\sigma_{z}\otimes\sigma_{z} ,\\
M(\Gamma_{2})&=&I\otimes\sigma_{y} , \\
M(\Gamma_{3})&=&I\otimes\sigma_{x} ,\\
M(\Gamma_{4})&=&\sigma_{x}\otimes\sigma_{z},
\Eeq
which satisfy the relation $\{M(\Gamma_{j}),M(\Gamma_{k})\}=2\delta_{jk}$.

Similar results can be generalized to the cases with more majorana zero modes, and the degeneracy of the ground states can be understood as the existence of projective Rep of the corresponding `symmetry group'.

Above picture is still valid even if there are interactions in the Hamiltonian $H$.

\subsubsection{Monodromy defects in bosonic SPT phases}

In 2D fermionic topological insulators, a magnetic $\pi$-flux (a symmetry defect) gives rise to extra degeneracy. Actually, this picture also holds for bosonic SPT phases.

Generally, symmetry defects can not be created locally if they carry fractional symmetry flux. These defects are the end points of strings and are called the Monodromy defects in literature\cite{Wen14}. In some SPT phases, monodromy defects may carry projective Reps of the symmetry group\cite{Wen14,Decorated,Zaletel, WangGuWen16}.

{\bf Bosonic topological insulators}. Bosonic topological insulators are SPT phases protected by $U(1)\rtimes Z_2^T$ symmetry\cite{LGW14}. In 2D bosonic topological insulators, a $\pi$-flux (gauging the  $U(1)$ symmetry) carries projective Rep of the remaining symmetry $Z_2^T$ which gives rise to Kramers' degeneracy. This nontrivial result comes from the fact that in the bosonic topological insulator (which is a nontrivial SPT phase), the vortex creation operator reverses its sign under time reversal $\widehat T b_v\widehat T^{-1} = -b_v^\dag$, then
\[
\widehat T|\pi\rangle = |-\pi\rangle = \hat b_v |\pi\rangle,
\]
\[
\widehat T |-\pi\rangle = \widehat T\hat b_v \widehat T^{-1}\widehat T|\pi\rangle=-b_v^\dag |-\pi\rangle=-|\pi\rangle,
\]
so \[\widehat T^2=-1,\] which is a projective Rep of $Z_2^T$. On the other hand, in the trivial bosonic insulator $\widehat T b_v\widehat T^{-1} = b_v^\dag$, which yields a linear Rep $\widehat T^2=1$. In this case creating $\pi$ fluxes will not cause degeneracy.

{\bf $Z_{N_1} \times Z_{N_2}\times  Z_{N_3}$ SPT phases}. Another example is the 2D bosonic SPT phases with symmetry group $G=Z_{N_1} \times Z_{N_2}\times  Z_{N_3}$ (where $N_1,N_2,N_3$ are integers) with generator $h_1,h_2,h_3$ respectively. The SPT phases are classified by the third group cohomology $\mathcal H^3(G,U(1))$ and each nontrivial phase corresponds to a nontrivial class of 3-cocycles\cite{CGLW1, CGLW2}. We will focus on the type-III SPT phases described by the type-III cocycles \cite{95thesis}. If we label a group element $A$ as $A=(a_1,a_2,a_3)=(h_1^{a_1},h_2^{a_2},h_3^{a_3})$ with $0\leq a_i\leq N_i-1$, then a type-III cocycle takes the following form\cite{95thesis},
\[
\omega_3(A,B,C)=e^{2\pi i{p^{III}\over N_{123}} a_1b_2c_3}, \ \ 1\leq p^{III}\leq N_{123},
\]
where $N_{123}$ is the greatest common devisor of $N_1, N_2, N_3$.

From above 3-cocyle, one can construct a 2-cocyle from the slant product $i_A$:
\Beq
\chi_A(B,C) &=& (i_A\omega_3)(B,C)\\
&=&\omega_3(A,B,C)\omega_3^{-1}(B,A,C)\omega_3(B,C,A)\\
&=&e^{2\pi i{p^{III}\over N_{123}}(a_1b_2c_3-b_1a_2c_3+b_1c_2a_3)}.
\Eeq
Since $i_A$ commutes with the coboundary operator $d$, $\chi_A(B,C)$ is a 2-cocycle: $(d \chi_A)(D,B,C)=d i_A \omega_3(D,B,C)=i_A d\omega_3(D,B,C)=1$. Furthermore, if $p^{III}\neq0$ then $\chi_A(B,C)$ is a nontrivial 2-cocyle since it agrees with the analytic solution in Eq.~(\ref{2cocycle-analytic}). Specially, if $A\in Z_{N_1}$ with $a_2=a_3=0$, then $\chi_A(B,C)$ is a nontrivial 2-cocyle of the subgroup $Z_{N_2}\times Z_{N_3}$.

The physical meaning of above result is that  in the SPT phases corresponding to nonzero $p^{III}$, the symmetry fluxes of $Z_{N_1}$ carry nontrivial projective Reps (thus give rise to degeneracy of the energy spectrum) of the group $Z_{N_2}\times Z_{N_3}$\cite{Decorated, Zaletel}. Similarly, the $Z_{N_2}$ (or $Z_{N_3}$) fluxes carry projective Reps of the group $Z_{N_1}\times Z_{N_3}$ (or $Z_{N_1}\times Z_{N_2}$). Similar conclusion also holds for certain SPT phases with non-Abelian symmetry groups.

\subsubsection{$2$D topological order/SET phases}\label{sec:SET}

We have shown that in SPT  phases, Monodromy defects may carry projective Reps of the symmetry group (or subgroup of the symmetry).
If the system carries intrinsic topological order, then point-like excitations, such as spinons (`electronic' excitations) or visons (`magnetic' excitations), can carry projective Reps of the symmetry group. Topologically ordered phases enriched by symmetry are called symmetry enriched topological (SET) phases.

An example is the projective symmetry group (PSG)\cite{ZhouWen02, WenRMP06} in quantum spin liquid with lattice space group symmetry. In the parton construction of spin-1/2 quantum spin liquids, the system has an $SU(2)$ gauge symmetry. At mean field level, this gauge symmetry may break down to $U(1)$ or $Z_2$ (called invariant gauge group, or IGG in brief) owning to paring between the spinons. Generally the space group is no longer the symmetry of the mean field ground state, but the mean field state is invariant under space group operation followed by a gauge transformation. The symmetry elements, which are combinations of space group operations and the corresponding gauge transformations, form the PSG. In PSG, the multiplication rule of two space group operations are twisted by gauge transformations in the IGG.  As a result, the PSG can be classified by the second group cohomology $\mathcal H^2(SG, IGG)$ (for details, see Refs. \onlinecite{WenRMP06, Ran13}).

PSG actually describes how the spinons (the `electronic' excitations) carry fractional charge of the symmetry group. On the other hand, visons (the `magnetic' excitations), or bond states of vison and spinon, may also carry projective Reps of the symmetry group\cite{QiChFa, LiuTu15}. For example, in the $SO(3)$ non-abelian chiral spin liquid, a vison not only carries non-Abelian anyon, but also carries spin-1/2 degrees of freedom, which is a projective Rep of the symmetry group $SO(3)$. 
As a result, a more complete classification of SET phases with symmetry group $G$ is $\mathcal H^2(G,\mathcal A)$\cite{ChengWang,Chen}, where $\mathcal A$ is the fusion group of abelian anyons in the raw topological order. A subtle issue is that obstructions may exist in realizing all the phases classified by $\mathcal H^2(G,\mathcal A)$ in pure 2D\cite{ChengWang,Chen}.

\subsection{Sign problems in Quantum Monte Carlo} 

{\bf QMC simulations.} The generalized Schur's lemma corresponding to projective Reps of anti-unitary groups can also be applied to search for models which are free of sign problems in quantum Monte Carlo (QMC) simulations\cite{CJWu05,Yao15,Yao16, Xiang16, WangLei15}. By introducing the Hubbard-Stratonovich field $\xi(x,\tau)$, interacting fermion models can be mapped to free fermion Hamiltonians. The Boltzmann weight $\rho(\xi)$ of the auxiliary field configuration $\xi$ can be obtained by tracing out the fermions. Sign problem occurs when $\rho(\xi)<0$, in which case one have trouble to simulate the quantum system in a Markov process.

When the fermion number is conserved in the Hubbard-Stratonovich Hamiltonian
\beq\label{HC}
H(\xi)=C^\dag \mathscr H(\xi)C
\eeq
where $C=(c_1,...,c_N)^T$ are fermion bases, then the Boltzmann weight in the QMC simulation is given by 
\[
\rho(\xi) = {\rm Tr}\ [{\mathbf T}e^{-\int_0^\beta  H(\xi) d\tau}] =  \det [I+\mathscr B(\xi)], 
\]
where 
$\mathscr B(\xi)
={\mathbf T}e^{-\int_0^\beta \mathscr H(\xi) d\tau}$,  $\mathbf T$ means time ordered integration, $I$ is an identity matrix, $1/\beta$ is the temperature and $\xi$ is space-time dependent Hubbard-Stratonovich field.  If the matrix $\mathscr H$, and then the matrix $(I+\mathscr B)$, are commuting with the projective representation of an anti-unitary symmetry group $Z_2^T=\{E,T\}$ with 
$\widehat T^2=-1$, 
then according to the generalized Schur's lemma and its corollary, the eigenvalues of $(I+\mathscr B)$ are either pairs of complex conjugate numbers, or even-fold degenerate (the Kramers degeneracy) real numbers, or a mixture of them.  As a consequence, the determinant must be a non-negative number\cite{CJWu05}, i.e. $\rho(\xi)$ is non-negative and the model is free of sign problem under QMC simulation.

If fermion pairing terms exist after the Hubbard-Stratonovich decomposition, one can use Majorana bases to write the decoupled Hamiltonian as $H(\xi)=\Gamma^T \mathscr H(\xi)\Gamma$, where $\Gamma=(\gamma_1,...,\gamma_{2N})^T$ are Majorana bases and $\mathscr H$ is a skew symmetric matrix $\mathscr H^T=-\mathscr H$ at half-filling (consequently all the nonzero eigenvalues of $\mathscr H$ appear in pairs of $\pm\lambda$) \cite{Yao15}. In this case, the Boltzmann weigh is given by
\[
\rho(\xi) =  [\det (\mathscr B^{-1}+\mathscr B)]^{1\over2},
\]
where $\mathscr B(\xi)=\mathbf Te^{-2\int_0^\beta \mathscr H(\xi)d\tau} = e^{-\beta \mathscr {\bar H(\xi)}}$, $\mathscr {\bar H}(\xi)$ is a skew symmetric matrix. The Boltzmann weigh can also be written in forms of the eigenvalues $\lambda_k$ of $\mathscr {\bar H}$ as $\rho(\xi) = \prod_k(e^{2\beta\lambda_k} +e^{-2\beta\lambda_k})$, where any pair of $(\lambda_k,-\lambda_k)$ are counted only once in $\prod_k$. If $\mathscr H$ (and then $\mathscr {\bar H}$) commute with the projective Rep of $Z_2^T$ symmetry with $\widehat T^2=-1$, both $(\lambda_k,  -\lambda_k)$ and $(\lambda_k^*, -\lambda_k^*)$ are eigenvalues of $\mathscr {\bar H}$ when $\lambda_k$ is not purely imaginary. However, if $\lambda_k$ is purely imaginary, then $\lambda_k^*=-\lambda_k$ and it is possible that only $(\lambda_k, -\lambda_k)$ appear in the eigenvalues. 
In that case, sign problem appears. To avoid this problem, one can enlarge the symmetry group.


The simplest case is to enlarge the symmetry group as $G=Z_2\times Z_2^T=\{E, P, T, PT\}$ and let $\mathscr H$ commute with a non-trivial projective Rep of $G$. In Majorana bases,  the representations are required to be real orthogonal matrices such that after the group action the bases are still Majorana fermions. Under this constraint, the projective Reps are classified by $\mathcal H^2(G, Z_2)=\mathbb Z_2^3$, which contains 7 nontrivial classes as shown in Table \ref{proj_real} 
(it should be noted that real orthogonality is a sufficient but not necessary condition for $Z_2$ coefficient projective Reps, namely,  real orthogonal projective Reps are classified by $\mathcal H^2(G, Z_2)$ but a $Z_2$ coefficient projective Rep is not necessarily real orthogonal). In the class $(-1,+1,-1)$ irreducible projective Rep where $\{\widehat T, \widehat {PT}\}=0$ and $\widehat T^2=-1$, $(\widehat{ PT})^2=1$, one cannot find a 2-dimensional (diagonal) block of $\bar{\mathscr H}$ which satisfies the following conditions: it is skew-symmetric and commutes with the Rep of $G$. The smallest block satisfying above conditions is at least 4-dimensional, i.e. it spans the direct sum space of at least two block of irreducible projective Reps $I\otimes M(g)K_{s(g)}$.
The 4-dimensional block takes the form $i\sigma_y\otimes (a I+b i\sigma_z)$ which has eigenvalues $b\pm ia$ and $-b\pm ia$, where $a,b$ are real numbers. This will not cause sign problem even if all the eigenvalues are purely imaginary.  Another way to see the absence of sign problem is to use the eigenstates of $P$ to block diagonalize $\bar{\mathscr H}$ into two parts that are mutually complex conjugate to each other\cite{Yao16}. The class $(+1,-1,-1)$ gives the same result since it defers from the class $(-1,+1,-1)$ only by switching the roles of $T$ and $PT$. These symmetry classes are called the Majorana class\cite{Yao16}.

In the class $(-1,-1,-1)$ where $\{\widehat T, \widehat {PT}\}=0$ and $\widehat T^2=-1$, $(\widehat{ PT})^2=-1$, the Rep is 4-dimensional. {\it It can be reduced into 2-dimensional irreducible Reps by unitary transformations} but cannot be reduced by real orthogonal transformations. Since $\widehat P$ is skew-symmetric and anti-commutes with $\widehat T$, we can use the eigenstates of $P$ to transform the Hamiltonian in forms of (\ref{HC}). It is free of sign problem owning to the $\widehat T^2=-1$ symmetry. This is called the Kramers class\cite{Yao16}.

\begin{table}[t]
\caption{Real orthogonal projective Reps of the group $G=Z_2\times Z_2^T=\{E,P,T,PT\}$ have classification $\mathcal H^{2}(G,Z_2)=\mathbb Z_2^3$, which is labeled by $\left(\omega_2(T,T), \omega_2(PT,PT),\omega_2(T,PT)\right)$ respectively. We have fixed $\omega_2(TP,T)=1$, therefore $\omega_2(T,PT)=\pm1$ means that $\widehat T$ and $\widehat {TP}$ are commuting/anti-commuting, respectively.} \label{proj_real}
\centering
\begin{tabular}{ |c||ccc|c| }
\hline
 $Z_{2}\times Z_{2}^{T}$ &$T$&$P$&$TP$&  classification label\\
 \hline
 &$I K$                & $i\sigma_y$            &$i\sigma_y K$ & ($+1,-1,+1$)\\
 &$i\sigma_y K$  & $i\sigma_y$            &$I K$               & ($-1,+1,+1$)\\
 &$i\sigma_{y}K$ & $-I$                          &$i\sigma_y K$  & ($-1,-1,+1$)\\
 &$\sigma_{z}K$ & $i\sigma_y$            &$-\sigma_x K$  & ($+1,+1,-1$)\\
 &$\sigma_{x}K$ & $-\sigma_{z}$        &$-i\sigma_y K$ &($+1,-1,-1$) \\
 &$-i\sigma_{y}K$ & $\sigma_{z}$        &$\sigma_x K$  &($-1,+1,-1$) \\%
 &$i\sigma_{y}\otimes IK$ & $\sigma_z\otimes i\sigma_y$&$-\sigma_x\otimes i\sigma_y K$& ($-1,-1,-1$)\\
 \hline
\end{tabular}
\end{table}


Above results can be generalized. If particle number is conserved in $H(\xi)$, the condition $\widehat T^2=-1$ can be relaxed. The essential point of avoiding the sign problem is that the Hubbard-Stratonovich Hamiltonian $\mathscr H(\xi)$ commutes with a nontrivial projective Rep of an anti-unitary group $G$, where all the irreducible projective Reps of this class are even-dimensional.  Even if for any anti-unitary operator $Tg$ (where $g\in G$ is unitary) $[M(Tg)K]^2\neq -1$, the sign problem can still be avoided.
Here we give two examples. For the type-I anti-unitary groups,  the class $(+1,-1)$ irreducible projective Rep of the group $(Z_2\times Z_2)\rtimes Z_2^T\simeq \mathscr D_{2d}^T$  in Table \ref{tb1} is an even dimensional Rep, where $[M(T)K]^2=[M(PT)K]^2=1$ and $[M(QT)K]^2=-[M(PQT)K]^2=-i\sigma_z$.  Especially, as an example of the type-II anti-unitary groups, the nontrivial irreducible projective Rep of the group $Z_4^T$ is an even dimensional Rep and is characterized by $[M(T)K]^4=-1$. Similar even-dimensional projective Reps can also be found in other anti-unitary groups.


If particle number is not conserved in $H(\xi)$, the real orthogonal condition of the Reps of the symmetry group $G$ (which is imposed in the Majorana class and Kramers symmetry class) can be relaxed. If an anti-unitary symmetry group $G$ has a 4-dimensional (or $4n$-dimensional, $n\in Z$) irreducible unitary projective Rep, and if $H(\xi)$ commutes with it, then the model is free of sign problem. For example, the group $Z_2\times Z_2\times Z_2^T\simeq \mathscr D_{2h}^T$ has four classes of 4-dimensional irreducible projective Reps (see  Table \ref{tb1}). If $\mathscr {\bar H}$ [or $(\mathscr B^{-1}+\mathscr B)$] commutes with one of these projective Reps, the degeneracy of the eigenvalues of $(\mathscr B^{-1}+\mathscr B)$ is increased: if an eigenvalue $\Lambda$ is real, then it is 4-fold degenerate; if $\Lambda$ is complex, then both $\Lambda$ and $\Lambda^*$ are 2-fold degenerate. Resultantly, the Boltzmann weight is non-negative.

These generalizations provide hint to find new classes of 
models which are free of sign problem in quantum Monte Carlo simulation.

\subsection{Space groups, Spectrum degeneracy and other applications}

Projective Reps can also be applied in many other fields, such as space groups, spectrum degeneracy and so on.

{\bf Non-symmorphic space groups.} In the Rep theory of non-symmorphic space groups, the `little co-group' (subgroup of the point group) is represented projectively at some symmetric wave vectors, where the factor system comes naturally from fractional translations. The representation theory of space groups was thoroughly studied in literature, for instance in Ref.~\onlinecite{ChenJQRMP85} and references therein, so we will not discuss in more detail here.

{\bf Spectrum degeneracy.} Similar to the Kramers degeneracy in time reversal symmetric systems with odd number of electrons, in some cases projective Reps can also explain the degeneracy of the ground states or excited states. For example, consider a spin-1/2 system respecting $Z_2\times Z_2=\{E,R_x\}\times\{E,R_y\}$ spin-rotation symmetry, where the two $Z_2$ subgroups are generated by $\pi$ rotation along $x$ and $y$ directions respectively,
\Beq
R_x(S_m)=e^{-i\pi S_x}S_m e^{i\pi S_x},\ \ R_y(S_m)=e^{-i\pi S_y}S_m e^{i\pi S_y}.
\Eeq
More explicitly,
\Beq
&&R_x(S_x)=S_x,\ R_x(S_y)=-S_y,\ R_x(S_z)=-S_z,\\
&&R_y(S_x)=-S_x,\ R_y(S_y)= S_y,\ R_y(S_z)=-S_z.
\Eeq
If the system contains an odd number of spins, then the operators carry linear Rep of the symmetry group $Z_2\times Z_2$ but the total Hilbert space forms a (reducible) projective Rep.  As a consequence all the energy levels are at least doubly degenerate as long as the $Z_2\times Z_2$ symmetry is not broken explicitly.

\section{Conclusion and Discussion}\label{sec:sum}

In summary, we generalized the eigenfunction method to obtain irreducible projective Reps of finite unitary or anti-unitary groups by reducing the regular projective Reps. To this end we first solved the 2-cocyle equations of a group $G$ to obtain the factor systems and their classification, specially if $G$ is anti-unitary we introduced the decoupled factor systems for convenience. Therefore regular projective representations are obtained. Then using class operators we constructed the complete set of commuting operators, and using their common eigenfunctions we transformed the regular projective representations into irreducible ones. Anti-unitary groups are generally complicated than unitary groups in the reduction process, where modified Schur's lemmas were used. We applied this method to a few familiar finite groups and gave their irreducible Reps.

We then discussed  applications of projective Reps in many-body physics, for instance, the edge states of one-dimensional Symmetry Protected Topological (SPT) phases, symmetry fluxes in 2-dimensional SPT phases and anyons in Symmetry Enriched Topological (SET) phases carry projective Reps of the symmetry groups. We also showed that recently discovered models which are free of sign problem under quantum Monte Carlo simulations are commuting with nontrivial projective Reps of certain anti-unitary groups. Other applications related to spectrum degeneracy were also summarized.

Our approach can be generalized to obtain irreducible projective Reps of infinite groups, such as space groups, magnetic space groups and Lie groups. Furthermore, it may shed light on 
the reduction of symmetry operations corresponding to higher order group cocycles, namely, in the boundary of 2D SPT phases which are classified by the third group cohomology.



\begin{acknowledgements}

We thank Zheng-Cheng Gu, Xie Chen, Meng Cheng,  Peng Ye, Xiong-Jun Liu, Shao-Kai Jian, Zhong-Chao Wei, Cong-Jun Wu, Tao Li and Li You for helpful discussions. We especially thank Xiao-Gang Wen for discussion about the numerical calculation of group cohomology, and thank Hong Yao and Zi-Xiang Li for comments and helpful discussion about the application of projective Reps in QMC. This work is supported by NSFC (Grant Nos. 11574392), the Ministry of Science and Technology of China (Grant No. 2016YFA0300504), and the Fundamental Research Funds for the Central Universities, and the Research Funds of Renmin University of China (No. 15XNLF19). J.Y. is supported by MOST (No. 2013CB922004) of the National Key Basic Research Program of China, by NSFC (No. 91421305) and by National Basic Research Program of China (2015CB921104).

\end{acknowledgements}

\appendix \section{Group cohomology}\label{app1}
In this section, we will briefly introduce the group cohomology theory\footnote{For an introduction to group cohomology, see wiki {\href{http://en.wikipedia.org/wiki/Group_cohomology}{http://en.wikipedia.org/wik/Group\_cohomology}} and
Romyar Sharifi,  {\href{http://math.arizona.edu/~sharifi/groupcoh.pdf}{AN INTRODUCTION TO GROUP COHOMOLOGY}}}.

%
%

Let $\omega_{n}(g_{1},\ldots,g_{n})$ be a $U(1)$ valued function of $n$ group elements.
In other words, $\omega_{n}: G^{n}\rightarrow U(1)$.
Let $\mathcal{C}^{n}(G,U(1))=\{\omega_{n}\}$ be the space of such functions. Note that $\mathcal{C}^{n}(G,U(1))$ is an Abelian
group under the function multiplication $\omega''_{n}(g_{1},\ldots,g_{n})=\omega_{n}(g_{1},\ldots,g_{n})\omega'_{n}(g_{1},\ldots,g_{n})$. We define a map $d$
from $\mathcal{C}^{n}(G,U(1))$ to $\mathcal{C}^{n+1}(G,U(1))$:
\begin{eqnarray}\label{ddef}
& & (d\omega_{n})(g_{1},\ldots,g_{n+1}) \nonumber\\
&& =[g_{1}\cdot\omega_{n}(g_{2},\ldots,g_{n+1})]\omega^{(-1)^{n+1}}_{n}(g_{1},\ldots,g_{n})\times  \nonumber\\
&&\prod^{n}_{i=1}\omega^{(-1)^{i}}_{n}(g_{1},\ldots,g_{i-1},g_{i}g_{i+1},g_{i+2},\ldots,g_{n+1}),
\end{eqnarray}
where $g\cdot \omega_{n}(g_{2},\ldots,g_{n+1})=\omega_{n}(g_{2},\ldots,g_{n+1})$ if $g$ is unitary and $g\cdot \omega_{n}(g_{2},\ldots,g_{n+1})=\omega_{n}^{-1}(g_{2},\ldots,g_{n+1})$ if $g$ is anti-unitary. In other words, $g\cdot \omega_{n}(g_{2},\ldots,g_{n+1})=\omega_{n}^{s(g)}(g_{2},\ldots,g_{n+1})$, where $s(g)=1$ if $g$ is unitary and $s(g)=-1$ otherwise. From above definition (\ref{ddef}), it is easily checked that $d^2\equiv1$.

Let
\begin{equation}
\mathcal{B}^{n}(G,M)=\{\omega_{n}\mid\omega_{n}=d\omega_{n-1}\mid\omega_{n-1}\in\mathcal{C}^{n-1}(G,M)\}
\end{equation}
and
\begin{equation}
\mathcal{Z}^{n}(G,M)=\{\omega_{n}\mid d\omega_{n}=1,\omega_{n}\in\mathcal{C}^{n}(G,M)\}
\end{equation}
$\mathcal{B}^{n}(G,M)$ and $\mathcal{Z}^{n}(G,M)$ are also Abelian groups which satisfy
$\mathcal{B}^{n}(G,M)\subset\mathcal{Z}^{n}(G,M)$. The $n$th-cohomology of $G$ is defined as
\begin{equation}
\mathcal{H}^{n}(G,M)=\mathcal{Z}^{n}(G,M)/\mathcal{B}^{n}(G,M).
\end{equation}

Now we give some examples. From
\begin{equation}\label{}
(d\omega_{1})(g_{1},g_{2})=\frac{\omega_{1}(g_{1})\omega_{1}^{s(g_{1})}(g_{2})}{\omega_{1}(g_{1}g_{2})},
\end{equation}
we find
\begin{equation}\label{1cocycle1}
\mathcal{Z}^{1}(G,U(1))=\{\omega_{1}\mid
 \frac{\omega_{1}^{s(g_{1})}(g_{2})\omega_{1}(g_{1})}{\omega_{1}(g_{1}g_{2})}=1
 \},
\end{equation}
and
\begin{equation}\label{}
\mathcal{B}^{1}(G,U(1))=\{\omega_{1}\mid
\omega_{1}(g_{1})=\frac{\omega_{0}^{s(g_{1})}}{\omega_{0}}\}.
\end{equation}
$\mathcal{H}^{1}(G,U(1))=\frac{\mathcal{Z}^{1}(G,U(1))}{\mathcal{B}^{1}(G,U(1))}$ is the
set of all inequivalent $1D$ Reps of $G$.

From
\begin{equation}\label{}
(d_{2}\omega_{2})(g_{1},g_{2},g_{3})=
\frac{\omega_{2}^{s(g_{1})}(g_{2},g_{3})\omega_{2}(g_{1},g_{2}g_{3})}{\omega_{2}(g_{1}g_{2},g_{3})\omega_{2}(g_{1},g_{2})},
\end{equation}
we find
\begin{equation}\label{2cocycle3}
\mathcal{Z}^{2}(G,U(1))=\{\omega_{2}\mid
\frac{\omega_{2}^{s(g_{1})}(g_{2},g_{3})\omega_{2}(g_{1},g_{2}g_{3})}{\omega_{2}(g_{1}g_{2},g_{3})\omega_{2}(g_{1},g_{2})}=1
 \}
\end{equation}
and
\begin{equation}\label{2coboundary3}
\mathcal{B}^{2}(G,U(1))=\{\omega_{2}\mid
\omega_{2}(g_{1},g_{2})=\frac{\omega_{1}(g_{1})\omega_{1}^{s(g_{1})}(g_{2})}{\omega_{1}(g_{1}g_{2})}\}.
\end{equation}

The $2$-cohomology of group $\mathcal{H}^{2}(G,U(1))=\frac{\mathcal{Z}^{2}(G,U(1))}{\mathcal{B}^{2}(G,U(1))}$
can classify the projective Reps of symmetry group $G$
. Correspondingly, it can classify
1-dimensional SPT phases with on-site symmetry group $G$.

\section{Canonical gauge choice and solutions of cocycles} \label{app2}
Once the cocycle equations and the coboundary functions are written in terms of linear equations (see section \ref{Factorsys}), the classification of cocycles is equivalent to solving some linear algebra. In the following, we will focus on 2-cocycles.

For a general group $G$, the 2-cocycle $\omega_2(g_1,g_2)=e^{i\theta_2(g_1,g_2)}$ has $G^{2}$ components ($G$ is the order of group) which satisfy the following equations ($g_1,g_2,g_3\in G$):
\begin{eqnarray*}
s(g_{1})\theta_{2}(g_{2},g_{3})+\theta_{2}(g_{1},g_{2}g_{3})-\theta_{2}(g_{1}g_{2},g_{3})-\theta_{2}(g_{1},g_{2})=0 .
\end{eqnarray*}
Notice that `=0' in above equation means ${\rm mod\ } 2\pi =0$ (the same below).

Above equations can be written in matrix form as
\begin{eqnarray}\label{CCeq}
\sum_n \left(C^{mn}\right)\theta_2^n=0,
\end{eqnarray}
where $C$ is an $G^3\times G^2$ matrix (the matrix elements are given above), and $\theta_2$ is a $G^2$-component vector.
We do not specify the coefficient space at the beginning, and will go back to it when discussing the classification of the cocycles.

The coboundaries $\Omega_2(g_1,g_2)=e^{i\Theta_2(g_1,g_2)}$ are given by:
\begin{eqnarray*}
\Theta_2(g_1,g_2)=s(g_{1})\theta_{1}(g_{2})+\theta_{1}(g_{1})-\theta_{1}(g_{1}g_{2}).
\end{eqnarray*}
The $\theta_1$ contains $G$ variables. Above equations can also be written in matrix form
\begin{eqnarray}\label{CBfunc}
\Theta_2^m= \sum_n \left(B^{mn}\right)\theta_1^n,
\end{eqnarray}
where $B$ is an $G^2\times G$ matrix.

The $C$ matrix in (\ref{CCeq}) defines a set of equations while the $B$ matrix in (\ref{CBfunc}) defines a set of functions. All the functions defined by B are solutions of the equations defined by $C$. If one can impose some constraints on both $C$ and $B$, the classification will be easier to obtain. For instance, in the extreme case, if there exists a certain `gauge' condition such that there is only one trivial function defined by $B$  (namely, all the 2-coboundaries $\Theta_2(g_1,g_2)$ equal zero), then each solution of $C$ will stand for an equivalent class, in which case, the classification reduces to solving the equations under the special gauge condition.

Actually, we can really fix some gauge degrees of freedom. It has been proved that the canonical gauge\cite{CGLW2} is a valid gauge condition by setting
\beq\label{gague1}
\theta_{2}(E,g_{1})=\theta_{2}(g_{1},E)=0,
\eeq
where $E$ is the unit of group $G$. Under this convention, there are only $(G-1)^2$ components remaining nonzero in $\theta_2(g_1,g_2)$ and the size of the new matrix $C$ is $(G-1)^3\times (G-1)^2$.
Similarly, for coboundaries, we have
\beq\label{gauge2}
\Theta_2(E,g_1)=\theta_1(E)=0.
\eeq
As a result, we only have $(G-1)^2$ nonzero variables for coboundaries and the size of the new matrix $B$ is $(G-1)^2\times (G-1)$. Although this canonical
convention does not fix all the `gauge' degrees of freedom, it does simplify the calculation to solve the cocycle equations. After this simplification, the rank of $2$-cocycle equations (\ref{CCeq}) is denoted as $R_{C}$, and $R_{B}$ is the rank of $2$-coboundary equations (\ref{CBfunc}).
For finite groups, the identity
\beq\label{}
(G-1)^{2}-R_{C}=R_{B}
\eeq
is always satisfied.

As usual linear problems, the cocycle equations can be solved by elimination method. To this end, we need to transform the matrix $C$ (and $B$) into partially diagonal form $C'$ ( and $B'$) by linear operations, namely, by adding or subtracting multiple of one row (or column) to some other row (or column). However, since the equation is defined mod $2\pi$, in each step of  linear operations, the coefficients must be integer numbers. As a result, in the final matrix all the entries are integer numbers. The (partially) diagonal matrix with integer entries is called Smith normal form\footnote{See wiki {\href{http://en.wikipedia.org/wiki/Smith_normal_form}{http://en.wikipedia.org/wik/Smith\_normal\_form}}}.  The diagonal entries of $C'$ and $B'$ completely determine the classification of cocycles.

As an example, we consider the group $D_{2}=Z_2\times Z_2=\{E,P,Q,PQ\}$ where $P^2=Q^2=E$. If we adopt the canonical gauge condition, there are $9$ nonzero variables in $\theta_{2}(g_1,g_2)$ and $3$ nonzero ones in $\theta_{1}(g_1)$. The cocycle equations and the coboundary functions correspond to two matrices $C$ ($27\times 9$, with rank $6$) and $B$ ($9\times 3$, with rank $3$), respectively:
{\small  \beq\label{C'_D2}
C=\left(
\begin{array}{ccccccccc}
    0 &   0 &   0 &   0 &   0 &   0 &   0 &   0 &   0\\
   -1 &   1 &   1 &   0 &   0 &   0 &   0 &   0 &   0\\
   -1 &   1 &   1 &   0 &   0 &   0 &   0 &   0 &   0\\
    0 &  -1 &   1 &   1 &   0 &   0 &  -1 &   0 &   0\\
    0 &  -1 &   0 &   0 &   1 &   0 &   0 &  -1 &   0\\
    1 &  -1 &   0 &   0 &   0 &   1 &   0 &   0 &  -1\\
    0 &   1 &  -1 &  -1 &   0 &   0 &   1 &   0 &   0\\
    1 &   0 &  -1 &   0 &  -1 &   0 &   0 &   1 &   0\\
    0 &   0 &  -1 &   0 &   0 &  -1 &   0 &   0 &   1\\
    1 &   0 &   0 &  -1 &   0 &   0 &  -1 &   0 &   0\\
    0 &   1 &   0 &  -1 &   0 &   1 &   0 &  -1 &   0\\
    0 &   0 &   1 &  -1 &   1 &   0 &   0 &   0 &  -1\\
    0 &   0 &   0 &   1 &  -1 &   1 &   0 &   0 &   0\\
    0 &   0 &   0 &   0 &   0 &   0 &   0 &   0 &   0\\
    0 &   0 &   0 &   1 &  -1 &   1 &   0 &   0 &   0\\
   -1 &   0 &   0 &   0 &   1 &  -1 &   1 &   0 &   0\\
    0 &  -1 &   0 &   1 &   0 &  -1 &   0 &   1 &   0\\
    0 &   0 &  -1 &   0 &   0 &  -1 &   0 &   0 &   1\\
    1 &   0 &   0 &  -1 &   0 &   0 &  -1 &   0 &   0\\
    0 &   1 &   0 &   0 &  -1 &   0 &  -1 &   0 &   1\\
    0 &   0 &   1 &   0 &   0 &  -1 &  -1 &   1 &   0\\
   -1 &   0 &   0 &   1 &   0 &   0 &   0 &  -1 &   1\\
    0 &  -1 &   0 &   0 &   1 &   0 &   0 &  -1 &   0\\
    0 &   0 &  -1 &   0 &   0 &   1 &   1 &  -1 &   0\\
    0 &   0 &   0 &   0 &   0 &   0 &   1 &   1 &  -1\\
    0 &   0 &   0 &   0 &   0 &   0 &   1 &   1 &  -1\\
    0 &   0 &   0 &   0 &   0 &   0 &   0 &   0 &   0\end{array}
 \right),
\eeq
}
\beq\label{B'_D2}
B=\left(
\begin{array}{ccc}
     2 &    0 &    0\\
     1 &    1 &   -1\\
     1 &   -1 &    1\\
     1 &    1 &   -1\\
     0 &    2 &    0\\
    -1 &    1 &    1\\
     1 &   -1 &    1\\
    -1 &    1 &    1\\
     0 &    0 &    2
\end{array}
 \right).
\eeq

After some linear operations as mentioned above, the matrices $C$ and $B$ can be written in the Smith normal form
\begin{eqnarray}\label{Smith}
C'&=&\left( \begin{array}{ccccccccc}
1 & 0 & 0 & 0 & 0 & 1 & -1 & 1& -1\\ 0 & 1 & 0 & 0 & 0 & 0 & -1& 1&0\\
0 & 0 & 1 & 0 & 0 & 1 & 0 & 0& -1\\ 0 & 0& 0 & 1 & 0 & -1 & -1 & 2&0\\
0 & 0 & 0 & 0 & 1 & 0 & 0 & 1&-1\\ 0 & 0 & 0 & 0 & 0 & 2 & 0 & -2&0\end{array} \right),\label{C'C'D2}\\
B'&=&\left( \begin{array}{ccc}
1 & -1 & -1 \\ 0 & 2 & 0\\
0 & 0 & 2\end{array} \right).\label{B'}
\end{eqnarray}
Notice that the number of variables (number of columns) in (\ref{C'C'D2}) is greater than the number of equations (number of rows), and the difference is 3.
If the coefficient space is $U(1)$, the solutions form a 3-dimensional continuous space $U(1)\times U(1)\times U(1)$. Furthermore, the last row of (\ref{C'C'D2}) indicates the two variables corresponding to $2$ is determined upto $\frac{2m\pi}{2}(m=0,1)$, which gives an extra $Z_2$ group. So the solution space of $\theta_{2}(g_1,g_2)$ is $U(1)\times U(1)\times U(1) \times Z_2$,while the coboundary space is $U(1)\times U(1)\times U(1)$. The quotient gives the classification of cocycles:
\begin{equation}\label{}
\mathcal{H}^{2}(D_2,U(1))=\mathbb{Z}_2.
\end{equation}

If the coefficient space is $Z$, then the common factor 2 in the last row of (\ref{C'C'D2}) can be eliminated, and the solution space is $Z\times Z\times Z$. On the other hand, the common factor 2 in the last two rows of (\ref{B'}) indicates that the coboundary space is $Z\times 2Z\times 2Z$. Therefore, the classification is $\mathcal{H}^{2}(D_2,Z)=\mathbb{Z}_2\times \mathbb Z_2.$

If the coefficient space is $Z_2$, then the solution space of $\theta_{2}(g_1,g_2)$ is $Z_2\times Z_2\times Z_2\times Z_2$. The last two rows of (\ref{B'}) give trivial functions, so the coboundary space is $Z_2$. The quotient gives the classification of cocycles: $\mathcal H^2(D_2, Z_2)=\mathbb{Z}_2^3$.

Similarly, if the coefficient space is $Z_N$, then the classification of cocycles is $\mathcal H^2(D_2, Z_N)=\mathbb{Z}_{(N,2)}\times \mathbb{Z}_{(N,2)}\times \mathbb{Z}_{(N,2)}$, where $(N,2)$ stands for the greatest common divisor of $N$ and 2. The first $\mathbb{Z}_{(N,2)}$ is owing to the factor 2 in the Smith normal form $C'$, and the last two $\mathbb{Z}_{(N,2)}$ are owing to the two 2s in the Smith normal form of $B'$. This rule of classification is also true for  $\mathcal H^2(G, Z_N)$ with a general symmetry group $G$.

As another example, we consider the anti-unitary symmetry group $Z_2\times Z_2^T$. Under canonical gauge condition, there are $9$ nonzero variables in $\theta_{2}(g_1,g_2)$ and $3$ nonzero variables in $\theta_{1}(g_1)$. The cocycle matrix $C$ and coboundary matrix $B$ can be respectively transformed into the Smith normal form $C'$ and $B'$ after some linear operations,
\begin{eqnarray}
C'&=&\left( \begin{array}{ccccccccc}
1 & 0 & 0 & 0 & 0 & -1 & -1 & 1& 1\\ 0 & 1 & 0 & 0 & 0 & -1 & 0 & 0&1\\
0 & 0 & 1 & 0 & 0 & 0 & -1 & 1& 0\\ 0 & 0& 0 & 1 & 0 & -1 & 0 & -1&0\\
0 & 0 & 0 & 0 & 1 & 0 & -1 & 0&-1\\ 0 & 0 & 0 & 0 & 0 & 2 & 0 & 0&0\\
0 & 0 & 0 & 0 & 0 & 0 & 2 & 0&0\end{array} \right), \label{C'C'D2T}\\
B'&=&\left( \begin{array}{ccc}
1 & -1 & 1 \\ 0 & 2 & -2\end{array} \right).
\end{eqnarray}
If the coefficient space is $U(1)$, it can be seen that the solution space of $\theta_{2}(g_1,g_2)$ is $U(1)\times U(1) \times Z_2\times Z_2$ from the last two rows of (\ref{C'C'D2T}). And the coboundary space is still $U(1) \times U(1)$. The quotient gives the classification of cocycles:
$\mathcal{H}^{2}(Z_2\times Z_2^T,U(1))=\mathbb{Z}_2^2$.
If the coefficient space is $Z$,  the solution space of $\theta_{2}(g_1,g_2)$ is $Z\times Z$, while the coboundary space is $Z\times 2Z$. The quotient gives the classification of cocycles: $\mathcal H^2(Z_2\times Z_2^T, Z)=\mathbb{Z}_2$. Similarly, if the coefficient space is $Z_N$, then $\mathcal H^2(Z_2\times Z_2^T, Z_N)=\mathbb{Z}_{(N,2)}^3$.

From the Smith normal form, we can also obtain the solutions of the cocycles. For example, in the Smith normal form (\ref{Smith}) of the cocycle equations of group $D_2=Z_2\times Z_2=\{E,P,Q,PQ\}$, the columns correspond to variables $\theta_2(P,P)$, $\theta_2(P,Q)$, $\theta_2(P,PQ)$, $\theta_2(Q,P)$, $\theta_2(PQ,P)$, $\theta_2(Q,PQ)$, $\theta_2(Q,Q)$, $\theta(PQ,Q)$, $\theta_2(PQ,PQ)$, respectively. We note these variables as $X_1,X_2,..., X_9$. The last three variables
are coboundary degrees of freedom and can be fixed as $X_7=X_8=X_9=0$. Under this condition, the last equation becomes 
$2X_6=0$ mod $2\pi$, which has two solutions $X_6=0$
and $X_6=\pi$.
The first solution is trivial, from which we can obtain that all other components are zero; the second solution is nontrivial, from (\ref{Smith}) we can easily obtain that $X_1=-\pi, X_2=0, X_3=-\pi, X_4=\pi, X_5=0$.
Equivalently, the nontrivial solution is given by $\omega_2(P,P)=\omega_2(P,PQ)=\omega_2(Q,P)=\omega_2(Q,PQ)=-1$, other components are equal to 1.


In this paper we solve all the 2-cocycle equations using above method. However, for abelian unitary groups there exist analytic cocyle solutions. For example, for the group $G=Z_{N_1}\times Z_{N_2}\times ...\times Z_{N_k}$ with generators $h_1,...,h_k$, if we label the group elements as $A=(a_1,a_2,...,a_k)=(h_1^{a_1},h_2^{a_2},...,h_k^{a_k})=h_1^{a_1}h_2^{a_2}...h_k^{a_k}$, then the analytic 2-cocycle solutions take the following form:
\beq\label{2cocycle-analytic}
\omega_2 (A,B)=e^{2\pi i\sum_{i<j}{p^{ij}\over N_{ij}}a_ib_j},
\eeq
where $N_{ij}$ is the greatest common devisor of $N_i$ and $N_j$ and $0\leq p^{ij}\leq N_{ij}-1$. There are totally $\prod_{i<j} N_{ij}$ different classes of solutions.

\section{Anti-unitary groups: decoupled cocycles }\label{standardcyc} 

In this section, we consider anti-unitary symmetry group $G$ which contains a unitary subgroup $H$ as its invariant subgroup, and the quotient group is the time reversal group,
\[
G/H\simeq Z_2^T,
\]
where $Z_2^T=\{E,\mathbb T\}$ with $\mathbb T^2=E$.

It should be noted that {\it $\mathbb T$ is not necessarily an element of $G$}. Supposing $T$ is the anti-unitary group element in $G$ with the smallest order $2^m$ with $m\in\mathbb Z$ and $m\geq1$, namely, $T^{2^m}=E$ (the order of $T$ should not contain odd factors. The reason is that if $T^{2^p(2q+1)}=E$ where $p,q\in\mathbb Z$ and $p\geq1,q\geq1$, then we can find another anti-unitary group element $T'=T^{2q+1}$ with $T'^{2^p}=E$ such that the order of $T'$ is less than $T$), then there are roughly two types of anti-unitary groups:

{\bf Type-I}: $T^2=E$, namely, $m=1$. In this case, we can identify $\mathbb T$ with $T$, such that $\mathbb T\in G$.

{\bf Type-II}: $T^{2^m}=E$ with $m\geq2$. In this case, $\mathbb T$ can never be identified with any element of $G$, namely, $\mathbb T\notin G$. The simplest example with $m=2$ is the fermionic time reversal group $Z_4^T=\{E,T^2=P_f,T,T^3\}$, where $P_f$ is the fermion parity. We will only discuss the fermionic symmetry groups, and will not discuss the more complicated cases with $m>2$.


\subsection{Type-I anti-unitary groups: $T^2=E$}\label{TypeI} 

This kind of groups can be written in forms of direct product group $G=H\times Z_2^T$ or semi-direct product group $G=H\rtimes Z_2^T$. In the following we will note the generator of $Z_2^T$ as $T=\mathbb T$.

We can decouple the cocycles into two parts, namely, the $Z_2^T$ part $\omega_2(T,T)$ and the $H$ part $\omega_2(g_1,g_2)$, and illustrate that they determine the classification. The crossing terms, such as $\omega_2(Tg_1,g_2),\omega_2(g_1,Tg_2),\omega_2(Tg_1,Tg_2)$, can be either fixed or expressed in terms of $\omega_2(T,T)$ and $\omega_2(g_1,g_2)$. This can be shown in the following four steps:

{\bf (1) No gauge degrees of freedom in $\omega_2(T,T)=\pm1$.} Let us see $\omega_2(T,T)$ first. From the cocycle equation
\begin{eqnarray*}
(d\omega_2)(T,T,T)&=&\omega_2^{-1}(T,T)\omega_2^{-1}(E,T)\omega_2(T,E)\omega_2^{-1}(T,T) \\
&=&1,
\end{eqnarray*}
we get $\omega_2^2(T,T)=1$, so
\beq\label{OmgTT}
\omega_2(T,T)=\pm1,
\eeq
here we have adopted the canonical gauge. Since $(d\Omega_1)(T,T)=\Omega_1^{-1}(T)\Omega_1^{-1}(E)\Omega_1(T)=1$, there are no gauge degrees of freedom for $\omega_2(T,T)$, $\omega_2(T,T)=1$ and $\omega_2(T,T)=-1$ stand for different classes.

{\bf (2)  Tune coboundaries to fix $\omega_2(T, g)=1$ and $\omega_2(g,T)=1$.} Further, for unitary elements $g\in H$, we will show that the following gauge can be fixed:
\begin{eqnarray}\label{Tg1}
&&\omega_2(T, g)=\omega_2(T, \tilde g)=1,\nonumber\\
&&\omega_2(g, T)=\omega_2(\tilde g,T)=1,
\end{eqnarray}
where $Tg=\tilde gT$.

If $\tilde g=g$,  then we can tune the coboundary
\begin{eqnarray*}
&&(d\Omega_1) (T,g)=\Omega_1^{-1}(g) \Omega_1^{-1}(Tg)\Omega_1(T),\\
&&(d\Omega_1) ( g,T)=\Omega_1(T) \Omega_1^{-1}(gT)\Omega_1( g),
\end{eqnarray*}
such that $\omega_2'(T,g)=\omega_2(T,g)(d\Omega_1)(T,g)=1$, and $\omega_2'(g,T)=\omega_2(g,T)(d\Omega_1)(g,T)=1$. This can be done by setting
\begin{eqnarray}\label{Z2gauge}
&&\Omega_1(g)=\sigma\left({\omega_2(T,g)\over\omega_2(g,T)}\right)^{1/2},\\
&&\Omega_1(gT)=\sigma\Omega_1(T)\left[\omega_2(T,g)\omega_2(g,T)\right]^{1/2}. \nonumber
\end{eqnarray}
where $\sigma=\pm1$ is a $Z_2$ variable and $\Omega_1(T)$ is a $U(1)$ variable which can be set as 1. In the square root, $-1$ should be treated as $e^{i\pi}$ such that $\sqrt{(-1)\times(-1)}=-1$ and $\sqrt{(-1)/(-1)}=1$.

If $\tilde g\neq g$, then the illustration of the gauge choice (\ref{Tg1}) is a little complicated. From the cocycle equation $(d\omega_2)(\tilde g, T,T)=1$, we have
\begin{eqnarray}\label{gTT}
\omega_2(Tg,T)\omega_2(\tilde g,T)=\omega_2(T,T).
\end{eqnarray}
Similarly, from  $(d\omega_2)(T, T, \tilde g)=1$ we have
\begin{eqnarray}\label{TTg}
{\omega_2(T,gT)\over \omega_2(T,\tilde g)}=\omega_2(T,T).
\end{eqnarray}
On the other hand, $(d\omega_2)(T, g, T)=1$ gives
\begin{eqnarray}\label{TgT}
{\omega_2(T,gT)\over\omega_2(Tg,T)}=\omega_2(T,g)\omega_2(g,T).
\end{eqnarray}
Comparing (\ref{gTT}),(\ref{TTg}) and (\ref{TgT}), we obtain
\[
\omega_2(T,g)\omega_2(g,T)=\omega_2(T,\tilde g)\omega_2(\tilde g,T).
\]
Owing to this relation, the following gauge fixing equations have solutions: $\omega_2'(T,g)=\omega_2(T,g)(d\Omega_1)(T,g)=1$, $\omega_2'(T,\tilde g)=\omega_2(T,\tilde g)(d\Omega_1)(T,\tilde g)=1$, $\omega_2'(g, T)=\omega_2(g,T)(d\Omega_1)(g, T)=1$, $\omega_2'(\tilde g, T)=\omega_2(\tilde g,T)(d\Omega_1)(\tilde g, T)=1$, where
\begin{eqnarray*}
&&(d\Omega_1) (T,g)=\Omega_1^{-1}(g) \Omega_1^{-1}(Tg)\Omega_1(T),\\
&&(d\Omega_1) (T, \tilde g)=\Omega_1^{-1}(\tilde g) \Omega_1^{-1}(gT)\Omega_1(T),\\
&&(d\Omega_1) (\tilde g,T)=\Omega_1(T) \Omega_1^{-1}(Tg)\Omega_1(\tilde g),\\
&&(d\Omega_1) ( g,T)=\Omega_1(T) \Omega_1^{-1}(gT)\Omega_1( g),
\end{eqnarray*}
and the solutions are
\begin{eqnarray}
&&\Omega_1(\tilde g)=\Omega_1^{-1}(g)\omega_2(T,g)\omega_2^{-1}(\tilde g,T),\nonumber\\
&&\Omega_1(Tg)=\Omega_1(T)\Omega_1^{-1}(g)\omega_2(T,g),\label{musolution}\\
&&\Omega_1(gT)=\Omega_1(T)\Omega_1(g)\omega_2(g,T), \nonumber
\end{eqnarray}
where $\Omega_1(g)$ is an $U(1)$ variable and $\Omega_1(T)$ can be set as 1.  Once the gauge fixing condition (\ref{Tg1}) is satisfied, then in the remaining gauge degrees of freedom for each pair of $g$,$\tilde g$ with $\tilde g\neq g$, \[\Omega_{1}(g)\Omega_{1}(\tilde g)=1\] is required and there remains one free coboundary variable $\Omega_{1}(g)$.

In later discussion we will fix $\omega_2(T,g)=\omega_2(g,T)=1$ for any $g\in H$.


{\bf (3) The values of $\omega_2(Tg_1,Tg_2)$ are determined.}  From the equation
\Beq
(d\omega_2)(T,T,g)=\omega_2^{-1}(T,g)\omega_2^{-1}(E,g)\omega_2(T,Tg)\omega_2^{-1}(T,T)=1, 
\Eeq
we get
$\omega_2(T,Tg)=\omega_2(T,g)\omega_2(T,T)=\omega_2(T,T)$.
Similarly, from $(d\omega_2)(T,g,T)=1$ we have
 $\omega_2(Tg, T)=\omega_2(\tilde gT, T)=\omega_2(T,T).$ That is to say, $\omega_2(T,Tg)$ and $\omega_2(Tg,T)$ can be fixed as real
\Beq\label{TTg1}
\omega_2(T,Tg)= \omega_2(Tg,T)=\omega_2(T,T)=\pm1,
\Eeq
for any $g\in H$.

Furthermore, from the cocycle equation $(d\omega_2)(Tg_1,T,g_2)=\omega_2^{-1}(T,g_2)\omega_2^{-1}(\tilde g_1,g_2) \omega_2(Tg_1,Tg_2)\omega_2^{-1}(Tg_1,T)=1$ with $g_1,g_2\in H$, we obtain
\beq\label{Tg1_Tg2}
\omega_2(Tg_1,Tg_2)=\omega_2(\tilde g_1,g_2)\omega_2(T,T).
\eeq

{\bf (4) The values of $\omega_2(Tg_1,g_2)$ and $\omega_2(g_1,Tg_2)$ are determined.} Finally, considering the following cocycle equations
\begin{eqnarray}
& &(d\omega_2)(T, g_1,g_2)\nonumber\\
& & =\omega_2^{-1}(g_1,g_2)\omega_2^{-1}(Tg_1,g_2)\omega_2(T,g_1g_2)\omega_2^{-1}(T,g_1)\nonumber\\
& & =\omega_2^{-1}(g_1,g_2)\omega_2^{-1}(Tg_1,g_2)=1, \label{rlt1}\\
& & (d\omega_2)(g_1,T,g_2)\nonumber\\
& & =\omega_2(T,g_2)\omega_2^{-1}(g_1T,g_2)\omega_2(g_1,Tg_2)\omega_2^{-1}(g_1,T)\nonumber\\
&& =\omega_2^{-1}(g_1T,g_2)\omega_2(g_1,Tg_2)=1,\label{rlt2}\\
& & (d\omega_2)( g_1,g_2,T)\nonumber\\
&& =\omega_2(g_2,T)\omega_2^{-1}(g_1g_2,T)\omega_2(g_1,g_2T)\omega_2^{-1}(g_1,g_2)\nonumber\\
&& =\omega_2(g_1,g_2T)\omega_2^{-1}(g_1,g_2)=1,\label{rlt3}
\end{eqnarray}
we obtain the following relations
\beq\label{Tg1_g2}
&&\omega_2(Tg_1,g_2)=\omega_2^{-1}(g_1,g_2),\nonumber\\
&&\omega_2(g_1,Tg_2)=\omega_2(g_1,\tilde g_2),
\eeq
and an important constraint
\beq\label{Constraint}
\omega_2(g_1,g_2)\omega_2(\tilde g_1,\tilde g_2)=1, \label{ww1}
\eeq
where $g_1,g_2\in H$.

We call the cocycles satisfying relations (\ref{OmgTT}), (\ref{Tg1}), (\ref{Tg1_Tg2}),(\ref{Tg1_g2}) as `decoupled cocyles' or decoupled factor systems, in which case only $\omega_2(T,T)$ and $\omega_2(g_1,g_2)$ are important. Since $\omega_2(T,T)$ contributes $Z_2$ classification to $\mathcal H^2(G,U(1))$, in appendix~\ref{app:Calc_class} we will see how the constraint (\ref{Constraint}) influences the classification of $\omega_2(g_1,g_2)$.

\subsection{Type-II anti-unitary groups: $T^{2^m}=E, m\geq2$} \label{app:type-II}  


We focus on the case $m=2$, which corresponds to the anti-unitary symmetry group of fermions with half-integer spin.

{\bf Fermionic time reversal group $Z_4^T$.} The simplest example of the type-II anti-unitary group is the fermionic time reversal group $Z_4^T=\{E,P_f,T,P_f T\}$ with $T^4=E$ and $T^2=P_f$, $P_f$ is the fermion parity. The unitary normal subgroup of $Z_4^T$ is $H=Z_2^f=\{E,P_f\}$ with $Z_4^T/H=Z_2^T=\{E,\mathbb T\}$. Obviously $\mathbb T$ is not an element of $Z_4^T$.  In the following we will figure out the classification of 2-cocyles of $Z_4^T$ by gauge fixing procedure.

Firstly we adopt the canonical gauge by setting $\omega_{2}(E,g)=\omega_{2}(g,E)=1$. We still define the time reversal conjugation as $T^{-1}gT=\tilde g$, since $T^{-1}=TP_f$, we have $T\tilde gT^{-1}=T^{-1}\tilde gT=g$, namely,  $\tilde{\tilde g}=g$. Since $\tilde P_f=P_f$, similar to (\ref{Z2gauge}) we can fix
\beq\label{Z4gauge}
\omega_2(T,P_f)=\omega_2(P_f,T)=1.
\eeq
After above gauge fixing, from (\ref{Z2gauge}), there are still a $Z_2$ degree of freedom $\sigma$ and an U(1) degree of freedom $\Omega_1(T)$. If we set $\Omega_1(T)=1$, then there remains a $Z_2$ gauge degree of freedom $\Omega_1(P_f)=\sigma=\pm1$. Since $(d\Omega_1)(P_f,P_f)=\Omega_1(P_f)^2=1$, under the gauge fixing (\ref{Z4gauge}), $\omega_2(P_f,P_f)$ is completely fixed without any gauge degrees of freedom.  In other words, {\it different values of $\omega_2(P_f,P_f)$ stand for different classes of 2-cocycles.}  

From $(d\omega_2)(T,T,P_f)=1$ we have \[\omega_2(T,TP_f)={\omega_2(P_f,P_f) \omega_2(T,T)}.\] On the other hand, $(d\omega_2)(TP_f,T,TP_f)=1$ gives \[\omega_2(TP_f,T)\omega_2(T,TP_f)=1;\] while $(d\omega_2)(T,P_f,T)=1$ yields \[\omega_2(T,TP_f)=\omega_2(TP_f,T).\] Comparing these relations, we obtain
\[
\left[{\omega_2(P_f,P_f)  \omega_2(T,T)}\right]^2=\omega_2(P_f,P_f)^2=1,
\]
namely, $\omega_2(P_f,P_f)=\pm1$. Here we have used $\omega_2(T,T)^2={\omega_2(T,P_f)\over\omega_2(P_f,T)}=1$, which results from $(d\omega_2)(T,T,T)=1$. We have shown previously that $\omega_2(P_f,P_f)$ has no gauge degrees of freedom, therefore $\omega_2(P_f,P_f)=1$ (trivial) and $\omega_2(P_f,P_f)=-1$ (nontrivial) stand for two different classes. In the following we will show that all other components are dependent on $\omega_2(P_f,P_f)$ or can be fixed to be 1.

From $(d\omega_2)(P_f,P_f,T)=1$ and $(d\omega_2)(P_f,T,P_f)=1$ we obtain that $\omega_2(P_f,TP_f)=\omega_2(TP_f,P_f)=\omega_2(P_f,P_f)$.

Since $\omega_2(T,T)^2=1$ and $(d\Omega_1)(T,T)=\Omega_1(P_f)^{-1}=\sigma=\pm1$, the value of $\omega_2(T,T)$ only depends on the gauge degree of freedom $\sigma$. We can tune $\sigma$ such that $\omega_2(T,T)=\omega_2(P_f, P_f)$ and consequently $\omega_2(T,TP_f)=\omega_2(TP_f,T)=1$. Finally, the value of $\omega_2(TP_f,TP_f)=1$ can be derived from $(d\omega_2)(TP_f,TP_f,T)=1$.

%

Above we showed that the value of $\omega_2(P_f,P_f)=\pm1$ completely determines the classification $\mathcal H^2(Z_4^T,U(1))=\mathbb Z_2$. The nontrivial factor system is given as
\Beq
&&\omega_2(P_{f},P_{f})=-1,\\
&&\omega_2(P_{f},TP_{f})=\omega_2(TP_{f},P_{f})=\omega_2(P_f,P_f)=-1,\\
&&\omega_2(T,T)=-1,
\Eeq
and others are fixed to be 1.

Similar to the invariant $[M(T)K]^2=\pm1$ for the projective Reps of $Z_2^T$, the projective Reps of $Z_4^T$ also have an invariant
\beq\label{Z4T}
[M(T)K]^4=\pm1.
\eeq
Noticing $[M(T)K]^4=\omega_2(T,T)^2\omega_2(P_f,P_f)$, it is invariant under gauge transformation $M'(g)K_{s(g)}=M(g)\Omega_1(g)K_{s(g)}$ since
\beq\label{Z4TInv}
[M'(T)K]^4&=&\omega_2'(T,T)^2\omega_2'(P_f,P_f) \nonumber\\
&=&\omega_2(T,T)^2\omega_2(P_f,P_f) \Omega_2(T,T)^2\Omega_2(P_f,P_f)\nonumber\\
&=&\omega_2(T,T)^2\omega_2(P_f,P_f) {\Omega_1(P_f)^2\over \Omega_1(P_f)^2}\nonumber\\
&=&[M(T)K]^4.
\eeq



{\bf General fermionic anti-unitary groups.} 
In the following, we will discuss general fermionic groups $G_f$. Since physical operations usually conserve fermion parity (the fermion creation or annihilation operators do not conserve fermion parity, but they are not allowed in the Hamiltonian), we assume that $[P_f,g]=0$ for any $g\in G_f$. Therefore, the subgroup $Z_2^f=\{E,P_f\}$ is a center of $G_f$ and we have
\Beq G_f/Z_2^f=G_b,
\Eeq
where $G_b$ is the bosonic (namely, type-I) anti-unitary group.

If we note $H_f$ as the unitary normal subgroup of $G_f$ such that $G_f/H_f=Z_2^T$, and note the unitary normal subgroup of $G_b$ as $H_b$ such that $G_b/H_b=Z_2^T$,  then $H_f/Z_2^f=H_b$, there are two cases: (1)$H_f$ is a direct product group of $H_b$ and $Z_2^f$, $H_f=H_b\times Z_2^f$, then the group $G_f$ can be written as $G_f=H_b\times Z_4^T$ or $G_f=H_b\rtimes Z_4^T$; (2) $H_f$ is not a direct product group of $H_b$ and $Z_2^f$, then the group can be noted as $G_f=G_-^\pm(H_f,T)$, where the subscript $-$ means $T^2=P_f$, and the superscript $\pm$ means that $Th=h^{\pm1}T$ for $h\in H_f$.

After slight modification, the discussion about decoupled factor system in section \ref{TypeI} can also be applied for the fermionic anti-unitary groups.

For the cocycles of the $Z_4^T$ subgroup of $G$, we can adopt the previous gauge fixing procedure. In the following we try to decouple the factor system as we have done previously.

 Supposing $g\in H_f$, from the cocycle equations $(d\omega_2)(P_f,P_f,g)=1$ and $(d\omega_2)(g,P_f,P_f)=1$, we get
 \[
{\omega_2(g,P_f)\over \omega_2(P_f,g)}={\omega_2(P_f,P_f g)\over \omega_2(gP_f,P_f)}.
 \]
On the other hand, from the cocycle equation $(d\omega_2) (P_f,g,P_f)=1$ we have
\[
 {\omega_2(g,P_f)\over \omega_2(P_f,g)}={\omega_2(P_fg,P_f)\over\omega_2(P_f,gP_f)}.
\]
Since $gP_f=P_fg$, comparing above two equations we obtain
\[
\left({\omega_2(g,P_f)\over \omega_2(P_f,g)}\right)^2=1,
\]
namely, ${\omega_2(g,P_f)\over \omega_2(P_f,g)}=\pm1$ is a $Z_2$ variable, we note it as
\[{\omega_2(g,P_f)\over \omega_2(P_f,g)}=\sigma_{g},\ \ \ g\neq P_f.\]
We can fix the coboundary $\Omega_1(P_fg)=\omega_2(P_f,g)\Omega_1(g)\Omega_1(P_f)$ such that $\omega_2'(P_f,g)=\omega_2(P_f,g)(d\Omega_1)(P_f,g)=1$.
After this gauge fixing, we have ($g\neq P_f$)
\Beq
&&\omega_2(P_f,g)=1,\\
&&\omega_2(g,P_f)=\sigma_g.
\Eeq

Similar to (\ref{gTT}), (\ref{TTg}), (\ref{TgT}),  from the cocycle equations $(d\omega_2)(g,T,T)=1$, $(d\omega_2)(T,T,g)=1$, and $(d\omega_2)(T, \tilde g,T)=1$, we have
\beq\label{newgauge}
  {\omega_2( g,T) \omega_2(T, g) \over  \omega_2(T,\tilde g) \omega_2(\tilde g,T) } = \sigma_{ g} =\sigma_{\tilde g}^{-1}=\sigma_{\tilde g}
  =\omega_2(\tilde{g},P_f).
\eeq

In the following we fix the values of $\omega_2(T, g)$ and $\omega_2(g,T)$ and decouple the factor system as discussed in section \ref{TypeI}.

(1), If $\tilde g=g$, namely, $Tg=gT$, then $\sigma_g=1$. Similar to (\ref{Z2gauge}), we can fix
\[\omega_2(T,g) = \omega_2(g,T)=1\]
by tuning the values of $\Omega_1(g)$ and $\Omega_1(Tg)$.

(2),  If $\tilde g\neq g$, the gauge fixing depends on the value of $\sigma_g$:

\  A) $\sigma_g=1$. In this case, the discussion in equations (\ref{gTT}), (\ref{TTg}), (\ref{TgT}) is still valid, we can simultaneously fix $\omega_2(T,g)=\omega_2(g,T)=\omega_2(T,\tilde g)=\omega_2(\tilde g,T)=1$ by adopting the gauge in (\ref{musolution}).

\  B) $\sigma_g=-1$. Owing to (\ref{newgauge}), the values of $\omega_2(T,g), \omega_2(g,T), \omega_2(T,\tilde g), \omega_2(\tilde g,T)$ cannot be simultaneously set to be 1 (in this sense the factor system is not completely decoupled), and the coboundary in (\ref{musolution}) should be slightly modified. We can set
\[\omega_2(T,g)=\omega_2(T,\tilde g)=1,\] and set one of $\omega_2(g,T), \omega_2(\tilde g,T)$ to be 1. For example, we can choose
\Beq
&&\omega_2(\tilde g,T)=1,\\
&&\omega_2(g,T)=\sigma_g.
\Eeq
Since $g$ and $\tilde g$ are mutually time-reversal conjugate, there is still a gauge degree of freedom in above equation.
Above gauge fixing can be carried out by adopting the following coboundary
\Beq
&&\Omega_1(\tilde g)=\Omega_1^{-1}(g)\omega_2(T,g)\omega_2^{-1}(\tilde g,T),\nonumber\\
&&\Omega_1(Tg)=\Omega_1(T)\Omega_1^{-1}(g)\omega_2(T,g),\\
&&\Omega_1(gT)=\sigma_g\Omega_1(T)\Omega_1(g)\omega_2(g,T). \nonumber
\Eeq

We further discuss the gauge fixing of other components. The equations (\ref{Tg1_Tg2}),(\ref{Tg1_g2}) should be modified as follows:
\Beq
 \omega_2(Tg_1,Tg_2)=\omega_2(g_1,T) \omega_2(\tilde g_1,g_2)\omega_2(T,T) \sigma_{\tilde{g}_1}\sigma_{g_2}\sigma_{\tilde{g}_1g_2},
\Eeq
(here we have used $\omega_2(\tilde g_1P_f,g_2)=\omega_2(\tilde {g}_1,g_2)\sigma_{\tilde{g}_1}\sigma_{g_2}\sigma_{\tilde{g}_1g_2}$
) and 
\Beq
&&\omega_2(Tg_1,g_2) = \omega_2^{-1}(g_1,g_2),\\
&&\omega_2(g_1,Tg_2) =\omega_2(g_1,\tilde g_2)\omega_2^{-1}(\tilde g_2,T)\omega_2(g_1\tilde g_2,T).
\Eeq
And the constraint (\ref{Constraint}) becomes
\[
\omega_2(g_1,g_2)\omega_2(\tilde g_1,\tilde g_2) = \sigma_{g_1}\sigma_{g_2}\sigma_{g_1g_2}.
\]

It can be proved that $\sigma_{g_1}\sigma_{g_2}\sigma_{g_1g_2}=1$ and $\sigma_{\tilde{g}_1}\sigma_{g_2}\sigma_{\tilde{g}_1g_2}=1$,
so $\sigma_{g}$ carries a $Z_2$ valued linear representation of $H_f$. After the factor system being decoupled, the reduction of regular projective Reps can be performed following the procedure of type-I anti-unitary groups (slight modification may be necessary).


\section{Classification of some simple anti-unitary groups}\label{app:Calc_class}

In this section we only discuss type-I anti-unitary groups. In subsections \ref{sec:directg} and \ref{semiprod}, we discuss the groups where all the 2-cocycles can be set as real numbers. In subsection \ref{complexcocycle}, we give some examples where the 2-cocycles cannot be set as real numbers.


\subsection{Direct product group $H\times Z_2^T$}\label{sec:directg}

If the group is $G=H\times Z_2^T$, then $Tg=gT$ and
\[\tilde g=g.\]
From (\ref{ww1}) we have \[\omega_2^2(g_1,g_2)=1\] which constraints
\Beq\label{directg}
\omega_2(g_1,g_2)=\pm1.
\Eeq
Owing to the relations (\ref{rlt1}) $\sim$ (\ref{rlt3}), we have
\[
\omega_2(Tg_1,g_2)=\omega_2(g_1,Tg_2)=\omega_2(Tg_1,Tg_2)=\omega_2(g_1,g_2),
\]
so all the cocycles (and the remaining coboundary degrees of freedom) are fixed as {\it real numbers.}

Consequently, the classification of $H\times Z_2^T$ are determined by two parts, $\omega_2(T,T)=\pm1$ which is given by $\mathcal H^2(Z_2^T, U(1))$, and $\omega_2(g_1,g_2)=\pm1$ which is given by $\mathcal H^2(H, Z_2)$. In other words,
\Beq
\mathcal H^2(H\times Z_2^T, U(1))&=&\mathcal H^2(Z_2^T, U(1))\times\mathcal H^2(H, Z_2)\nonumber \\
&=& \mathbb Z_2\times \mathcal H^2(H, Z_2)\\
&=& \mathbb Z_2^n, \ \ \ {\rm with\ } n\geq1.\nonumber
\Eeq

\subsection{Semidirect product group $H\rtimes Z_2^T$ with $Tg=g^{-1}T$}\label{semiprod}

The general case of semidirect product groups $H\rtimes Z_2^T$ is a little complicated. Here we only consider the simple case where $H$ is abelian and $Tg=g^{-1}T$, or equivalently $\tilde g=g^{-1}$.

Noticing that
\begin{eqnarray}\label{21T}
T\tilde g_1\tilde g_2 = T(\tilde g_1\tilde g_2) =T\widetilde{g_2g_1}= (g_2g_1 )T,
\end{eqnarray}
on the other hand,
\begin{eqnarray}\label{12T}
T\tilde g_1\tilde g_2 = g_1T\tilde g_2= g_1g_2T=(g_1g_2)T ,
\end{eqnarray}
so we have
\[
g_1g_2 = g_2g_1,
\]
namely, {\it the subgroup $H$ is essentially Abelian.}

We first show that $\omega_2(g, \tilde g)$ must be real.
From equation
\[
(d\omega_2)(\tilde g, g,\tilde g)=\omega_2(g,\tilde g) \omega_2^{-1}(E,\tilde g) \omega_2(\tilde g,E) \omega_2^{-1} (\tilde g,g)=1
\]
we obtain $\omega_2(g,\tilde g)=\omega_2(\tilde g,g)$. On the other hand, from (\ref{ww1}), $\omega_2(g,\tilde g)\omega_2(\tilde g, g)=1$, so we have
$
\omega_2^2(g,\tilde g)=1,
$ which yields the constraint,
\begin{eqnarray}\label{const1}
\omega_2(g,\tilde g)=\pm1.
\end{eqnarray}
Notice that there are no gauge degrees of freedom for $\omega_2(g,\tilde g)$, because the value of the coboundary
$(d\Omega_{1})(g,\tilde g)=\Omega_{1}(g)\Omega_{1}(\tilde g)=1$ has been fixed after (\ref{musolution}).

Furthermore, from the relations
\begin{eqnarray*}
&&M(g_1)M(g_2) = M(g_1g_2) \omega_2(g_1,g_2),\\
&&M(\tilde g_2)M(\tilde g_1) = M(\widetilde{g_1g_2}) \omega_2(\tilde g_2, \tilde g_1)\\
\end{eqnarray*}
and $M(g)M(\tilde g)=\omega_2(g,\tilde g)$, we can obtain
\begin{eqnarray*}\label{wwbar}
\omega_2(g_1,g_2) \omega_2(\tilde g_2,\tilde g_1)=\omega_2(g_1,\tilde g_1)\omega_2(g_2,\tilde g_2)\omega_2(g_1g_2,\widetilde{g_1g_2}).
\end{eqnarray*}
Comparing with (\ref{ww1}), we have
$
\omega_2(\tilde g_1,\tilde g_2) =\omega_2(\tilde g_2,\tilde g_1)\omega_2(g_1,\tilde g_1)\omega_2(g_2,\tilde g_2)\omega_2(g_1g_2,\widetilde{g_1g_2}),
$
or equivalently
\begin{eqnarray}\label{const2}
\omega_2(g_1,g_2) &=&\omega_2(g_2,g_1)\omega_2(g_1,\tilde g_1)\nonumber\\
&&\times\omega_2(g_2,\tilde g_2)\omega_2(g_1g_2,\widetilde{g_1g_2}).
\end{eqnarray}

Since $H$ is abelian, it must be a cyclic group or a direct product of cyclic groups. We will consider the two cases separately.

\subsubsection{When $H=Z_N$}

We first consider the case $H$ is a cyclic group $Z_N$ generated by $g$ with $g^N=E$. Since $Z_2\rtimes Z_2^T=Z_2\times Z_2^T$, we assume $N\geq 3$.

Suppose that in the projective Rep of $H\rtimes Z_2^T$, the Rep matrix of $g$ is $M(g)$. We tune the phase factor of $M(g)$ such that \[[M(g)]^N=1.\]  Obviously, $M(g^n)$ is proportional to $[M(g)]^n$ by a phase factor. Suppose
\begin{eqnarray*}
M(g^n)= [M(g)]^n \mu(n),
\end{eqnarray*}
where $\mu(n)$ is an $U(1)$ phase factor and obviously $\mu(0)=\mu(N)=1$. From
\begin{eqnarray*}
M(g^m)M(g^n)&=&M(g^n)M(g^m)\\
&=&M(g)^{m+n}\mu(m)\mu(n)\\
&=&M(g^{m+n})\mu(m)\mu(n)/\mu(m+n)
\end{eqnarray*}
we have
\begin{eqnarray}\label{omgOMG}
\omega_2(g^m,g^n)=\omega_2(g^n,g^m)=\frac{\mu(m)\mu(n)}{\mu(m+n)}.
\end{eqnarray}

From $\omega_2(g^m,g^n)=\omega_2(g^n,g^m)$ and (\ref{const2}) we have
\begin{eqnarray}\label{Z2Rep}
\omega_2(g^m,\tilde g^m)\omega_2(g^n,\tilde g^n)\omega_2(g^{m+n},\tilde g^{m+n})=1
\end{eqnarray}
From (\ref{const1}) and (\ref{Z2Rep}), we can see that $\omega_2(g^m,\tilde g^m)$ form a $Z_2$ Rep of the group $H$, namely, they are real numbers and are classified by $\mathcal H^1(H,Z_2)$.

In the following we discuss $N$=even and $N$=odd separately.

{\bf 1), $N$ is odd: $N=2Q+1, \ Q\in Z$}

Since $\mathcal H^{1}(Z_{2Q+1},Z_2)=Z_1$, there are no nontrivial solutions for $\omega_2(g^m,\tilde g^m)$, we can safely set
\[\omega_2(g^m,\tilde g^m)=1\] in the following discussion.

Since $\mu(m)\mu(-m)=\omega_2(g^m,\tilde g^{m})=1$, for each pair of $g^m,\tilde g^m$, only one $\mu(m)$ is free. Remembering that for each pair of $g^m, \tilde g^m$ we have a free coboundary degrees of freedom, so we can further gauge $\mu(m)$ away by introducing
\begin{eqnarray*}
&&M'(g^m)=M(g^m)\Omega_1(g^m),\\
&&M'(g^{-m})=M(g^{-m})\Omega_1^{-1}(g^m)
\end{eqnarray*}
with $\Omega_1(g^m)=\mu^{-1}(m)$ for $1\leq m\leq Q$,  such that
\begin{eqnarray*}
&&\mu'(m)=\mu(m)\Omega_1(g^m)=1,\\
&&\mu'(-m)=\mu(-m)\Omega_1^{-1}(g^m)=\omega_2(g^m,\tilde g^m)=1,\\
&&\mu'(0)=\mu(0)=1,\\
&&\mu'(2Q+1)=\mu(2Q+1)=1,
\end{eqnarray*}
then we have
\[
M'(g^m)M'(g^n)=M'(g^{m+n})\omega'_2(g^m,g^n),
\]
\[\omega'_2(g^m,g^n)=\mu'(m)\mu'(n)/\mu'(m+n),\]
where $\omega'_2(g^m,g^n)=1$ is a trivial cocycle. So $M'(g^m)$ form a linear Rep of $H=Z_{2Q+1}$.

Until now, all the coboundary degrees of freedom are fixed (except for $\Omega_1(T)$, which is not important since it does not change any cocycle values, so it can be set as 1). So we conclude that the classification of projective Reps of $Z_{2Q+1}\rtimes Z_2^T$ is purely determined by $\mathcal H^2(Z_2^T,U(1))=\mathbb Z_2$. Namely,
\[
\mathcal H^2(Z_{2Q+1}\rtimes Z_2^T,U(1))=  \mathcal H^2(Z_2^T,U(1))=\mathbb Z_2.
\]

{\bf 2), $N$ is even: $N=4Q, \ Q\in Z$.}

Since
\begin{eqnarray*}
&&M(g^m)M(g^m)M(\tilde g^m)M(\tilde g^m)\\
 &\ \ &= \omega_2(g^m,g^m)\omega_2(\tilde g^m,\tilde g^m)\omega_2(g^{2m},\tilde g^{2m})\\
&\ \ &=\omega_2(g^m,\tilde g^m)^2=1,
\end{eqnarray*}
owing to (\ref{ww1}) we have
\[\omega_2(g^{2m},\tilde g^{2m})=1.\]

Further, from (\ref{omgOMG}), since $g^{N}=E$ and $g^{N/2}=\tilde g^{N/2}$ , $\mu(N)=\mu(N/2)^2/\omega_2(g^{N/2},\tilde g^{N/2})=1$, we have
$\mu^2(N/2)=\mu(2Q)^2=1$(here we have used $\omega_2(g^{N/2},\tilde g^{N/2})=\omega_2(g^{2Q},\tilde g^{2Q})=1$), which yields
\[
\mu(N/2)=\mu(2Q)=\pm1.
\]

Since there is remaining $Z_2$ coboundary degrees of freedom for $\Omega_1(g^{2Q})$ after fixing $\omega_2(T,g^{2Q})=\omega_2(g^{2Q},T)=1$ (see (\ref{Z2gauge})), we can choose $\Omega_1(g^{2Q})=\mu^{-1}(2Q)$ and redefine $M'(g^{2Q})=M(g^{2Q})\Omega_1(g^{2Q})=M(g)^{2Q}$ such that $\mu'(2Q)=\mu(2Q)\Omega_1(g^{2Q})=1$.

Similar to the discussion in case (1), we can introduce the following conboundary to gauge away $\mu(m)$, here we constrain $-2Q< m< 2Q$:
\begin{eqnarray*}
&&M'(g^m)=M(g^m)\Omega_1(g^m),\\
&&M'(g^{-m})=M(g^{-m})\Omega_1^{-1}(g^m)
\end{eqnarray*}
with $\Omega_1(g^m)=\mu^{-1}(m)$ such that
\begin{eqnarray*}
&&\mu'(m)=\mu(m)\Omega_1(g^m)=1,\\
&&\mu'(-m)=\mu(-m)\Omega_1^{-1}(g^m)=\omega_2(g^m,\tilde g^m),\\
&&\mu'(0)=\mu'(4Q)=1,\\
&&\mu'(2Q)=1,
\end{eqnarray*}
then
\[
M'(g^m)M'(g^n)=M'(g^{m+n})\omega'_2(g^m,g^n),
\]
where \[\omega'_2(g^m,g^n)=\mu'(m)\mu'(n)/\mu'(m+n)\] with $\mu'(m\pm4Q)=\mu'(m)$.

Until now, all the cocycles are set to be real and all the coboundary degrees of freedom are fixed (except for $\Omega_1(T)$, which is not important since it does not change any cocycle values, so it can be set as 1). This means that the $H$ part of the 2-cocycles are classified by $\mathcal H^2(H,Z_2)$. Notice that all the nontrivial 2-cocycles are related to $\omega_2(g,\tilde g)$, which is classified by $\mathcal H^1(H,Z_2)$.  So we conclude that for the group $Z_{4Q}\rtimes Z_2^T$,
\begin{eqnarray}
\mathcal H^2(H\rtimes Z_2^T, U(1))&=&\mathcal H^2(Z_2^T, U(1))\times\mathcal H^2(H, Z_2)\nonumber \\
&=& \mathbb Z_2\times \mathcal H^1(H, Z_2)\\
&=& \mathbb Z_2^2.\nonumber
\end{eqnarray}

{\it Remark: When $\omega_2(T,T)=1$, the nontrivial sign $\omega_2(g^m,\tilde g^m)=-1$ stands for a Kramers doublet. The reason is the following. Notice that $(Tg^m)^2=E$ and
\[
\omega_2(Tg^m,Tg^m)=\omega_2(Tg^m,\tilde g^m T) =\omega_2^{-1}(g^m,\tilde g^m)=-1,
\]
we have
\[[M(Tg^m)K]^2 = - M[(Tg^m)^2]=-1.\] So the nontrivial sign $\omega_2(g^m,\tilde g^m)=-1$ corresponds to a Kramers doublet.
}

{\bf 3), $N$ is even: $N=4Q+2, \ Q\in Z.$}

The discussion is similar to case (2), except that $\mu(N/2)=\mu(2Q+1)$ is not necessarily $\pm1$, it can be complex:
\[
\mu^2(2Q+1)=\omega_2(g^{2Q+1},\tilde g^{2Q+1})=\pm1.
\]
But the conclusion is the same, since we can introduce the following coboundary for $-Q\leq m\leq Q$:
with $\Omega_1(g^{2m+1})=\mu^{-1}(2m+1)\mu(2Q+1)$ and $\Omega_1(g^{2m})=\mu^{-1}(2m)$
such that
\begin{eqnarray*}
&&\mu'(2m+1)=\mu(2m+1)\Omega_1(g^{2m+1})=\mu(2Q+1),\\
&&\mu'(-2m-1)=\mu(-2m-1)\Omega_1^{-1}(g^{2m+1})\\&&\ \ \ \ \ \ \ \ \ \ \ \ \ \ \ \ \ \ =\omega_2(g^{2m+1},\tilde g^{2m+1})\mu^{-1}(2Q+1),\\
&&\mu'(2m)=\mu(2m)\Omega_1(g^{2m})=1,\\
&&\mu'(-2m)=\mu(-2m)\Omega_1^{-1}(g^{2m})=\omega_2(g^{2m},\tilde g^{2m}),\\
&&\mu'(0)=\mu'(4Q+2)=1,\\
&&\mu'(2Q+1)=\mu(2Q+1),
\end{eqnarray*}
then
\[
M'(g^m)M'(g^n)=M'(g^{m+n})\omega'_2(g^m,g^n),
\]
where \[\omega'_2(g^m,g^n)=\mu'(m)\mu'(n)/\mu'(m+n)\] with $\mu'(m\pm(4Q+2))=\mu'(m)$. Notice that all $\omega_2(g^m,g^n)$ are real numbers which are classified by $\mathcal{H}^{2}(H,Z_{2})$.

So, similar to case (2), we conclude that, for the group $Z_{4Q+2}\rtimes Z_2^T$,
\begin{eqnarray}
\mathcal H^2(H\rtimes Z_2^T, U(1))&=&\mathcal H^2(Z_2^T, U(1))\times\mathcal H^2(H, Z_2)\nonumber \\
&=& \mathbb Z_2\times \mathcal H^1(H, Z_2)\\
&=& \mathbb Z_2^2.\nonumber
\end{eqnarray}

{\it Remark: When approaching $U(1)$ by taking the limit $\lim_{N\rightarrow\infty}Z_{N}=U(1)$, the result depends on if $N$ is even or odd. Actually, careful analysis shows that $\mathcal{H}^{2}(U(1)\rtimes Z_{2}^{T},U(1))=\mathbb{Z}_{2}$. The reason is the following.

Suppose that the generator of projective Rep of $U(1)$ is an Hermitian matrix $N$, then we can assume
\begin{equation}
M(U_{\theta})=e^{iN\theta}f(\theta)
\end{equation}
where $U_{\theta}$, $\theta\in[0,2\pi)$ is an element in $U(1)$, and $|f(\theta)|=1$ to ensure that $M(U_{\theta})$ is unitary. Obviously $f(0)=1$. We further assume that $T$ is represented by $M(T)K$, then the gauge condition $\omega_{2}(T,\theta)=\omega_{2}(-\theta,T)=1$ and $TU_{\theta}=U_{-\theta}T$ indicate that
\begin{eqnarray*}
M(T)Ke^{iN\theta}f(\theta)&=&M(T)e^{-iN^{\ast}\theta}f^{\ast}(\theta)K \\
  &=&e^{-iN\theta}f(-\theta)M(T)K.
\end{eqnarray*}
Above equation is true for arbitrary $\theta$, so we have
\[M(T)N^{\ast}=NM(T)~,~f^{\ast}(\theta)=f(-\theta).
\]
Noticing that
\[\omega_{2}(\theta,-\theta)=\frac{f(\theta)f(-\theta)}{f(0)}=f^{\ast}(\theta)f(\theta)=1,
\]
 which is under the same condition with the group $Z_{2Q+1}$.So the classification is purely determined by $\mathcal{H}^{2}(Z_{2}^{T},U(1))=\mathbb{Z}_{2}$.

However, for the group $U(1)\times Z_{2}^{T}$, we have $TU_{\theta}=U_{\theta}T$ and
\begin{eqnarray*}
M(T)Ke^{iN\theta}f(\theta)&=&M(T)e^{-iN^{\ast}\theta}f^{\ast}(\theta)K \\
  &=&e^{iN\theta}f(\theta)M(T)K,
\end{eqnarray*}
which means
\[-M(T)N^{\ast}=NM(T)~,~f^{\ast}(\theta)=f(\theta).
\]
So $f(\theta)$ is real $f(\theta)=\pm1$, and $\omega_{2}(\theta,\theta')=\frac{f(\theta)f(\theta')}{f(\theta+\theta')}=\pm1$. This is consistent with the discussion in section \ref{sec:directg},where it is shown that all the cocycles $\omega_{2}(\theta,\theta')$ can be fixed as real and is classified by $\mathcal H^2(U(1),Z_2)$. The complete classification of $U(1)\times Z_{2}^{T}$ is then given by
\begin{equation}\label{}
\mathcal H^2(U(1)\times Z_{2}^{T},U(1))=\mathcal H^2(Z_{2}^{T},U(1))\times\mathcal H^2(U(1),Z_2)=\mathbb Z_2^2.
\end{equation}
}

\subsubsection{When $H$ is a direct product of cyclic groups}

Now suppose the group $H=Z_M\times Z_N$ has more than one generator, noted as $h_1,h_2, ...$. For each cyclic subgroup, the argument in the previous section applies, and all cocycles $\omega_2(h_1^m, h_1^n)$ and $\omega_2(h_2^m, h_2^n)$ are constrained to be $\pm1$ by introducing gauge transformation $\Omega_1(h_1^m)$ and $\Omega_1(h_2^m)$ as discussed previously.

In the following we will discuss about the cocycles $\omega_2(h_1^m,h_2^n)$. In the coboundary
\[
(d\Omega_1)(h_2^n,h_1^m) =  \Omega_1(h_1^m) \Omega_1(h_2^n)\Omega_1^{-1}(h_2^n h_1^m),
\]
if we set $\Omega_1(h_2^n h_1^m)= \Omega_1(h_2^n) \Omega_1(h_1^m)\omega_2( h_2^n,h_1^m)$, where the values of $\Omega_1(h_2^n)$ and $\Omega_1(h_1^m)$ have been fixed in previous gauge fixing processes, then $\omega_2( h_2^n,h_1^m)$ can be fixed:
\[
\omega_2'( h_2^n,h_1^m) =\omega_2( h_2^n,h_1^m) (d\Omega_1)(h_2^n,h_1^m) =1.
\]

From Eqs.~(\ref{const2}) and (\ref{const1}), we have
\begin{eqnarray*}
\omega_2(h_1^m, h_2^n)&=&\omega_2( h_2^n,h_1^m) \omega_2(h_1^m,\tilde h_1^m) \omega_2(h_2^n,\tilde h_2^n) \\
&& \cdot\omega_2(h_1^mh_2^n,\widetilde{h_1^mh_2^n})\\
&=&\omega_2(h_1^m,\tilde h_1^m) \omega_2(h_2^n,\tilde h_2^n) \omega_2(h_1^mh_2^n,\widetilde{h_1^mh_2^n})\\
&=&\pm1.
\end{eqnarray*}
Thus all the components with the form $\omega_2(h_1^m, h_2^n)$ and $\omega_2(h_2^m, h_1^n)$ are set to be real numbers.

Owing to the cocycle equations $(d\omega_2)(h_1^m,h_2^n,h_2^p)={\omega_2(h_2^n,h_2^p)\omega_2(h_1^m,h_2^{n+p})\over\omega_2(h_1^mh_2^n,h_2^p)\omega_2(h_1^m,h_2^n)}=1$, we have $\omega_2(h_1^mh_2^n,h_2^p)={\omega_2(h_2^n,h_2^p)\omega_2(h_1^m,h_2^{n+p})\over \omega_2(h_1^m,h_2^n)}$, which is a real number. Similarly, owning to cocycle equations, all the components $\omega_2(h_1^mh_2^n,h_1^qh_2^p)$ take values $\pm1$ under above gauge fixing. Now it is clear that the cocycles $\omega_2(g_1,g_2)$ (where $g_1,g_2\in H$) are classified by $\mathcal H^2(H,Z_2)$. This conclusion can be generalized to the case where $H$ is a direct product of more cyrclic groups $H=Z_M\times Z_N\times...$.

In conclusion, the  complete classification is given by
\begin{eqnarray}
\mathcal H^2(H\rtimes Z_2^T, U(1))&=&\mathcal H^2(Z_2^T, U(1))\times\mathcal H^2(H, Z_2)\nonumber \\
&=& \mathbb Z_2\times \mathcal H^2(H, Z_2)
\end{eqnarray}

\subsection{The case of complex cocycles}\label{complexcocycle}

For a general group $G=H\rtimes Z_2^T$, the cocycles may not be tuned into real numbers. For example,  $\mathcal H^2(Z_3\times (Z_3\rtimes Z_2^T), U(1))=\mathbb Z_6$, 2 classes of the cocycles can be set as real numbers, and the remaining 4 classes are complex numbers.

It can be shown that for $G=Z_m\times (Z_n\rtimes Z_2^T)$,
\Beq
&&\mathcal H^2(G,U(1)) = \mathcal H^2(Z_m,Z_2)\times \mathcal H^2(Z_n,Z_2)\\&&\ \ \ \ \ \ \ \ \ \ \ \ \ \ \ \ \ \ \ \
 \times \mathcal H^2(H,U(1))\times \mathcal H^2(Z_2^T,U(1)),
\Eeq
where $H=Z_m\times Z_n$.


\section{Proof of some statements} \label{app: prove}

\subsection{The Dimension of irreducible projective Rep is a multiple of the period of its factor system}\label{app:Dim_Period}

For a symmetry group $G$, its irreducible projective Rep matrix is $M(g)$, then
\begin{equation}\label{irredrep}
M(g_{1})M_{s(g_{1})}(g_{2})=\omega_{2}(g_{1},g_{2})M(g_{1}g_{2}).
\end{equation}
We denote the determinant of Rep matrix as following
\Beq
&&\det [ M(g_{1})]=\varepsilon_{1},\\
&&\det [M(g_{2})]=\varepsilon_{2},\\
&&\det [M(g_{1}g_{2})]=\varepsilon_{12}.
\Eeq
From (\ref{irredrep}) we have $\varepsilon_{1}\cdot (\varepsilon_{2})_{s(g_1)} = \omega_{2}^{D}(g_{1},g_{2})\varepsilon_{12}$, where $(\varepsilon_{2})_{s(g_1)}= \varepsilon_{2}$ if $g_1$ is unitary and $(\varepsilon_{2})_{s(g_1)} = \varepsilon_{2}^*$ if $g_1$ is anti-unitary. This yields
\begin{eqnarray*}
\omega_{2}(g_{1},g_{2}) = e^{i\frac{2\pi n}{D}}\times \left[\frac{\varepsilon_{1}\cdot (\varepsilon_{2})_{s(g_1)}}{\varepsilon_{12}}\right]^{\frac{1}{D}},
\end{eqnarray*}
where $D$ is the lowest dimension of the irreducible projective Reps corresponding to the factor system $\omega_2(g_1,g_2)$ and $n=0,1,\cdots,D-1$. We may choose a coboundary $\Omega_{2}(g_{1},g_{2})=\left[\frac{\varepsilon_{12}}{\varepsilon_{1}\cdot (\varepsilon_{2})_{s(g_1)}}\right]^{\frac{1}{D}}$
to get a different factor system
\begin{eqnarray}\label{}
\omega_{2}'(g_{1},g_{2})=  \Omega_{2}(g_{1},g_{2})\omega_{2}(g_{1},g_{2})=e^{i\frac{2\pi n}{D}},
\end{eqnarray}
which satisfies
\beq\label{D}
\omega_{2}'(g_{1},g_{2})^D=1.
\eeq

Now we define the period $P$ of a factor system $\omega_2(g_1,g_2)$ as the smallest positive integer number such that $\omega_2(g_1,g_2)^P$ is a trivial class of 2-cocycle.
$P$ can be read out from the group structure of the classification of 2-cocycles, i.e. the second group cohomology of $G$\cite{WangGuWen16}. For example, if the second group cohomology of $G$ is $\mathcal H^2(G, U(1))=\mathbb Z_m\times \mathbb Z_n$, and the cocycle $\omega_2(g_1,g_2)$ belongs to the $(a,b)$th class, where $a,b\in Z$ and $1\leq a\leq m$ and $1\leq b\leq n$ and the $(m,n)$th class stands for the trivial class.
Then the period of cocycle $\omega_2(g_1,g_2)$ is \[P=\left[ {[a,m]\over a}, {[b,n]\over b}\right],\] where $[a,m]$ is the least common multiple of $a$ and $m$.


From the relation (\ref{D}), we conclude that the dimension $D$ of any projective Rep with factor system $\omega_2(g_1,g_2)$ must be integer times of $P$,
\[D= NP,\]
 where $N\in   Z$ and $N\geq1$.

\subsection{Completeness of irreducible bases}\label{app6}
Since
\beq
|g^r\rangle=\hat g|\alpha_i^{(\nu)}\rangle=\sum_{j=1}^{n_\nu}M(g)_{ji}K_{s(g)}|\alpha_j^{(\nu)}\rangle,
\eeq
the complete relation $\sum_g |g^{r}\rangle\langle g^{r}|\propto I$ can be written as
\begin{equation}
\sum_{g}\sum_{k,j=1}^{n_\nu}M(g)_{ki}|\alpha_{k}^{\nu}\rangle\langle \alpha_{j}^{\nu}|M^{\dagger}(g)_{ij}\propto1,
\end{equation}
which is equivalent to
\begin{equation}\label{comprel1}
\sum_{g}M(g)_{ki}M^{\dag}(g)_{ij}\propto \delta_{kj},
\end{equation}
where we assume $M(g)$ represents equivalent irreducible unitary projective Reps.
To prove (\ref{comprel1}), we introduce the matrix
\beq\label{matrixM}
 Y&=&\sum_{g}M(g)K_{s(g)}X K_{s(g)}M^\dag (g)\nonumber\\
&=&\sum_{g}M(g)X M^\dag (g),
\eeq
where the matrix $X$ is real.

For any $g_n\in G$, from (\ref{MMProj}), the matrix $Y$ satisfies
\beq
&&M(g_{n})K_{s(g_n)}Y\nonumber\\
&=&\sum_{g}M(g_{n})K_{s(g_n)}M(g)K_{s(g)}X K_{s(g)}M^\dag (g) \nonumber\\
&=&\sum_{g}M(g_{n}g)\omega_2(g_n,g)K_{s(g_n g)}X K_{s(g)}M^\dag(g) K_{s(g_n)}\nonumber\\& &\cdot M^\dag(g_n)
M(g_n)K_{s(g_n)}\nonumber\\
&=&\sum_{g}M(g_{n}g)\omega_2(g_n,g)K_{s(g_n g)}X\nonumber\\& &\cdot
 [M(g_n)K_{s(g_n)}M(g)K_{s(g)}]^{\dag}M(g_n)K_{s(g_n)}\nonumber\\
&=&\sum_{g}M(g_{n}g)\omega_2(g_n,g)K_{s(g_n g)}X \nonumber\\& &
\cdot[\omega_2(g_n,g)M(g_ng)K_{s(g_n g)}]^\dag M(g_n)K_{s(g_n)}\nonumber\\
&=&\sum_{g}M(g_{n}g)\omega_2(g_n,g)K_{s(g_n g)}X K_{s(g_n g)}\omega_2^\ast(g_n,g)\nonumber\\& &\cdot
M^\dag(g_n g)M(g_n)K_{s(g_n)}\nonumber\\
&=&\sum_{g}M(g_{n}g)K_{s(g_n g)}X K_{s(g_n g)}M^\dag(g_n g)\cdot
M(g_n)K_{s(g_n)}\nonumber\\
&=&YM(g_{n})K_{s(g_n)},\label{MMgn}
\eeq
which means $ [\hat g_n, Y]=0$.

Supposing $X$ is a diagonal matrix with only one nonzero element $X_{ii}=1$, then from Eq.~(\ref{matrixM}), $Y$ is an Hermitian matrix with
\Beq
Y_{kj}
&=&\sum_{g}M(g)_{ki}M^{\dag}(g)_{ij},\\
\Eeq

Since the projective representation $M(g_n)K_{s(g_n)}$ is irreducible, according to the generalized Schur lemma, $Y$ must be a constant matrix, i.e. $Y=\lambda I$, where $\lambda $ is a real number. Therefore, $Y_{kj}=\lambda \delta_{kj}$,
and the relation (\ref{comprel1}) is proved.

 \subsection{Two relations between class operators of anti-unitary groups}

Here we prove the following two relations between the class operators.

(1) $C_i=\bar C_i$ for anti-unitary groups, where $C_i$ is the class operator corresponding to the unitary element $g_i$.

Suppose $G$ is anti-unitary, $g_3\in G$ is an arbitrary element, and $C_1$ is a class operator corresponding to a unitary element $\hat g_1$. Then
\Beq
\bar C_1|g_3\rangle &=& \left[ \sum_{g_2\in G}\hat {\bar g}_2\hat {\bar g}_1\hat {\bar g}_2^{-1}\right] \hat{\bar g}_3|E\rangle \\
&=&\sum_{g_2\in G}e^{-is({g_3})\theta_2(g_2^{-1},g_2)}\left\{\left[(\hat {\bar g}_2\hat {\bar g}_1)\widehat {\overline {g_2^{-1}}}\right] \hat{\bar g}_3\right\}|E\rangle
\Eeq
\Beq
			       &=&\sum_{g_2\in G}e^{-is({g_3})\theta_2(g_2^{-1},g_2)}e^{is({g_2g_3})\theta_2(g_1,g_2)}\nonumber\\&&\times
			       e^{is({g_3})\theta_2(g_2^{-1},g_1g_2)} e^{i\theta_2(g_3,g_2^{-1}g_1g_2)}|g_3g_2^{-1}g_1g_2\rangle,
\Eeq
here we have used $\widehat {\overline {g_2^{-1}}}|g_3\rangle=e^{is({g_3})\theta_2(g_2^{-1},g_2)} \hat {\bar g}_2^{-1}|g_3\rangle$.

 On the other hand, from
\Beq
\widehat{g_{2}^{-1}}\hat{g}_{2}=M(g_{2}^{-1})M_{s(g_2)}(g_2)=e^{i\theta_{2}(g_{2}^{-1},g_2)},
\hat{g}_{2}=M(g_2)K_{s(g_2)},
\Eeq
we get
\Beq
\hat{g}_{2}^{-1}&=&M_{s(g_2)}^{-1}(g_2)K_{s(g_2)}=e^{-i\theta_{2}(g_{2}^{-1},g_2)}M(g_{2}^{-1})K_{s(g_2)}\\
&=&e^{-i\theta_{2}(g_{2}^{-1},g_2)}\widehat{g_{2}^{-1}}.
\Eeq

So, we have
\Beq
C_1|g_3\rangle &=& \left[ \sum_{g_2\in G}\hat {g}^{-1}_2\hat { g}_1\hat { g}_2\right] \hat{ g}_3|E\rangle\nonumber\\
			&=& \hat{ g}_3\left[ \sum_{g_2\in G}\hat {g}^{-1}_2 \hat { g}_1\hat { g}_2\right]|E\rangle\nonumber\\ 
			&=& \hat{ g}_3\left[ \sum_{g_2\in G} e^{-i\theta_2(g_2^{-1},g_2)}\widehat {g^{-1}_2}\hat { g}_1\hat { g}_2\right]|E\rangle\\
&=&\sum_{g_2\in G}e^{-is({g_3})\theta_2(g_2^{-1},g_2)}e^{i\theta_2(g_3,g_2^{-1}g_1g_2)}\nonumber\\&&\times
e^{is({g_3})\theta_2(g_2^{-1},g_1g_2)} e^{is({g_2g_3})\theta_2(g_1,g_2)}|g_3g_2^{-1}g_1g_2\rangle.
\Eeq
Therefore, $C_1=\bar C_1$.

(2) $i(C_i^H-C_{\tilde i}^H) = (\bar C_i^H-\bar C_{\tilde i}^H)S$, where $S=i(P_u-P_a)$.

Under the gauge transformation $\Omega_1(g)=e^ { i{\pi\over2}(\delta_{g,g_{1}} -\delta_{g,\tilde g_{1}})} $  ($g_1$ is unitary), from the relations (\ref{U_transf}) and  (\ref{intr U trans}), the regular projective Reps will be changed into
\Beq
&&M(g_1) \to U^\dag iM(g_1) U, \\
&&M(\tilde g_1) \to -U^\dag iM(\tilde g_1) U, \\
&&M(\bar g_1) \to U^\dag M(\bar g_1)  [i(P_u-P_a)] U=U^\dag M(\bar g_1) S U, \\
&&M(\bar {\tilde g}_1) \to U^\dag M(\bar {\tilde g}_1)  [-i(P_u-P_a)]U= -U^\dag M(\bar {\tilde g}_1)  SU, \\
&&M(g_2) \to U^\dag M(g_2) U,  \ \ \  g_2\neq g_1,\tilde g_1\\
&&M(\bar g_2) \to U^\dag M(\bar g_2) U,  \ \ \  g_2\neq g_1,\tilde g_1
\Eeq
where $U_{g_i,g_j}=\delta_{g_i,g_j}\Omega_1(g_i)$ is a diagonal matrix.

Consequently, the class operators $C_1= C_1^H + C_{\tilde 1}^H,\ \bar C_1=\bar C_1^H + \bar C_{\tilde 1}^H$ are transformed into
\Beq
&&C_1\to U^\dag i(C_1^H - C_{\tilde 1}^H)U,
\Eeq
\Beq
&&\bar C_1\to U^\dag (\bar C_1^H - \bar C_{\tilde 1}^H)SU.
\Eeq

After the gauge transformation, above two quantities are still equal, namely, $U^\dag i(C_1^H - C_{\tilde 1}^H)U= U^\dag (\bar C_1^H - \bar C_{\tilde 1}^H)SU$. This yields
\[
i(C_1^H - C_{\tilde 1}^H) = (\bar C_1^H - \bar C_{\tilde 1}^H)S.
\]


\begin{table*}[htbp]
\caption{Irreducible linear Reps of some anti-unitary groups. The symbols $\sigma_{x,y,z}$ are Pauli matrices, and $\omega=e^{i\frac{2\pi}{3}}$, $N_{uc}$ is the number of independent unitary class operators, namely,  half of the total number of class operators. The multiplicity is the number of times each irreducible Rep occurs in the regular Rep. }\label{tb2}
\centering
\begin{tabular}{ |c||ccc|c| }
\hline
 $Z_{2}\times Z_{2}^{T}$ &$T$&$P$& &multiplicity\\
 ($N_{uc}=2$)&&&&\\
 \hline
 & $1K$ & 1     & & 2 \\
 & $1K$ & $-1$& & 2 \\
 \hline
$Z_{4}^{T}$ &$T$& & & \\
($N_{uc}=2$)&&&&\\
 \hline
 & $1K$& & &2$$\\
 & $\sigma_{y}K$& & &$1$\\
\hline
$Z_{4}\times Z_{2}^{T}$ &$T$ &$P$&&  \\
($N_{uc}=3$)&&&&\\
\hline
&$1K$ & 1& &2 \\
&$1K$ & $-1$& &2 \\
&$\sigma_{x}K$ &$i\sigma_{z}$ & &  2 \\
\hline
$Z_{4}\rtimes Z_{2}^{T}$ &$T$ &$P$ & &\\ %
($N_{uc}=4$)&&&&\\
\hline
&$1K$&  $i$ & & 2 \\ %
&$1K$& $-1$ & & 2 \\ %
&$1K$& $-i$ & & 2 \\ %
&$1K$& $1$ & & 2 \\ %
\hline
$Z_{3}\times(Z_{3}\rtimes Z_{2}^{T})\simeq$ &$T$ & $P$&$Q$ &\\
$(Z_3\times Z_3)\rtimes Z_2^T$\ \ ($N_{uc}=6$ ) & &&&  \\
\hline
&$1K$ &1 &1 &2 \\
&$1K$ &1&$\omega$ &2 \\
&$1K$ &1&$\omega^2$ &2 \\
& $\sigma_{x}K$ &$\left(\begin{array}{cc} \omega&0\\0&\omega^2\end{array}\right)$ &$ I$&2 \\%
& $\sigma_{x}K$ &$\left(\begin{array}{cc} \omega&0\\0&\omega^2\end{array}\right)$ &$ \omega I$&2 \\%
& $\sigma_{x}K$ &$\left(\begin{array}{cc} \omega&0\\0&\omega^2\end{array}\right)$ &$\omega^2I$ &2 \\%
\hline
$A_4\rtimes Z_2^T\simeq S_{4}^{T}\simeq \mathscr T_{d}^{T}$ &$T=(12)K$&$P=(123)$ &$Q=(124)$& \\%
($N_{uc}=4$ )&&&&\\
\hline
&$1K$ &1&1& 2\\
&$1K$ &$\omega^2$&$\omega$&2 \\
&$1K$ &$\omega$&$\omega^2$& 2\\
& $\left(\begin{array}{ccc}
1& 0&0 \\0 &1&0\\0&0&1\end{array}\right)K$&$\left(\begin{array}{ccc}1& 0&0 \\0 &\omega^2&0\\0&0&\omega\end{array}\right)$& $\frac{1}{3}\left(\begin{array}{ccc}-1& 2\omega^{\frac{1}{2}} &2\omega^{-\frac{1}{2}} \\2\omega^{\frac{1}{2}}  &\omega^{-\frac{1}{2}} &2\\2\omega^{-\frac{1}{2}} &2&\omega^{\frac{1}{2}} \end{array}\right)$& $6$\\%
\hline
\end{tabular}
\end{table*}

\section{Reduction of regular projective Reps}\label{app5}
In this section, we use the procedure in section \ref{Reduce} to reduce the regular projective Reps of some important groups, including anti-unitary groups. We only present results of generators. 
\subsection{Unitary groups} 

For unitary groups, it is known that if a Rep ($\nu$) is irreducible, then its characters must satisfy
\[
\sum_{g\in G}|\chi^{(\nu)}_g|^2 =G,
\]
where $\chi^{(\nu)}_g$ is the character of element $g\in G$. This result also holds for projective Reps.

In the following, we will illustrate the reduction of regular projective Reps via an example $Z_{2}\times Z_{2}\times Z_{2}$. The Abelian group $Z_{2}\times Z_{2}\times Z_{2}$ has eight group elements, its canonical subgroup chain is $Z_{2}\times Z_{2}\times Z_{2}\supset Z_{2}\times Z_{2}\supset Z_{2}$. From $2$-cocycle and $2$-coboundary equations (\ref{2cocycle}),(\ref{2coboundary}), we find there are seven classes of $8$-dimensional non-trivial regular projective
Reps of $Z_{2}\times Z_{2}\times Z_{2}$. We only present how to reduce one class ($+1,-1,-1$)
. The nontrivial cocycle solutions are
\Beq
&&\omega_{2}(P,P)=\omega_{2}(P,Q)=\omega_{2}(P,R)=\omega_{2}(P,PQR)=-1,
\\
&&\omega_{2}(Q,P)=\omega_{2}(Q,Q)=\omega_{2}(Q,R)=\omega_{2}(Q,PQR)=-1,\\
&&\omega_{2}(PR,P)=\omega_{2}(PR,Q)=\omega_{2}(PR,R)\\&&
=\omega_{2}(PR,PQR)=-1,
\Eeq
\Beq
&&\omega_{2}(QR,P)=\omega_{2}(QR,Q)=\omega_{2}(QR,R)\\&&
=\omega_{2}(QR,PQR)=-1.
\Eeq
The regular projective Rep matrices of three generators of the group are
\Beq
M(P)&=&\left( \begin{array}{cccccccc}
0 & -1 & 0 & 0 & 0 & 0 & 0 & 0\\ 1 & 0 & 0 & 0 & 0 & 0 & 0 & 0\\
0 & 0 & 0 & 1 & 0 & 0 & 0 & 0\\ 0 & 0& -1 & 0 & 0 & 0 & 0 & 0\\
0 & 0 & 0 & 0 & 0 & 1 & 0 & 0\\ 0 & 0 & 0 & 0 & -1 & 0 & 0 & 0\\
0 & 0 & 0 & 0 & 0 & 0 & 0 & -1\\ 0 & 0& 0 & 0 & 0 & 0 & 1 & 0 \end{array} \right),
\Eeq
\Beq
M(Q)&=&\left( \begin{array}{cccccccc}
0 & 0 & -1 & 0 & 0 & 0 & 0 & 0\\ 0 & 0 & 0 & 1 & 0 & 0 & 0 & 0\\
1 & 0 & 0 & 0 & 0 & 0 & 0 & 0\\ 0 & -1& 0 & 0 & 0 & 0 & 0 & 0\\
0 & 0 & 0 & 0 & 0 & 0 & 1 & 0\\ 0 & 0 & 0 & 0 & 0 & 0 & 0 & -1\\
0 & 0 & 0 & 0 & -1 & 0 & 0 & 0\\ 0 & 0& 0 & 0 & 0 & 1 & 0 & 0 \end{array} \right),\\
M(R)&=& \left( \begin{array}{cccccccc}
0 & 0 & 0 & 0 & 1 & 0 & 0 & 0\\ 0 & 0 & 0 & 0 & 0 & 1 & 0 & 0\\
0 & 0 & 0 & 0 & 0 & 0 & 1 & 0\\ 0 & 0& 0 & 0 & 0 & 0 & 0 & 1\\
1 & 0 & 0 & 0 & 0 & 0 & 0 & 0\\ 0 & 1 & 0 & 0 & 0 & 0 & 0 & 0\\
0 & 0 & 1 & 0 & 0 & 0 & 0 & 0\\ 0 & 0& 0 & 1 & 0 & 0 & 0 & 0 \end{array} \right).
\Eeq
From Eqs. (\ref{CSCO1}) and (\ref{CSCO2}), the matrices of all the class operators are given below:
\Beq
C&=&\left( \begin{array}{cccccccc}
0 & 0 & 0 & 1 & 0 & 0 & 0 & 0\\ 0 & 0 & 1 & 0 & 0 & 0 & 0 & 0\\
0 & 1 & 0 & 0 & 0 & 0 & 0 & 0\\ 1 & 0& 0 & 0 & 0 & 0 & 0 & 0\\
0 & 0 & 0 & 0 & 0 & 0 & 0 & 1\\ 0 & 0 & 0 & 0 & 0 & 0 & 1 & 0\\
0 & 0 & 0 & 0 & 0 & 1 & 0 & 0\\ 0 & 0 & 0 & 0 & 1 & 0 & 0 & 0 \end{array} \right),\\
C(s_{1})&=&\left( \begin{array}{cccccccc}
0 & 0 & -4 & 0 & 0 & 0 & 0 & 0\\ 0 & 0 & 0 & 4 & 0 & 0 & 0 & 0\\
4 & 0 & 0 & 0 & 0 & 0 & 0 & 0\\ 0 & -4& 0 & 0 & 0 & 0 & 0 & 0\\
0 & 0 & 0 & 0 & 0 & 0 & 4 & 0\\ 0 & 0 & 0 & 0 & 0 & 0 & 0 & -4\\
0 & 0 & 0 & 0 & -4 & 0 & 0 & 0\\ 0 & 0& 0 & 0 & 0 & 4 & 0 & 0 \end{array} \right),\\
C(s_{2})&=& \left( \begin{array}{cccccccc}
0 & -2 & 0 & 0 & 0 & 0 & 0 & 0\\ 2 & 0 & 0 & 0 & 0 & 0 & 0 & 0\\
0 & 0 & 0 & 2 & 0 & 0 & 0 & 0\\ 0 & 0& -2& 0 & 0 & 0 & 0 & 0\\
0 & 0 & 0 & 0 & 0 & 2 & 0 & 0\\ 0 & 0 & 0 & 0 & -2 & 0 & 0 & 0\\
0 & 0 & 0 & 0 & 0 & 0 & 0 & -2\\ 0 & 0& 0 & 0 & 0 & 0 & 2 & 0 \end{array} \right),
\Eeq
\Beq
\bar C(s_{1})&=&\left( \begin{array}{cccccccc}
0 & 0 & -4 & 0 & 0 & 0 & 0 & 0\\ 0 & 0 & 0 & 4 & 0 & 0 & 0 & 0\\
4 & 0 & 0 & 0 & 0 & 0 & 0 & 0\\ 0 & -4& 0 & 0 & 0 & 0 & 0 & 0\\
0 & 0 & 0 & 0 & 0 & 0 & -4 & 0\\ 0 & 0 & 0 & 0 & 0 & 0 & 0 & 4\\
0 & 0 & 0 & 0 & 4 & 0 & 0 & 0\\ 0 & 0& 0 & 0 & 0 & -4 & 0 & 0 \end{array} \right),\\
\bar{C}(s_{2})&=& \left( \begin{array}{cccccccc}
0 & -2 & 0 & 0 & 0 & 0 & 0 & 0\\ 2 & 0 & 0 & 0 & 0 & 0 & 0 & 0\\
0 & 0 & 0 & 2 & 0 & 0 & 0 & 0\\ 0 & 0& -2& 0 & 0 & 0 & 0 & 0\\
0 & 0 & 0 & 0 & 0 & -2 & 0 & 0\\ 0 & 0 & 0 & 0 & 2 & 0 & 0 & 0\\
0 & 0 & 0 & 0 & 0 & 0 & 0 & 2\\ 0 & 0& 0 & 0 & 0 & 0 & -2 & 0 \end{array} \right).
\Eeq
Practically, we can diagonalize the linear combination $\widehat{O}=C+aC(s_{1})+bC(s_{2})+c\bar{C}(s_{1})+d\bar{C}(s_{2})$, here $a$,$b$,$c$ and $d$ are arbitrary nonzero real constants ensuring all the eigenvalues of $\widehat{O}$ are non-degenerate. From the orthonormal eigenvectors (column vectors) of $\widehat{O}$, we can obtain
a unitary transformation matrix
\begin{equation}\label{}
U=\left( \begin{array}{cccccccc}
\frac{1}{2}& 0 & \frac{1}{2}i & 0 &-\frac{1}{2}i& 0 & 0 &\frac{1}{2}i\\
-\frac{1}{2}i & 0 &-\frac{1}{2}& 0 &\frac{1}{2}& 0 & 0 &\frac{1}{2}\\
\frac{1}{2}i & 0 &\frac{1}{2}& 0 &\frac{1}{2}& 0 & 0 &\frac{1}{2}\\
-\frac{1}{2} & 0 &-\frac{1}{2}i& 0 &-\frac{1}{2}i& 0 & 0 &\frac{1}{2}i\\
 0 &\frac{1}{2}i & 0 &-\frac{1}{2}i & 0 &\frac{1}{2}&\frac{1}{2}i & 0\\
0 &\frac{1}{2} & 0 &\frac{1}{2} & 0 &\frac{1}{2}i&\frac{1}{2} & 0\\
0 &-\frac{1}{2} & 0 &-\frac{1}{2} & 0 &\frac{1}{2}i&\frac{1}{2} & 0\\
0 &-\frac{1}{2}i & 0 &\frac{1}{2}i & 0 &\frac{1}{2}&\frac{1}{2}i & 0 \end{array} \right).
\end{equation}
The above three generators can be transformed into,respectively:
\beq
U^{\dag}M(P)U&=&\left( \begin{array}{cccccccc}
i & 0 & 0 & 0 & 0 & 0 & 0 & 0\\ 0 & -i & 0 & 0 & 0 & 0 & 0 & 0\\
0 & 0 & -i & 0 & 0 & 0 & 0 & 0\\ 0 & 0& 0 & i & 0 & 0 & 0 & 0\\
0 & 0 & 0 & 0 & -i & 0 & 0 & 0\\ 0 & 0 & 0 & 0 & 0 & i & 0 & 0\\
0 & 0 & 0 & 0 & 0 & 0 & -i & 0\\ 0 & 0& 0 & 0 & 0 & 0 & 0 & i \end{array} \right) ,\\
U^{\dag}M(Q)U&=&\left( \begin{array}{cccccccc}
-i & 0 & 0 & 0 & 0 & 0 & 0 & 0\\ 0 & i & 0 & 0 & 0 & 0 & 0 & 0\\
0 & 0 & i & 0 & 0 & 0 & 0 & 0\\ 0 & 0& 0 & -i & 0 & 0 & 0 & 0\\
0 & 0 & 0 & 0 & -i & 0 & 0 & 0\\ 0 & 0 & 0 & 0 & 0 & i & 0 & 0\\
0 & 0 & 0 & 0 & 0 & 0 & -i & 0\\ 0 & 0& 0 & 0 & 0 & 0 & 0 & i \end{array} \right),
\eeq
\beq
U^{\dag}M(R)U&=&\left( \begin{array}{cccccccc}
0 & i & 0 & 0 & 0 & 0 & 0 & 0\\ -i & 0 & 0 & 0 & 0 & 0 & 0 & 0\\
0 & 0 & 0 & -1 & 0 & 0 & 0 & 0\\ 0 & 0& -1 & 0 & 0 & 0 & 0 & 0\\
0 & 0 & 0 & 0 & 0 & i & 0 & 0\\ 0 & 0 & 0 & 0 & -i & 0 & 0 & 0\\
0 & 0 & 0 & 0 & 0 & 0 & 0 & 1\\ 0 & 0& 0 & 0 & 0 & 0 & 1 & 0 \end{array} \right).
\eeq

\subsection{Anti-unitary groups}
For anti-unitary groups, as mentioned in section \ref{sec:antiU}, there is a one-to-one correspondence between the anti-unitary class operator $C_{i^{\pm}T}$ and the unitary class operator $C_{i^{\pm}}$,  where the unitary class operators can be obtained solely from the unitary normal subgroup $H$ by the relation (\ref{classG1}) and (\ref{classG2}). Moreover, the anti-unitary class operators do not provide meaningful quantum numbers. So the independent class operators include only the unitary class operators $C_{i^{+}}$ and $C_{i^{-}}$. We may use their linear combination $C=\sum_i (k_i C_{i^{+}}+k'_i C_{i^-})$ to form CSCO-I of the regular projective Rep, where $k_i$ and $k'_i$ are arbitrary nonzero real constants.

Notice that the characters for the unitary subgroup $H$ are still good quantities. Depending on the Rep matrices of the normal subgroup $H$, there are three different situations for the irreducible Rep $(\nu)$ of anti-unitary group $G$:

1) the induced Rep of $H$ in $(\nu)$ [we will note it as $(\nu_H)$] is irreducible.  In this case, the characters of $H$ satisfy,
\[
\sum_{g\in H}|\chi^{(\nu)}_g|^2 =H,
\]
where $H$ is the order the normal subgroup $H$;

2) the induced Rep of $H$ in $(\nu)$ is a direct sum of the induced irreducible {\it linear} Rep $(\nu_H)$ and its complex conjugate $(\nu_H^*)$, where $(\nu_H)$ is not equivalent to $(\nu_H^*)$, then the characters of $H$ satisfy,
\[
\sum_{g\in H}|\chi^{(\nu)}_g|^2 = 2H;
\]

3) the Rep matrices of $H$ in $(\nu)$ is a direct sum of two copies of the induced irreducible Rep ($\nu_H$) [in this case all the characters of ($\nu_H$) are real],
 then the characters of $H$ satisfy,
\[
\sum_{g\in H}|\chi^{(\nu)}_g|^2 = 4H.
\]

In case 3), the eigenvalues of the unitary class operators in CSCO-I are real, and there exists at least one intrinsic class operator of anti-unitary elements in CSCO-III whose eigenvalues are complex numbers.  It should be carefully checked if 
there exists another set of CSCO-III where all the eigenvalues are real; if true, then $(\nu)$ can be further reduced. For example, as shown in Sec.\ref{Z2Z2Z2T}, the projective Rep of class  ($+1,-1,+1,+1$) of the group $Z_{2}\times Z_{2}\times Z_{2}^{T}$ is reducible though the eigenvalues of $\widehat{\overline T}$ are complex numbers.

\subsubsection{Abelian anti-unitary group $Z_{2}\times Z_{2}\times Z_{2}^{T}$}\label{Z2Z2Z2T}
As for the Abelian group $Z_2\times Z_2\times Z_2^T = \{E, P\}\times \{E, Q\}\times \{E, T\}$, where $T^{2}=E$. Solving $2$-cocycle and $2$-coboundary equations (\ref{2cocycle}),(\ref{2coboundary}), there are $15$ classes of $8$-dimensional projective Reps of this group. Because the group contains anti-unitary time reversal transformation, we have the following mapping:
\begin{equation}\label{}
T\rightarrow M(T)K ~,~ P\rightarrow M(P) ~,~ Q\rightarrow M(Q),
\end{equation}
where $K$ is the complex-conjugate operator defined in section \ref{sec1}. The normal subgroup $H$ is $D_2=Z_2\times Z_2$. Since $T^{-1} g_i T=\tilde{g}_i=g_i$, it is clear that $C_{i^-}=0$ from (\ref{classG2}). Therefore, we can only use $C=\sum_i k_i C_{i^+}$ to construct CSCO-I. The CSCO-II are $(C, C(s))$, where $C(s)$ are class operators of the subgroup $Z_2 \subset H$. We discuss only two classes which can be reduced into 2-dimensional Rep. For the class ($+1,+1,+1,-1$) in Table \ref{tb1}, the nontrivial cocycle solutions in the decoupled factor systems (\ref{fixgauge}) are
\Beq
&&\omega_{2}(P,P)=\omega_{2}(P,Q)=\omega_{2}(P,PT)=\omega_{2}(P,QT)=-1, \\
&& \omega_{2}(Q,P)=\omega_{2}(Q,Q)=\omega_{2}(Q,PT)=\omega_{2}(Q,QT)=-1,\\
&& \omega_{2}(T,T)=\omega_{2}(T,PT)=\omega_{2}(T,QT)  =\omega_{2}(T,PQT)=-1,\\
&& \omega_{2}(PT,P)=\omega_{2}(PT,Q)=\omega_{2}(PT,T)\\&&
=\omega_{2}(PT,PQT)=-1,\\
&& \omega_{2}(QT,P)=\omega_{2}(QT,Q)=\omega_{2}(QT,T)\\&&
=\omega_{2}(QT,PQT)=-1,\\
&& \omega_{2}(PQT,T)=\omega_{2}(PQT,PT)=\omega_{2}(PQT,QT)\\&&
=\omega_{2}(PQT,PQT)=-1.
\Eeq
The regular projective Rep matrices of these generators are given below:
\Beq
M(T)K&=&\left( \begin{array}{cccccccc}
0 & 0 & 0 & 0 & -1 & 0 & 0 & 0\\ 0 & 0 & 0 & 0 & 0 & -1 & 0 & 0\\
0 & 0 & 0 & 0 & 0 & 0 & -1 & 0\\ 0 & 0& 0 & 0 & 0 & 0 & 0 & -1\\
1 & 0 & 0 & 0 & 0 & 0 & 0 & 0\\ 0 & 1 & 0 & 0 & 0 & 0 & 0 & 0\\
0 & 0 & 1 & 0 & 0 & 0 & 0 & 0\\ 0 & 0& 0 & 1 & 0 & 0 & 0 & 0 \end{array} \right)K,
\Eeq
\Beq
M(P)&=&\left( \begin{array}{cccccccc}
0 & -1 & 0 & 0 & 0 & 0 & 0 & 0\\ 1 & 0 & 0 & 0 & 0 & 0 & 0 & 0\\
0 & 0 & 0 & 1 & 0 & 0 & 0 & 0\\ 0 & 0& -1 & 0 & 0 & 0 & 0 & 0\\
0 & 0 & 0 & 0 & 0 & -1 & 0 & 0\\ 0 & 0 & 0 & 0 & 1 & 0 & 0 & 0\\
0 & 0 & 0 & 0 & 0 & 0 & 0 & 1\\ 0 & 0& 0 & 0 & 0 & 0 & -1 & 0 \end{array} \right),
\Eeq
\Beq
M(Q)&=& \left( \begin{array}{cccccccc}
0 & 0 & -1 & 0 & 0 & 0 & 0 & 0\\ 0 & 0 & 0 & 1 & 0 & 0 & 0 & 0\\
1 & 0 & 0 & 0 & 0 & 0 & 0 & 0\\ 0 & -1& 0 & 0 & 0 & 0 & 0 & 0\\
0 & 0 & 0 & 0 & 0 & 0 & -1 & 0\\ 0 & 0 & 0 & 0 & 0 & 0 & 0 & 1\\
0 & 0 & 0 & 0 & 1 & 0 & 0 & 0\\ 0 & 0& 0 & 0 & 0 & -1 & 0 & 0 \end{array} \right).
\Eeq

When $g_i=E,P,Q,PQ$, we can get $2$ independent class operators $C_{i^{+}}$, then $C=\sum_{i=1}^{2}k_i C_{i^{+}}$. The class operator of subgroup $Z_2=\{E,P\}\subset H$
can be calculated from Eq.(\ref{CSCO2}):
\Beq
C(s)&=& 2\left( \begin{array}{cccccccc}
0 & -1 & 0 & 0 & 0 & 0 & 0 & 0\\ 1 & 0 & 0 & 0 & 0 & 0 & 0 & 0\\
0 & 0 & 0 & 1 & 0 & 0 & 0 & 0\\ 0 & 0& -1 & 0 & 0 & 0 & 0 & 0\\
0 & 0 & 0 & 0 & 0 & -1 & 0 & 0\\ 0 & 0 & 0 & 0 & 1 & 0 & 0 & 0\\
0 & 0 & 0 & 0 & 0 & 0 & 0 & 1\\ 0 & 0& 0 & 0 & 0 & 0 & -1 & 0 \end{array} \right).
\Eeq
As emphasized in section \ref{sec:antiU}, we can adopt $\widehat{\overline T}$ and the class operator of
intrinsic subgroup $\overline{Z_2}=\{\overline{E},\overline{P}\}$ as members of $\overline{ C}(s')$  because of the accidental symmetry $[\widehat{\overline T}, \widehat{\overline P}] = 0$: 
\Beq
\overline{C}_{\overline{P}}&=&2\left( \begin{array}{cccccccc}
0 & -1 & 0 & 0 & 0 & 0 & 0 & 0\\ 1 & 0 & 0 & 0 & 0 & 0 & 0 & 0\\
0 & 0 & 0 & 1 & 0 & 0 & 0 & 0\\ 0 & 0& -1 & 0 & 0 & 0 & 0 & 0\\
0 & 0 & 0 & 0 & 0 & -1 & 0 & 0\\ 0 & 0 & 0 & 0 & 1 & 0 & 0 & 0\\
0 & 0 & 0 & 0 & 0 & 0 & 0 & 1\\ 0 & 0& 0 & 0 & 0 & 0 & -1 & 0 \end{array} \right),\\
M(\overline{T})&=&\left( \begin{array}{cccccccc}
0 & 0 & 0 & 0 & -1 & 0 & 0 & 0\\ 0 & 0 & 0 & 0 & 0 & -1 & 0 & 0\\
0 & 0 & 0 & 0 & 0 & 0 & -1 & 0\\ 0 & 0& 0 & 0 & 0 & 0 & 0 & -1\\
1 & 0 & 0 & 0 & 0 & 0 & 0 & 0\\ 0 & 1 & 0 & 0 & 0 & 0 & 0 & 0\\
0 & 0 & 1 & 0 & 0 & 0 & 0 & 0\\ 0 & 0& 0 & 1 & 0 & 0 & 0 & 0 \end{array} \right).
\Eeq
We need to diagonalize the linear combination of above class operators $\widehat{O}=C+aC(s)+b\overline{C}_{\overline{P}}+cM(\overline{T})$, where $a,b,c$ are arbitrary nonzero real constants.  
The eigenvalues of operator $\widehat{O}$ are non-degenerate,
thus
the orthonormal eigenvectors constitute the unitary transformation matrix
\beq\label{}
U=\frac{\sqrt{2}}{4}\left( \begin{array}{cccccccc}
-i&i&-i&i&-1& -1 &-i &i\\
-1&-1&1&1&i&-i&1&1\\
1&1&-1&-1&i&-i&1&1\\
i&-i&i&-i&-1&-1&-i&i\\
1&1&1&1&-i&i&1&1\\
-i&i&i&-i&-1&-1&i&-i\\
i&-i&-i&i&-1&-1&i&-i\\
-1&-1&-1&-1&-i&i&1&1\end{array} \right).
\eeq
The above Rep matrices of three generators can be transformed into:
\beq
U^{\dag}M(T)KU&=&\left( \begin{array}{cccccccc}
0 & -i & 0 & 0 & 0 & 0 & 0 & 0\\ i & 0 & 0 & 0 & 0 & 0 & 0 & 0\\
0 & 0 & 0 & -i & 0 & 0 & 0 & 0\\ 0 & 0& i & 0 & 0 & 0 & 0 & 0\\
0 & 0 & 0 & 0 & 0 & -i & 0 & 0\\ 0 & 0 & 0 & 0 & i & 0 & 0 & 0\\
0 & 0 & 0 & 0 & 0 & 0 & 0 & -i\\ 0 & 0& 0 & 0 & 0 & 0 & i & 0 \end{array} \right)K ,\nonumber\\
\\
U^{\dag}M(P)U&=&\left( \begin{array}{cccccccc}
i & 0 & 0 & 0 & 0 & 0 & 0 & 0\\ 0 & -i & 0 & 0 & 0 & 0 & 0 & 0\\
0 & 0 & -i & 0 & 0 & 0 & 0 & 0\\ 0 & 0& 0 & i & 0 & 0 & 0 & 0\\
0 & 0 & 0 & 0 & i & 0 & 0 & 0\\ 0 & 0 & 0 & 0 & 0 & -i & 0 & 0\\
0 & 0 & 0 & 0 & 0 & 0 & -i & 0\\ 0 & 0& 0 & 0 & 0 & 0 & 0 & i \end{array} \right),\\
U^{\dag}M(Q)U&=&\left( \begin{array}{cccccccc}
-i & 0 & 0 & 0 & 0 & 0 & 0 & 0\\ 0 & i & 0 & 0 & 0 & 0 & 0 & 0\\
0 & 0 & i & 0 & 0 & 0 & 0 & 0\\ 0 & 0& 0 & -i & 0 & 0 & 0 & 0\\
0 & 0 & 0 & 0 & i & 0 & 0 & 0\\ 0 & 0 & 0 & 0 & 0 & -i & 0 & 0\\
0 & 0 & 0 & 0 & 0 & 0 & -i & 0\\ 0 & 0& 0 & 0 & 0 & 0 & 0 & i \end{array} \right) .
\eeq

Now we consider another class ($+1,-1,+1,+1$), under the decoupled factor systems (\ref{fixgauge}), the nontrivial cocycle solutions are
\Beq
&&\omega_{2}(P,P)=\omega_{2}(P,PQ)=\omega_{2}(P,PT)=\omega_{2}(P,PQT)=-1, \\
&& \omega_{2}(Q,P)=\omega_{2}(Q,Q)=\omega_{2}(Q,PT)=\omega_{2}(Q,QT)=-1,\\
&& \omega_{2}(PQ,Q)=\omega_{2}(PQ,PQ)=\omega_{2}(PQ,QT)\\&&
=\omega_{2}(PQ,PQT)=-1,\\
&& \omega_{2}(T,T)=\omega_{2}(T,PT)=\omega_{2}(T,QT)\\&&
=\omega_{2}(T,PQT)=-1,\\
&& \omega_{2}(PT,P)=\omega_{2}(PT,PQ)=\omega_{2}(PT,T)\\&&
=\omega_{2}(PT,QT)=-1,\\
&& \omega_{2}(QT,P)=\omega_{2}(QT,Q)=\omega_{2}(QT,T)\\&&
=\omega_{2}(QT,PQT)=-1,\\
&& \omega_{2}(PQT,T)=\omega_{2}(PQT,PT)=\omega_{2}(PQT,Q)\\&&
=\omega_{2}(PQT,PQ)=-1.
\Eeq
The regular projective Rep matrices of these generators can be obtained:
\Beq
M(T)K&=&\left( \begin{array}{cccccccc}
0 & 0 & 0 & 0 & -1 & 0 & 0 & 0\\ 0 & 0 & 0 & 0 & 0 & -1 & 0 & 0\\
0 & 0 & 0 & 0 & 0 & 0 & -1 & 0\\ 0 & 0& 0 & 0 & 0 & 0 & 0 & -1\\
1 & 0 & 0 & 0 & 0 & 0 & 0 & 0\\ 0 & 1 & 0 & 0 & 0 & 0 & 0 & 0\\
0 & 0 & 1 & 0 & 0 & 0 & 0 & 0\\ 0 & 0& 0 & 1 & 0 & 0 & 0 & 0 \end{array} \right)K,\\
\Eeq
\Beq
M(P)&=&\left( \begin{array}{cccccccc}
0 & -1 & 0 & 0 & 0 & 0 & 0 & 0\\ 1 & 0 & 0 & 0 & 0 & 0 & 0 & 0\\
0 & 0 & 0 & -1 & 0 & 0 & 0 & 0\\ 0 & 0& 1 & 0 & 0 & 0 & 0 & 0\\
0 & 0 & 0 & 0 & 0 & -1 & 0 & 0\\ 0 & 0 & 0 & 0 & 1 & 0 & 0 & 0\\
0 & 0 & 0 & 0 & 0 & 0 & 0 & -1\\ 0 & 0& 0 & 0 & 0 & 0 & 1 & 0 \end{array} \right),\\
M(Q)&=& \left( \begin{array}{cccccccc}
0 & 0 & -1 & 0 & 0 & 0 & 0 & 0\\ 0 & 0 & 0 & 1 & 0 & 0 & 0 & 0\\
1 & 0 & 0 & 0 & 0 & 0 & 0 & 0\\ 0 & -1& 0 & 0 & 0 & 0 & 0 & 0\\
0 & 0 & 0 & 0 & 0 & 0 & -1 & 0\\ 0 & 0 & 0 & 0 & 0 & 0 & 0 & 1\\
0 & 0 & 0 & 0 & 1 & 0 & 0 & 0\\ 0 & 0& 0 & 0 & 0 & -1 & 0 & 0 \end{array} \right).
\Eeq

When $g_i=E,P,Q,PQ$, we can get only one independent class operator $C_{i^{+}}=E$. The class operator of subgroup $Z_2=\{E,P\}\subset H$ can be calculated from Eq.(\ref{CSCO2}):
\Beq
C(s)&=&2 \left( \begin{array}{cccccccc}
0 & -1 & 0 & 0 & 0 & 0 & 0 & 0\\ 1 & 0 & 0 & 0 & 0 & 0 & 0 & 0\\
0 & 0 & 0 & -1 & 0 & 0 & 0 & 0\\ 0 & 0& 1 & 0 & 0 & 0 & 0 & 0\\
0 & 0 & 0 & 0 & 0 & -1 & 0 & 0\\ 0 & 0 & 0 & 0 & 1 & 0 & 0 & 0\\
0 & 0 & 0 & 0 & 0 & 0 & 0 & -1\\ 0 & 0& 0 & 0 & 0 & 0 & 1 & 0 \end{array} \right).
\Eeq
As for the class operators of intrinsic subgroup $ \overline{C}(s')$, if we follow the original procedure to use the class operator $\widehat{\overline T}$ and class operator $\overline{C}_{\overline{P}}$ of intrinsic unitary subgroup $\overline{Z_2}$ as members of $\overline{C}(s')$:
\Beq
&&M(\overline{T})=\left( \begin{array}{cccccccc}
0 & 0 & 0 & 0 & -1 & 0 & 0 & 0\\ 0 & 0 & 0 & 0 & 0 & -1 & 0 & 0\\
0 & 0 & 0 & 0 & 0 & 0 & -1 & 0\\ 0 & 0& 0 & 0 & 0 & 0 & 0 & -1\\
1 & 0 & 0 & 0 & 0 & 0 & 0 & 0\\ 0 & 1 & 0 & 0 & 0 & 0 & 0 & 0\\
0 & 0 & 1 & 0 & 0 & 0 & 0 & 0\\ 0 & 0& 0 & 1 & 0 & 0 & 0 & 0 \end{array} \right),\\
&&\overline{C}_{\overline{P}}= 2\left( \begin{array}{cccccccc}
0 & -1 & 0 & 0 & 0 & 0 & 0 & 0\\ 1 & 0 & 0 & 0 & 0 & 0 & 0 & 0\\
0 & 0 & 0 & 1 & 0 & 0 & 0 & 0\\ 0 & 0& -1 & 0 & 0 & 0 & 0 & 0\\
0 & 0 & 0 & 0 & 0 & -1 & 0 & 0\\ 0 & 0 & 0 & 0 & 1 & 0 & 0 & 0\\
0 & 0 & 0 & 0 & 0 & 0 & 0 & 1\\ 0 & 0& 0 & 0 & 0 & 0 & -1 & 0 \end{array} \right),
\Eeq
the eigenvalues (quantum numbers) of $aM(\overline{T})+b\overline{C}_{\overline{P}}$ with $a=2, b=0.5$ are ($-i, -i, i, i, -3i, -3i, 3i, 3i$). From these quantum numbers, it seems that there are two 4-dimensional irreducible Reps, one is labeled by $-i, -i, i, i$ and the other by $-3i, -3i, 3i, 3i$.




However, we can choose another set of class operators such that the eigenvalues are all real numbers.  Noticing $\widehat{\overline{PQT}}$ anti-commutes with $M(\overline{PT})$, we adopt the following `class' operators:
\Beq
M(\overline{PT})&=&\left( \begin{array}{cccccccc}
0 & 0 & 0 & 0 & 0 & 1 & 0 & 0\\ 0 & 0 & 0 & 0 & -1 & 0 & 0 & 0\\
0 & 0 & 0 & 0 & 0 & 0 & 0 & -1\\ 0 & 0& 0 & 0 & 0 & 0 & 1 & 0\\
0 & -1 & 0 & 0 & 0 & 0 & 0 & 0\\ 1 & 0 & 0 & 0 & 0 & 0 & 0 & 0\\
0 & 0 & 0 & 1 & 0 & 0 & 0 & 0\\ 0 & 0& -1 & 0 & 0 & 0 & 0 & 0 \end{array} \right),\\
SM(\overline{PQT})&=&\left( \begin{array}{cccccccc}
0 & 0 & 0 & 0 & 0 & 0 & 0 & i\\ 0 & 0 & 0 & 0 & 0 & 0 & -i & 0\\
0 & 0 & 0 & 0 & 0 & i & 0 & 0\\ 0 & 0& 0 & 0 & -i & 0 & 0 & 0\\
0 & 0 & 0 & i & 0 & 0 & 0 & 0\\ 0 & 0 & -i & 0 & 0 & 0 & 0 & 0\\
0 & i & 0 & 0 & 0 & 0 & 0 & 0\\ -i & 0& 0 & 0 & 0 & 0 & 0 & 0 \end{array} \right)
\Eeq
as members of $\bar C(s')$, where $S=i(P_u-P_a)={\rm diag}(\underbrace{i,...i}_{G/2}, \underbrace{-i,...-i}_{G/2})$ is a diagonal matrix. Due to the accidental symmetry $[S\widehat{\overline {PQT}}, \widehat{\overline{PT}}]=0$ , the above `class' operators commute with each other (the accidental invariant quantity $S\widehat{\overline {PQT}}$ is not a true class operator since it does not commute with $\widehat {\overline T}$).
The eigenvalues (quantum numbers) of $aM(\overline{PT})+bS\widehat{\overline {PQT}}$ with $a=1, b=2$ are ($-1,-1, -3,-3, 1,1, 3,3$). Thus we obtain four 2-dimensional irreducible Reps.
Diagonalizing the linear combination $\widehat{O}=C_{i^{+}}+aC(s)+bM(\overline{PT})+cS\widehat{\overline {PQT}}$ with $a,b,c$ arbitrary nonzero real constants, we obtain the unitary transformation matrix,
\Beq
U=\frac{\sqrt{2}}{4}\left( \begin{array}{cccccccc}
-i&i&1& -1 &i& i &1 &1\\
-1&-1&-i&-i&-1&1&-i&i\\
-1&1&i&-i&1&1&i&i\\
i&i&1&1&i&-i&1&-1\\
1&1&-i&-i&-1&1&i&-i\\
-i&i&-1&1&-i&-i&1&1\\
i&i&-1&-1&-i&i&1&-1\\
1&-1&i&-i&1&1&-i&-i\end{array} \right).
\Eeq
The above Rep matrices of three generators can be transformed into irreducible Reps with $U$:
\Beq
U^{\dag}M(T)KU&=&\left( \begin{array}{cccccccc}
0 & -i & 0 & 0 & 0 & 0 & 0 & 0\\ i & 0 & 0 & 0 & 0 & 0 & 0 & 0\\
0 & 0 & 0 & -i & 0 & 0 & 0 & 0\\ 0 & 0& i & 0 & 0 & 0 & 0 & 0\\
0 & 0 & 0 & 0 & 0 & i & 0 & 0\\ 0 & 0 & 0 & 0 & -i & 0 & 0 & 0\\
0 & 0 & 0 & 0 & 0 & 0 & 0 & -i\\ 0 & 0& 0 & 0 & 0 & 0 & i & 0 \end{array} \right)K,
\Eeq
\Beq
U^{\dag}M(P)U&=&\left( \begin{array}{cccccccc}
i & 0 & 0 & 0 & 0 & 0 & 0 & 0\\ 0 & -i & 0 & 0 & 0 & 0 & 0 & 0\\
0 & 0 & i & 0 & 0 & 0 & 0 & 0\\ 0 & 0& 0 & -i & 0 & 0 & 0 & 0\\
0 & 0 & 0 & 0 & -i & 0 & 0 & 0\\ 0 & 0 & 0 & 0 & 0 & i & 0 & 0\\
0 & 0 & 0 & 0 & 0 & 0 & i & 0\\ 0 & 0& 0 & 0 & 0 & 0 & 0 & -i \end{array} \right), \\
U^{\dag}M(Q)U&=&\left( \begin{array}{cccccccc}
0 & -i & 0 & 0 & 0 & 0 & 0 & 0\\ -i & 0 & 0 & 0 & 0 & 0 & 0 & 0\\
0 & 0 & 0 & i & 0 & 0 & 0 & 0\\ 0 & 0& i & 0 & 0 & 0 & 0 & 0\\
0 & 0 & 0 & 0 & 0 & i & 0 & 0\\ 0 & 0 & 0 & 0 & i & 0 & 0 & 0\\
0 & 0 & 0 & 0 & 0 & 0 & 0 &-i\\ 0 & 0& 0 & 0 & 0 & 0 & -i & 0 \end{array} \right) .
\Eeq

\subsubsection{Non-Abelian anti-unitary group  $S_{4}^{T} $}

For the group $A_4\rtimes Z_2^T\simeq S_{4}^{T}\simeq \mathscr T_d^T$, all the odd-permutation operations are anti-unitary, and all the even-permutation operations (including identity element $E$) form a unitary normal subgroup $H=A_4$.

By solving $2$-cocycle and $2$-coboundary equations (\ref{2cocycle}),(\ref{2coboundary}), we find there are $3$ classes nontrivial projective Reps of this group. Under the decoupled factor systems (\ref{fixgauge}), the $24$-dimensional regular projective Rep matrices of all the  group elements can be obtained.

The unitary independent class operators $C_{i^{+}}$ and $C_{i^{-}}$ can be calculated from (\ref{classG1}) and (\ref{classG2}), where $g_i\in H=A_4$. The linear combination $C=\sum_i (k_i C_{i^{+}}+k'_i C_{i^{-}})$ is exactly the CSCO-I. Together with the class operators $C(s)$ of the subgroup $Z_3$ generated by $P=(123)$, 
we obtain the CSCO-II $(C, C(s))$.

For the intrinsic subgroups, we adopt the anti-unitary class operator $\widehat{\overline{T}}$ and the class operators [of the form (\ref{BarCs})] of the intrinsic subgroup $\overline{ S_3^T}={\overline {Z_3} \rtimes \overline {Z_2^T}}$ with two generators $\overline{P}=\overline{(123)}, \overline{T}=\overline{(12)K}$.
It can be verified that all the class operators commute with each other.

With above CSCO-III, we can completely reduce the regular irreducible Reps. The results about three generators are given in Table \ref{tb1}.




\end{document}